\documentclass{ws-ijmpe}
\usepackage[super,compress]{cite}
\usepackage{graphicx}
\usepackage{amstext,amsfonts}
\usepackage{bm}
\usepackage{xcolor}
\usepackage{hyperref}
\usepackage{braket}
\usepackage{slashed}
\def\slashchar#1{\setbox0=\hbox{$#1$}
   \dimen0=\wd0 \setbox1=\hbox{/} \dimen1=\wd1
   \ifdim\dimen0>\dimen1 \rlap{\hbox to \dimen0{\hfil/\hfil}} #1
   \else  \rlap{\hbox to \dimen1{\hfil$#1$\hfil}} / \fi}

\def\D{\slashchar{D}}
\graphicspath{{./eps/}}
\begin{document}

\markboth{A. Fatima, Z. Ahmad Dar, M. Sajjad Athar, and S. K. Singh}{Photon induced $K\Lambda $ production on the proton in the low energy region}

\catchline{}{}{}{}{}

\title{Photon induced $K\Lambda $ production on the proton in the low energy region}

\author{A. Fatima$^{1}$, Z. Ahmad Dar$^{1,2}$, M. Sajjad Athar$^{1,*}$, and S. K. Singh$^{1}$}
\address{$^{1}$ Department of Physics, \\ Aligarh Muslim University, Aligarh-202002, India }

\address{$^{2}$ Fermi National Accelerator Laboratory, \\
Batavia, Illinois 60510, USA \\
$^{*}$ sajathar@gmail.com}

\maketitle

\begin{history}
\received{Day Month Year}
\revised{Day Month Year}
\end{history}

\begin{abstract}
The associated photoproduction of $K\Lambda$ from the proton in the low energy region is studied using an isobar model in 
which the non-resonant contributions are obtained from the non-linear sigma model with chiral SU(3) symmetry which predicts, in 
a natural way, the contact term with its coupling strength along with the coupling strengths of the various Born terms predicted 
by the non-linear sigma model. The present model is an extension of the non-linear sigma model with chiral SU(2) symmetry, used 
earlier to study the photo, electro, and neutrino productions of pions. In the resonance sector, the contributions from the well 
established nucleon resonances~($R$) in the $s$ channel, the hyperon resonances~($Y^{*}$) in the $u$ channel, and the kaon 
resonances~($K^{*}$ and $K_{1}$) in the $t$ channel having spin $\le \frac{3}{2}$ and mass $<2$ GeV with a significant branching 
ratio in $K\Lambda$ decay mode, have been considered. The strong and electromagnetic couplings of the $s$ channel nucleon 
resonances are taken from experiments while the couplings for the resonances in the $t$ and $u$ channels are fitted to reproduce 
the current data on the associated photoproduction of $K\Lambda$ in this energy region. The numerical results are presented for 
the total and differential cross sections and are compared with the available experimental data from CLAS and SAPHIR 
as well as with some of the recent theoretical models.
\end{abstract}


\ccode{PACS numbers: 13.40.-f, 13.60.−r, 13.60.Le, 13.60.Rj, 14.20.−c, 14.20.Gk, 14.20.Jn,14.40.Aq}

\section{Introduction}\label{Intro}
The theoretical and the experimental study of the associated photoproduction of kaon-hyperon $K Y;~(Y=\Lambda,\Sigma)$ system 
on the proton was started almost 60 years ago~\cite{Kawaguchi, McDaniel:1959zz, McDaniel:1959zz1, Donoho:1958zz, Capps, Groom}. Out of the three 
isospin channels of the kaon-hyperon photoproduction from the proton, {\it viz.} $\gamma + p \longrightarrow K^{+}+ \Lambda,~ 
K^{+}+ \Sigma^{0},~ K^{0} + \Sigma^{+}$, the $K\Lambda$ production is the most studied one. The experimental measurements of 
the cross section and the polarization observables, although initially being scarce with large uncertainties~\cite{Groom,
McDaniel:1959zz, McDaniel:1959zz1, Donoho:1958zz}, are now available with improved statistics and better precision~\cite{ABBHHM, Tran:1998qw, 
SAPHIR, CLAS05, CLAS051, CLAS10, Sumihama:2005er, Lleres:2007tx, Lleres:2007tx1, Jude:2013jzs}. Since the threshold for the $K\Lambda$ production is 1.61 GeV, therefore, 
the study of this process~(and in general, the study of $KY$ production), gives important information about the nucleon 
resonances, lying even in the third and higher resonance regions, which is not available from the study of $N \pi$ and $N \eta$ 
production processes. Unlike the $N \pi$ and $N \eta$ productions where $P_{33}(1232)$~\cite{Tiator:2011pw} and 
$S_{11}(1535)$~\cite{ZPA, Tiator:2018heh, Chiang:2002vq} resonances, respectively, make the dominant contribution, there is no dominant resonance 
contributing to the $K\Lambda$ production and a large number of resonances may couple to this channel~\cite{Mart:2017xtf, 
Skoupil:2016ast, Skoupil:2018vdh, Jan01A, MartSu,Mart:2017mwj,Mart:2019mtq}. There are many resonances predicted in various quark models which are not observed 
in the pion-nucleon or electron-nucleon scattering processes~\cite{CapRob00, Bonn, Bonn1}, and one may get information about these 
resonances from the study of the $K \Lambda$ production process. Along with the real photons, the $K \Lambda$ production has also 
been studied using electrons, where the virtual photon interacts with the proton. Experimentally, the measurements of the cross sections, response functions and the polarization observables for the electron induced $K\Lambda$ production 
process have been done by CLAS~\cite{Ambrozewicz:2006zj, Gabrielyan:2014zun, Carman:2012qj}, MAINZ A1~\cite{Achenbach:2011rf}, 
JLab Hall-C~\cite{Gogami:2013dua} collaborations and several theoretical calculations for this process exist in the 
literature~\cite{SL, Williams:1992tp, SLA, Mart:2010ch, Janssen:2003kk, Skoupil:2018vdh,Mart:2017mwj,Mart:2019mtq}. 

With the availability of high intensity photon and electron beams at the Electron Stretcher System~(ELSA) Germany, Thomas 
Jefferson National Accelerator Facility~(TJNAF) US, Super Photon Ring -- 8 GeV~(SPring-8) Japan, European Synchrotron Radiation 
Facility~(ESRF) France and Mainz Microtron~(MAMI) Germany, it has been possible to precisely measure the cross sections and the 
polarization observables of the $K\Lambda$ channel in SAPHIR~\cite{SAPHIR, Tran:1998qw}, CLAS~\cite{CLAS05, CLAS051, CLAS10}, LEPS~\cite{Sumihama:2005er}, 
GRAAL~\cite{Lleres:2007tx, Lleres:2007tx1} and MAMI-C~\cite{Jude:2013jzs} experiments. There exists some disagreement between the CLAS and the SAPHIR 
data in the differential cross section especially in the forward angle region as well as in the total
cross section in the center of mass~(CM) energy range $W \ge 1.7$ GeV. Moreover, the forward angle data from CLAS 
2006~\cite{CLAS05} and CLAS 2010~\cite{CLAS10} also do not agree with each other in the kinematic region of $W < 1.84$ GeV. 
On the other hand, the data from SAPHIR 1998~\cite{Tran:1998qw}, SAPHIR 2004~\cite{SAPHIR} and MAMI-C~\cite{Jude:2013jzs} are fairly 
consistent with each other in this energy region. 

 The experiments at LEPS~\cite{Sumihama:2005er} and GRAAL~\cite{Lleres:2007tx, Lleres:2007tx1}  have studied the associated production of the strange particles
with polarized photon beams and made measurements on the beam asymmetry and other polarization observables of the final hyperon. Notwithstanding, the
importance of studying these observables in which a considerable amount of data is available, we have not included them in the present work. This is due to our immediate aim of finding a simple model with a minimal number of parameters to describe the total cross section and angular distributions in the photoproduction, which can be extended to weak production of strange particles induced by (anti)neutrinos in $\Delta S=0$ sector by benchmarking the contribution of the vector currents. It is,  therefore, appropriate time that the cross sections of these processes are calculated to complement the current efforts to model the neutrino nucleon cross sections in the few GeV energy region. A theoretical understanding of the total cross section and angular distributions in these weak processes induced by charged current~(CC) and neutral current~(NC) is currently of immense topical interest in modeling the (anti)neutrino cross section in the analysis of the present day neutrino oscillation experiments~\cite{Solomey:2005rs, MINERvA:2006aa} and no recent work has been done 
on these processes in the last 40 years~\cite{Shrock:1975an,Mechlenburg:1976, Amer:1977fy,Dewan:1981ab} except the work of
Adera {\it et al.}~\cite{Adera:2010zz}. However, keeping in mind, the important role of the measurements made on the various polarization observables in
 the study of associated production of strange particles induced by the unpolarized and polarized photons,
we plan to study them in future.

Theoretically $\gamma (\gamma^{*}) p \longrightarrow K^+\Lambda$ process has been studied in various models, for example, the 
quark model~\cite{CapRob00, Bonn, Bonn1, ZpL95, ZpL951, LLP95, FHZ91, JuliaDiaz:2005qj}, chiral perturbative model~\cite{Chpt}, chiral unitary model~\cite{Borasoy},
 coupled channel model~\cite{Giessen, Giessen1, JDiaz, Anisovich:2007bq, JuliaDiaz:2005ju, JuliaDiaz:2005qj}, isobar model~\cite{Thom, Janssen:2003kk, 
 SL, SLA, Mart:1999ed, Mart:2000jv, KM2, Jan01A, Williams:1992tp, Jan01B, Jan02, Mart:2010ch, SF, PSPV, PSPV1, 
Mart:2017xtf, Maxwell, Maxwell1, Skoupil:2016ast, Skoupil:2018vdh, Adelseck:1986fb, Adelseck:1989zt, Adelseck:1990ch, Saghai:1994mm, Williams:1991tw, Williams:1990hh,  Mart:2019mtq}, isobar-Regge hybrid model~\cite{GhentRPR07, GhentRPR071, GhentRPR072, Bydzovsky:2019hgn}, or purely 
Regge models~\cite{Guidal:1997hy}. Among these models, 
one of the widely studied model in recent times is the isobar model developed by various groups, for example, Saclay-Lyon 
(SL)~\cite{SL,SLA}, Kaon-MAID~(KM)~\cite{KM2}, Ghent-Isobar~\cite{Jan01A, Jan01B, Jan02}, BS1~\cite{Skoupil:2016ast}, 
BS3~\cite{Skoupil:2018vdh}, Mart~\cite{Mart:2000jv, MartSu, Mart:2017mwj, Mart:1999ed, Mart:2017xtf, Mart:2010ch} and others~\cite{Adelseck:1986fb, Adelseck:1989zt, 
Adelseck:1990ch, Williams:1991tw, Williams:1990hh, SF, Thom, Maxwell, Maxwell1}. The quark model is based on the quark degrees of freedom and assumes the 
extended structure of the baryons, in which the resonance contribution is taken through the excited states of the quarks. Hence, 
the quark model requires limited number of parameters. In the chiral models, the application of the chiral symmetry treats the 
pseudoscalar meson as the Goldstone boson and the Lagrangians for the meson-baryon system are obtained in the chiral limit. These 
models are best suited to calculate the $K\Lambda$ production in the threshold region but can be extended to higher energies 
using chiral unitary models. In the coupled channel models, the meson-baryon final state interactions are also 
included. For example, the photoproduction of $K\Lambda$ may take place through the primary production of the intermediate states {\it 
i.e.} $\gamma p \rightarrow \pi N,~\eta N,$ etc., leading to the $K \Lambda$ in the final state through the rescattering process. Therefore, the  intermediate state can be any strangeness conserving meson-baryon 
system like $N \pi,~ N \eta, ~ K\Lambda,~ K\Sigma,$ etc. In the isobar models, mostly using an effective Lagrangian approach, 
the hadronic current consists of the non-resonant Born terms~($s$, $t$ and $u$ channels) and the resonance exchanges in $s$, $t$ 
and $u$ channels. In some versions of the isobar models, in which the pseudovector coupling is used for describing the 
meson-nucleon interactions, the contact term also appears. The final state interactions are not considered in most of the isobar 
models as they are based on the effective Lagrangians, except in a few calculations. This is because most of the isobar models 
make use of the phenomenological values for the various electromagnetic and strong couplings which are assumed to simulate the 
effect of the final state interaction. However, in some versions of the isobar models in which a coupled channel analysis is used to treat the final particles, the final state interactions are taken into account~\cite{Anisovich:2007bq, Giessen, Giessen1, JDiaz}. The various isobar models are different from each other in many 
ways and are classified on the basis of their treatment of the non-resonant terms and the resonance terms. 

In the case of non-resonant terms considered in the $s$, $t$ and $u$ channels, the various models based on the effective 
Lagrangian differ in describing the meson-nucleon-hyperon interactions using either the pseudoscalar or the pseudovector 
coupling and the way in which the requirement of gauge invariance is implemented. The extensive studies made in the photo- and 
electro- productions of pions in a wide energy range extending from the threshold to high energies have demonstrated that the 
pseudovector coupling is to be preferred over the pseudoscalar coupling as it reproduces the low energy theorems~(LET) predicted 
by the partially conserved axial vector current~(PCAC) hypothesis and current algebra as a consequence of the chiral symmetry of 
strong interactions and are consistent with the experimental observations~\cite{Adler:1968}. Moreover, the choice of the 
pseudovector coupling generates a contact term in the presence of electromagnetic interactions in a natural way, which facilitates the understanding of LET and helps to 
implement the requirement of gauge invariance. However, in the presence of hadronic form factors at the strong vertex, the 
implementation of gauge invariance necessitates additional assumptions about the momentum dependence of the hadronic form 
factors. On the other hand, in the case of the associated photoproduction of $KY$, both the pseudoscalar and pseudovector 
couplings have been used in many calculations in the absence of any theoretical preference for the pseudovector coupling. This 
is due to the inadequacy of the low energy theorems implied by the pseudovector coupling arising from the slow convergence of 
the low energy expansion~\cite{Sakinah:2019zbd}.

In the case of resonance terms, the difference between various calculations arises mainly due to the number of resonances taken 
into account in the intermediate states and the determination of their electromagnetic couplings to the photons and their strong 
couplings to the meson-nucleon-hyperon systems {\it i.e.} $RKY$. For example, the Saclay-Lyon model~\cite{SL, SLA} has taken into 
consideration, spin $\frac{1}{2}$, $\frac{3}{2}$ and $\frac{5}{2}$ nucleon resonances in the $s$ channel, $K^{*}$ and $K_{1}$ in 
the $t$ channel and spin $\frac{1}{2}$ $\Lambda^{*}$ and $\Sigma^{*}$ resonances in the $u$ channel. The Kaon-MAID 
model~\cite{KM2} uses spin $\frac{1}{2}$ and $\frac{3}{2}$ nucleon resonances in the $s$ channel, $K^{*}$ and $K_{1}$ in the $t$ channel and 
no hyperon resonance in the $u$ channel. The Ghent model~\cite{Jan01A, Jan01B, Jan02} uses three different ways to fit the 
experimental data from SAPHIR~\cite{SAPHIR} for the $K\Lambda$ channel: (i)~assuming SU(3) symmetry and without considering the 
hyperon resonances, (ii)~assuming SU(3) symmetry and with hyperon resonances, and (iii)~without assuming SU(3) symmetry and 
without hyperon resonances. In all the three prescriptions, spin $\frac{1}{2}$ and $\frac{3}{2}$ nucleon resonances are taken in 
the $s$ channel. In BS1 and BS3 models~\cite{Skoupil:2016ast, Skoupil:2018vdh}, spin $\frac{1}{2}$, $\frac{3}{2}$ and $\frac{5} 
{2}$ nucleon resonances are taken in the $s$ channel, $K^{*}$ and $K_{1}$ in the $t$ channel and spin $\frac{1}{2}$ and 
$\frac{3}{2}$ $\Lambda^{*}$ and $\Sigma^{*}$ resonances in the $u$ channel. 

Other than these models, Regge model~\cite{Guidal:1997hy, Guidal:1999qi}, where particles are replaced by their Regge trajectories to 
extend the model to higher energies, is also used to study the $K\Lambda$ production, but its applicability is restricted to 
higher energies~($3 \text{ GeV} \le E_{\gamma} \le 16$ GeV). Also, there are hybrid models that combine the Regge and resonance 
models to study the photo- and electro- productions of strange particles, which describes the data both in the resonance region as well as at high energies~\cite{RPR11, RPR111, Bydzovsky:2019hgn}. 

In this work, we present an isobar model to study the photon induced $K \Lambda$ production on the proton. In this model, an 
effective Lagrangian based on the chiral SU(3) symmetry has been used to obtain the non-resonant terms consisting of $s$, $t$ 
and $u$ channel diagrams and the Lagrangian also generates the contact term as a requirement of the underlying symmetry. The 
electromagnetic couplings are described in terms of the charge and magnetic moment of the baryons like $p$, $\Lambda$ and 
$\Sigma$ occurring in the $s$, $t$ and $u$ channel diagrams. The strong couplings of the meson-nucleon-baryon system like $g_{K 
\Lambda p}$, $g_{K \Sigma p}$, $g_{\gamma K \Lambda p}$, $g_{\gamma K \Sigma p}$ are described in terms of $f_{\pi}$, $D$ and 
$F$ which are determined from the electroweak phenomenology of nucleons and hyperons, where $f_{\pi}$ is the pion decay constant 
and $D$ and $F$, respectively, are the axial vector current couplings of the baryon octet in terms of the symmetric and 
antisymmetric couplings. Therefore, the contribution of the non-resonant terms is calculated without any free parameters except 
the cut-off parameter used to define the form factors at the strong vertex which is taken to be the same for all the background 
terms i.e. non-resonant terms and the resonance terms in the $t$ and $u$ channels. The form factor of the contact term is fixed in terms of the other form factors according to the well known prescription 
given by Davidson and Workman~\cite{DW}. The non-linear sigma model with chiral SU(2) symmetry 
has been earlier used in the calculations of single pion production induced by electron, neutrino and 
antineutrino~\cite{Hernandez:2007qq, Hernandez:2013jka, Alam:2015gaa}, and has been extended to the chiral SU(3) symmetry to 
calculate the single kaon production induced by electron and neutrino~\cite{RafiAlam:2010kf, Alam:2012ry}, single antikaon 
production induced by positron and antineutrino~\cite{Alam:2012ry, Alam:2011xq}, eta production induced by neutrino and 
antineutrino~\cite{Alam:2015zla}. 

 In the resonance sector, we have considered various nucleon, hyperon and kaon resonances giving rise to $K \Lambda$ in the final 
state. Only those nucleon resonances $R$ are taken in the $s$ channel, which are well established and are referred by $****$ and 
$***$ status in the particle data group~(PDG), having spin $\le \frac{3}{2}$, mass in the range $1.6-1.9$ GeV and non-vanishing~($>4-5\%$) 
branching ratio in the $K\Lambda$ decay mode~(see Table~\ref{tab:included_resonances}). In the case of nucleon resonances, the 
electromagnetic couplings $\gamma NR$, are determined in terms of the helicity amplitudes and the strong $RK \Lambda$ couplings are 
determined by the partial decay width of the resonance decaying to $K \Lambda$ using an effective Lagrangian. A form factor of the general dipole form
with a cut-off parameter $\Lambda_R$ taken to be the same for all nucleon resonances in the s channel has been used to describe the $R\Lambda K$ vertex.
 
 In the $u$ channel, two spin $\frac{1}{2}$ hyperon resonances $viz.$ 
$\Lambda^{*}(1405)$ and $\Lambda^{*} (1800)$ and in the $t$ channel, two kaon resonances of spin 1 {\it viz.} $K^{*}(892)$ and $K_{1}(1270)$ are taken into account. The $t$ and $u$ channel resonances along with the non-resonant contributions constitute the background 
part of the hadronic current which are calculated using the effective Lagrangians. Due to the lack of the experimental data on the kaon and hyperon 
resonances, the strong and 
electromagnetic couplings of $u$ and $t$ channel resonances are not well determined phenomenologically and are, therefore, 
varied for fitting the data from CLAS~\cite{CLAS05, CLAS10} and SAPHIR~\cite{Tran:1998qw, SAPHIR} experiments. While doing this fitting, a form factor is taken into account, to be of a general dipole form with a cut-off parameter $\Lambda_B$, to describe the strong  
$R\Lambda K$ vertex. This cut-off parameter $\Lambda_{B}$ is taken to be the same as that has been considered for the Born terms as both contribute to the background terms.

 The calculation of various terms contributing to the background term is done in the lowest order tree-level approximation using the effective
Lagrangians for the non-resonant $s$, $t$ and $u$ channel diagrams and the contact terms as well as the resonance contribution 
 from all the $t$ channel and $u$ channel resonances while the calculations of the different resonance terms is done using the effective Lagrangians for all the $s$ channel nucleon resonances. The present calculation and the many earlier calculations~\cite{Bennhold:1998ib, SL, Williams:1992tp, Bennhold:1996, Ohta:1989ji, Haberzettl:1997jg}
done for this process in the tree level approximation are known to suffer from lack of unitarity as they do not consider the rescattering effects
in the $K\Lambda$ channel or other channels produced in the $\gamma p$ interaction. There are some prescriptions described in the literature to restore the unitarity~\cite{Benmerrouche:1989uc}, in the multichannel coupled channel models~\cite{Kamano:2016bgm, Kamano:2009im, JuliaDiaz:2009ww, Chiang:2004ye, Kamano:2012id, Nakamura:2015rta} and the Watson's treatment method~\cite{Watson:1952ji, Sato:2003rq, Alvarez-Ruso:2015eva}. We have examined the effect 
of restoring unitarity using the energy dependent width of the resonances weighted by the branching ratios of the various decay channels of the
considered resonances following the prescription of Bennhold {\it et al.}~\cite{Bennhold:1998ib}, Mart and Bennhold~\cite{Mart:1999ed, Mart:2000jv} and Skoupil and Bydzovsky~\cite{Skoupil:2018vdh}. The numerical results for the total and differential cross sections with fixed as well as energy dependent decay widths of the nucleon resonances are presented and compared with the experimental 
results available from SAPHIR 1998~\cite{Tran:1998qw}, SAPHIR 2004~\cite{SAPHIR}, CLAS 2006~\cite{CLAS05} and CLAS 2010~\cite{CLAS10}. We have 
also compared the results of the present work for the total and differential cross sections with the various theoretical models available in the current literature, like the 
Regge model~\cite{Guidal:1997hy,  Guidal:1999qi}, chiral perturbation model~\cite{Chpt}, Saclay-Lyon model~\cite{SL, SLA}, Kaon-MAID 
model~\cite{KM2}, Ghent model~\cite{Jan01A, Jan01B, Jan02}, BS1 model~\cite{Skoupil:2016ast}, BS3 model~\cite{Skoupil:2018vdh}, Bonn-Gatchina model~\cite{Anisovich:2012ct, Anisovich:2007bq, Anisovich:2014yza, Anisovich:2017bsk, Anisovich:2017ygb, Nikonov:2007br}, Bonn-Julich model~\cite{Ronchen:2018ury}, and KSU 
model~\cite{Wang:2017cfp, Hunt:2018mrt}.

The major advantage of the formalism developed in the present model is that, it makes use of many physics inputs available from 
various experimental observations on the electroweak and strong interaction phenomenology of mesons and baryons and involves 
very few parameters to reproduce the data. Specifically, the model has the following features:
\begin{itemize}
 \item [i)] The contact term in the non-resonant contribution occurs naturally in the model with the strength of its coupling 
 predicted by the model.
 
 \item [ii)] A general dipole form is used in all the form factors appearing at the strong 
 meson-nucleon-hyperon vertices for all the background terms with a common cut-off parameter $\Lambda_B$. The background terms consist of the $s$, $t$ and $u$ channel Born 
 terms, contact term as well as the $t$ and $u$ channel resonance terms. 
 
 \item [iii)] All the resonances included in the $s$ channel, {\it viz.} $S_{11}(1650)$, $P_{11}(1710)$, $P_{13}(1720)$, $P_{11} 
 (1880)$, $S_{11}(1895)$ and $P_{13} (1900)$, are the well established resonances with definite mass, decay width, branching 
 ratio in $K\Lambda$ channel given in PDG~\cite{PDG}.
 
 \item [iv)] The partial decay width of the resonances~($R$) for decaying into $K\Lambda$ channel from the PDG~\cite{PDG} is 
 used to determine the strength of the strong couplings of the resonance~($R$) to the $K\Lambda$ channel, using an effective 
 Lagrangian approach. We have chosen those resonances which have a branching ratio for decaying in the $K\Lambda$ channel greater than $4-5\%$.
 
 \item [v)] The helicity amplitudes of the resonances $S_{11} (1650)$ and $P_{13}(1720)$ are taken from 
 MAID~\cite{Tiator:2011pw}, and for the rest of the $s$ channel resonances, the helicity amplitudes are taken from 
 PDG~\cite{PDG}~(Table~\ref{tab:param-p2}). These amplitudes are used to determine the strength of the electromagnetic couplings 
 at the $\gamma NR$ vertex.
 
 \item [vi)] A common cut-off parameter $\Lambda_{R}$ is used to describe the hadronic form factors at the strong $RK\Lambda$ 
 vertex in the case of the nucleon resonances constituting the resonance terms in the $s$ channel.
 
 \item [vii)] The coupling strengths of the $t$ and $u$ channel resonances~(see Table~\ref{hyperon_resonances}) are fitted to 
 reproduce the experimental results, keeping the same cut-off parameter $\Lambda_{B}$. The cut-off parameters $\Lambda_{B}$ and $\Lambda_{R}$ are varied to reproduce the experimental results on the total cross sections specially in the low energy region where the data from the SAPHIR and CLAS agree with each other. The total cross sections at higher energies ($W>$1.72 GeV) as well as the angular
distributions as function of $W$ and $\cos \theta_{K}^{CM}$ are predictions of the model.
\end{itemize}

In Sect.~\ref{Formalism}, the formalism for the $K\Lambda$ production induced by the real photons on the proton has been 
presented, where we discuss the contribution to the hadronic current arising due to the non-resonant and resonance diagrams. 
Sect.~\ref{NRB} focuses on the non-resonant terms, determined by the non-linear sigma model assuming the chiral SU(3) symmetry. 
The structure of the nucleon, hyperon and kaon resonances and their couplings are discussed in Sect.~\ref{RES}. The results and 
their discussions are presented in Sect.~\ref{results}, and Sect.~\ref{summary} gives a summary and concludes the present 
findings.

\section{Formalism}\label{Formalism}
In this work, we have studied the $K\Lambda$ photoproduction on the proton,
\begin{equation}\label{reaction}
 \gamma (q) + p (p) \longrightarrow K^{+} (p_{k}) + \Lambda (p^{\prime}),
\end{equation}
where the quantities in the parentheses represent the four momenta of the corresponding particles. In Sect.~\ref{xsec}, we give 
the general discussion for the evaluation of the transition matrix element and cross section in the CM frame. The transition 
matrix element is written in terms of the photon polarization state vector and the hadronic current. The hadronic current receives 
contribution from the background and resonance terms. Following the standard terminology, the background terms consist of all 
the non-resonant terms contributing in the $s$, $t$ and $u$ channels and the contact term as well as the contributions from the 
resonance terms in the $t$ and $u$ channels. The non-resonant terms are determined using the non-linear sigma model and the 
chiral SU(3) symmetry, discussed in Sect.~\ref{NRB} while the contributions from the hyperon and kaon resonances are discussed 
in Sects.~\ref{spin12Y} and \ref{spin1K}, respectively. The nucleon resonances with spin $\frac{1}{2}$ and $\frac{3}{2}$ in the 
$s$ channel constitute the resonance contribution of the hadronic current and are discussed in Sects.~\ref{spin12N} and 
\ref{spin32N}, respectively.

\subsection{Matrix element and cross section}\label{xsec}
The differential cross section for the photoproduction process given in Eq.~(\ref{reaction}) is written as
\begin{eqnarray}\label{eq:sigma_gen}
d\sigma &=& \frac{1}{4 (q\cdot p)} (2 \pi)^{4} \delta^{4}(q+p-p_{k}-p^{\prime}) \frac{d{\vec{p}_{k}}}{(2 \pi)^{3} (2 
E_{k})} \frac{d{\vec p\,}^{\prime}}{(2 \pi)^{3} (2 E_{\Lambda})} \overline{\sum_{r}} \sum | \mathcal M^{r} |^2,
\end{eqnarray}
where $E_{k}$ and $E_{\Lambda}$, respectively, are the energies of the outgoing kaon and lambda. $ \overline{\sum} \sum | 
\mathcal M^{r} |^2$ is the square of the transition matrix element $\mathcal{M}^{r}$, for photon polarization state $r$, 
averaged and summed over the initial and final spin states. $\mathcal{M}^{r}$ is written in terms of the real photon 
polarization vector $\epsilon_{\mu}^{r}$ and the matrix element of the electromagnetic current taken between the hadronic states 
of $\ket{p}$ and $\ket{K\Lambda}$, {\it i.e.}
\begin{equation}
\mathcal{M}^{r} = e \epsilon_{\mu}^{r} (q) \bra{\Lambda(p^{\prime}) K^{+}(p_{k})} {J}^{\mu} \ket{p},
\end{equation}
where $e = \sqrt{4\pi \alpha}$ is the strength of the electromagnetic interaction, with $\alpha = \frac{1}{137}$ being the fine-structure constant. In the case when the photon polarization remains undetected, the summation over all the polarization states 
is performed which gives
\begin{equation}\label{lep}
 \sum_{r=\pm1} \epsilon^{*(r)}_{\mu}\epsilon^{(r)}_{\nu} \longrightarrow - g_{\mu \nu}. 
\end{equation}
In the case when the polarization states of the initial and the final baryon also remain unmeasured, the hadronic tensor ${\cal J}^{\mu 
\nu}$ is written in terms of the hadronic current $J^\mu$ as
\begin{equation}\label{had}
  {\cal J}^{\mu \nu} = \overline{\sum} \sum_{spins} {J^{\mu}}^{\dagger} J^{\nu} = \rm{Tr} \left[(\slashchar{p}+M) \tilde 
  J^{\mu}(\slashchar p^{\prime}+M_{\Lambda})J^\nu\right], \qquad \tilde J^\mu=\gamma_0(J^\mu)^{\dagger}\gamma_0,
\end{equation}
where $M$ and $M_{\Lambda}$ are the masses of the proton and lambda, respectively. The hadronic matrix element of the 
electromagnetic current $J^{\mu}$ receives the contribution from the background terms and resonance terms. 

Using Eqs.~(\ref{lep}) and (\ref{had}), the transition matrix element squared is obtained as
\begin{equation}\label{mat}
\overline{\sum_{r}} \sum_{spin} |\mathcal{M}^{r}|^2 = -\frac{1}{4}g_{\mu \nu} {\cal J}^{\mu \nu}.
\end{equation}

 \begin{figure}
 \begin{center}
 \includegraphics[height=4.5cm,width=8cm]{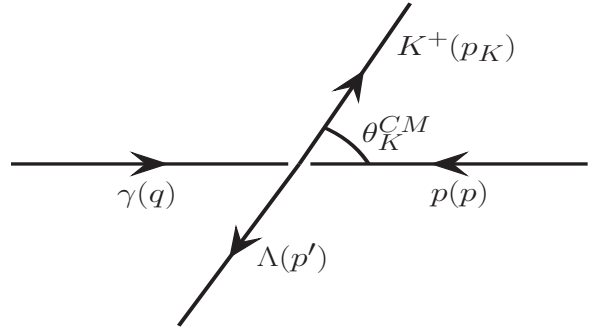}
 \caption{Diagrammatic representation of the process $ \gamma (q) + p(p) \rightarrow K^{+}(p_{k}) + \Lambda(p^{\prime})$ in the 
 center of mass frame. The quantities in the parentheses represent the four momenta of the corresponding particles. 
 $\theta_{k}^{CM}$ is the angle between photon and kaon in the CM frame.}\label{CM}
 \end{center}
 \end{figure}
Following the above expressions, the differential cross section $\frac{d\sigma}{d\Omega}$ in the CM frame is written 
as
\begin{equation}\label{dsig}
\left. \frac{d \sigma}{d \Omega}\right|_{CM} = \frac{1}{64 \pi^{2} s} \frac{|\vec{p}\;^{\prime}|}{|\vec{p}|} 
\overline{\sum_{r}} \sum_{spin} |\mathcal{M}^{r}|^2,
\end{equation}
where $s$ is the CM energy squared obtained as
\begin{equation}\label{s}
 s = W^2 = (q + p)^2 = M^{2} + 2M E_{\gamma} ,
\end{equation}
$E_{\gamma}$ is the energy of the incoming photon in the laboratory frame. The center of mass energies of the initial and the 
final particles are obtained as 
\begin{eqnarray}
 E_{\gamma}^{CM} &=& \frac{s - M^2}{2 \sqrt{s}}, \qquad ~~~ ~\qquad \qquad E_{p}^{CM} = \frac{s + M^2}{2 \sqrt{s}}, 
 \nonumber \\
 E_{k}^{CM} &=& \frac{s + M_{k}^2 - M_{\Lambda}^2}{2 \sqrt{s}}, \qquad \qquad E_{\Lambda}^{CM} = \frac{s + M_{\Lambda}^2 - 
 M_{k}^2}{2 \sqrt{s}}.
\end{eqnarray}
 
In the CM frame as shown in Fig.~\ref{CM}, $|\vec{q}| = |\vec{p}|$ and $|\vec{p}_{k}| = |\vec{p}^{\,\prime}|$ which are given 
by~\cite{PDG}:
\begin{equation}
 |\vec{q}| = \frac{\lambda^{1/2}(s,0,M^2)}{2 \sqrt{s}}, \qquad \qquad \qquad |\vec{p}^{\,\prime}| = 
 \frac{\lambda^{1/2}(s,M_{k}^{2},M_{\Lambda}^{2})}{2 \sqrt{s}},
\end{equation}
with $\lambda(a,b,c)$ being the Callan function, expressed as
\begin{equation*}
 \lambda (a,b,c) = a^{2} + b^{2} +c^{2} - 2ab - 2bc - 2ca.
\end{equation*}

Assuming the incoming photon to be along the $z$-axis, the energy and three momentum of the incoming and the outgoing particles 
are expressed as:
 \begin{eqnarray*}
  \text{photon}&:&  (E_{\gamma}^{CM},0,0,|\vec{q}|)  \\ \text{proton}&:& 
  (E_{p}^{CM},0,0, -|\vec{q}|) \\
  \text{kaon}&~~~:~~~& (E_{k}^{CM},0,|\vec{p}_{k}| \sin \theta_{k}^{CM}, |\vec{p}_{k}| \cos \theta_{k}^{CM}) \\
  \text{lambda} &~~~:~~~&  (E_{\Lambda}^{CM},0,-|\vec{p}_{k}| \sin \theta_{k}^{CM}, -|\vec{p}_{k}| \cos \theta_{k}^{CM}) ,
 \end{eqnarray*}
where $\theta_{k}^{CM}$ is the angle between the photon and kaon measured in the CM frame.
 
\subsection{Non-resonant contribution}\label{NRB}
The non-resonant contributions are obtained using the non-linear sigma model assuming the chiral SU(3) symmetry, which involves 
the low-lying baryons and mesons. This model implements spontaneous breaking of chiral symmetry~\cite{Scherer:2002tk, 
Scherer:2012xha, Koch:1995vp, Koch:1997ei}. In the SU(3) version of the model, it generates the octet of pseudoscalar mesons 
$\pi$, $K$ and $\eta$ as well as the interaction Lagrangians for the meson-meson and meson-baryon 
interactions~\cite{Scherer:2002tk, Scherer:2012xha}.
 
In order to get the Lagrangian which describes the dynamics of these pseudoscalar mesons, we need continuous fields which are 
described in terms of these Goldstone modes. The elements of $SU(3)$ pseudoscalar meson fields are written in terms of a unitary matrix 
\begin{equation}
 U(\Theta) = \exp\left( -i \Theta_k \frac{\lambda_k}{2} \right)\;,
\end{equation}
where $\Theta_k;~(k=1-8)$ are the real set of parameters and $\lambda_k$ are the traceless, Hermitian $3 \times 3$ Gell-Mann 
matrices. 

Each Goldstone boson corresponds to the $x$-dependent Cartesian component of the fields, $\phi_k (x) $, which in turn, is 
expressed in terms of the physical fields as 
\begin{eqnarray}\label{eq2:ps_matrix_final}
 \Phi(x) =\sum_{k=1}^{8} \phi_k(x) \lambda_k =
\left(\begin{array}{ccc}
\pi^0+\frac{1}{\sqrt{3}}\eta &\sqrt{2}\pi^+&\sqrt{2}K^+\\
\sqrt{2}\pi^-&-\pi^0+\frac{1}{\sqrt{3}}\eta&\sqrt{2}K^0\\
\sqrt{2}K^- &\sqrt{2}\bar{K}^0&-\frac{2}{\sqrt{3}}\eta
\end{array}\right).
\end{eqnarray}
For the baryons, we follow the same procedure as we do for the mesons. However, unlike the pseudoscalar mesons where the fields 
are real, in the case of baryon fields, represented by a $B$ matrix, each entry is a complex-field and the general representation is given by,
\begin{eqnarray}\label{eq2:bmatrix2}
B(x)=\sum_{k=1}^{8} \frac{1}{\sqrt2} b_k(x) \lambda_k = \left(\begin{array}{ccc}
\frac{1}{\sqrt{2}}\Sigma^0+\frac{1}{\sqrt{6}}\Lambda&\Sigma^+&p\\
\Sigma^-&-\frac{1}{\sqrt{2}}\Sigma^0+\frac{1}{\sqrt{6}}\Lambda&n\\
\Xi^-&\Xi^0&-\frac{2}{\sqrt{6}}\Lambda
\end{array}\right).
\end{eqnarray}

After getting the parameterization of pseudoscalar meson fields octet $\Phi(x)$ in Eq.(~\ref{eq2:ps_matrix_final}) and baryon fields octet  
$B(x)$ in Eq.~(\ref{eq2:bmatrix2}), we now discuss the construction of Lagrangian for meson-meson, baryon-meson interactions and 
their interaction with the external fields. 

\subsubsection{Meson - Meson Interaction}\label{Sec2:MMinter}
The lowest-order $SU(3)$ chiral Lagrangian describing the pseudoscalar mesons in the presence of an external current is obtained 
as~\cite{Scherer:2002tk, Scherer:2012xha}
\begin{equation}\label{eq2:lagM}
{\cal L}_M=\frac{f_\pi^2}{4}\mbox{Tr}[D_\mu U (D^\mu U)^\dagger],
\end{equation}
where $f_\pi(=92.4 \text{ MeV})$ is the pion decay constant obtained from the weak decay of pions, {\it i.e.,} $\pi^{\pm} 
\rightarrow \mu^{\pm} \nu_{\mu}(\bar{\nu}_{\mu})$. The covariant derivatives $D^{\mu} U$ and $D^{\mu} U^{\dagger}$ appearing in 
Eq.~(\ref{eq2:lagM}) are expressed in terms of the partial derivatives as
\begin{eqnarray}\label{eq2:coDer}
 D^\mu U &\equiv& \partial^\mu U - i r^\mu U + i U l^\mu, \nonumber \\
  D^\mu U^\dagger &\equiv& \partial^\mu U^\dagger + i U^\dagger r^\mu - i l^\mu U^\dagger,
\end{eqnarray}
where $U$ is the SU(3) unitary matrix given as
\begin{equation}
 U(x) = \exp\left(i\frac{\Phi(x)}{ f_\pi } \right), 
\end{equation}
where $\Phi(x)$ is given by Eq.~(\ref{eq2:ps_matrix_final}) and the left-($l^\mu$) and right-($r^\mu$) handed currents appearing in Eq.~(\ref{eq2:coDer}) are expressed as 
\begin{equation}\label{eq2:lmurmu}
 \begin{array}{ c c}
l_{\mu} = -e \hat Q {\cal A}_\mu, \qquad &\qquad r_{\mu} = -e \hat Q {\cal A}_\mu . 
 \end{array}
\end{equation}
${\cal A}^\mu$ is the electromagnetic four-vector potential and $\hat Q$ is the $SU(3)$ quark charge.

\subsubsection{Baryon - Meson Interaction}
To incorporate baryons in the theory, we have to take care of their masses which do not vanish in the chiral 
limit~\cite{Kubis:2007iy}. However, if we take nucleons as massive matter fields which couples to external currents and the 
pseudoscalar mesons, we have to then expand the Lagrangian according to their increasing number of momenta. Here, we shall 
present in brief the extension of the formalism to incorporate the heavy matter fields. 

The lowest-order chiral Lagrangian for the baryon octet in the presence of an external current may be written in terms of the 
$SU(3)$ matrix $B$ as~\cite{Scherer:2002tk, Scherer:2012xha},
\begin{equation}\label{eq2:lagB}
{\cal L}_{MB}=\mbox{Tr}\left[\bar{B}\left(i\D
-M\right)B\right]
-\frac{D}{2}\mbox{Tr}\left(\bar{B}\gamma^\mu\gamma_5\{u_\mu,B\}\right)
-\frac{F}{2}\mbox{Tr}\left(\bar{B}\gamma^\mu\gamma_5[u_\mu,B]\right),
\end{equation}
where $M$ denotes the mass of the baryon octet, $D=0.804$ and $F=0.463$ are the axial vector coupling constants for the baryon 
octet determined from the semileptonic decays of neutron and hyperons~\cite{Cabibbo:2003cu}, the matrix $B$ is given in 
Eq.~(\ref{eq2:bmatrix2}) and the Lorentz vector $ u^\mu$ is given by~\cite{Scherer:2012xha}:
\begin{equation}\label{eq2:vielbein}
u^\mu = i \left[ u^\dagger ( \partial^\mu - i r^\mu) u - u ( \partial^\mu - i l^\mu) u^\dagger \right].
\end{equation} 
In the case of meson-baryon interactions, the unitary matrix for the pseudoscalar field is expressed as 
$$
u = \sqrt U \equiv \rm{exp} \left( i \frac{\Phi(x)}{ 2 f_\pi } \right),
$$
and the covariant derivative of $B$ is given by
\begin{equation}\label{dmuB}
D_\mu B=\partial_\mu B +[\Gamma_\mu,B], \qquad \text{with} \qquad \Gamma^\mu=\frac{1}{2}\left[u^\dagger(\partial^\mu-
ir^\mu)u
+u(\partial^\mu-il^\mu)u^\dagger\right].
\end{equation}

Using Eqs.~(\ref{eq2:ps_matrix_final}), (\ref{eq2:bmatrix2}), (\ref{eq2:vielbein}) and (\ref{dmuB}) in the general expression of 
the Lagrangian given in Eq.~(\ref{eq2:lagB}), the Lagrangians for the desired vertices involved in the meson-baryon interactions 
among themselves and with the external fields are obtained. Some of the Lagrangians using chiral SU(3) symmetry relevant for the 
present work, are derived to be:
\begin{eqnarray}\label{ppgamma}
  {\cal L}_{\gamma pp} &=& - e e_{p} \bar{\psi}_{p} \gamma_{\mu} \psi_{p} A^{\mu}  
  \end{eqnarray}
  \begin{eqnarray}
  \label{LLgamma}
  {\cal L}_{\gamma \Lambda \Lambda} &=& - e e_{\Lambda} \bar{\psi}_{\Lambda} \gamma_{\mu} \psi_{\Lambda} A^{\mu} \\
  \label{KLp}
  {\cal L}_{K\Lambda p} &=& \left(\frac{D + 3F}{2 \sqrt{3} f_{\pi}} \right) \bar{\psi}_{\Lambda} \gamma_{\mu} \gamma_{5} 
  \psi_{p} \partial^{\mu} K^{\dagger}  \\
  \label{gammaKLp}
  {\cal L}_{\gamma K \Lambda p} &=& -ie \left(\frac{D + 3F}{2 \sqrt{3} f_{\pi}} \right) \bar{\psi}_{\Lambda} \gamma_{\mu} 
  \gamma_{5} \psi_{p} K^{\dagger} A^{\mu} \\
  \label{gammaKK}
  {\cal L}_{\gamma KK} &=& -ie \left(K^{\dagger} \partial_{\mu} K - K \partial_{\mu} K^{\dagger} \right) A^{\mu}
\end{eqnarray}
where $e_{p}$ and $e_{\Lambda}$, respectively, represents the electric charge of proton and lambda, $\bar{\psi}_{p}$ and 
$\bar{\psi}_{\Lambda}$ represent the outgoing proton and lambda fields, $\psi_{p}$ and $\psi_{\Lambda}$ represent the incoming 
proton and lambda fields, $A_{\mu}$ represents the electromagnetic field with $e$ being the strength of the electromagnetic field, and $K^{\dagger}$ and 
$\partial_{\mu} K^{\dagger}$ represent the kaon field and covariant derivative of kaon field, respectively.

The above Lagrangians are obtained assuming the baryons to be point particles. Since the baryons are composite particles, 
therefore, there is a charge distribution and the magnetic coupling appears due to the  structure of the baryons. Moreover, in the case of virtual photons, these 
electric and magnetic couplings acquire $q^2$ dependence.

\subsubsection{Current for the non-resonant terms}\label{NRB1}
   \begin{figure}
 \begin{center}
    \includegraphics[height=3cm,width=3.9cm]{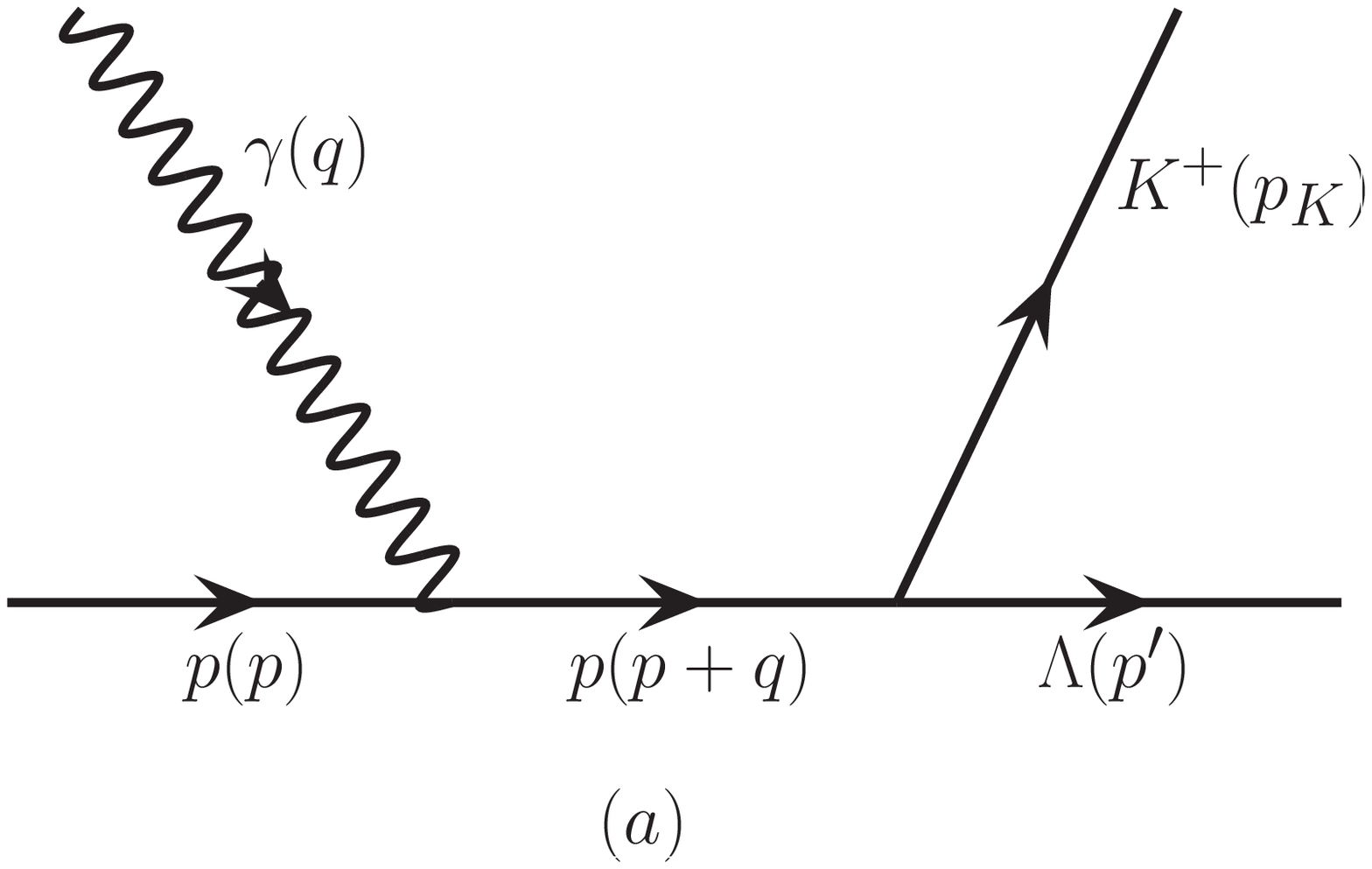}
    \hspace{5mm}
    \includegraphics[height=3.5cm,width=3.3cm]{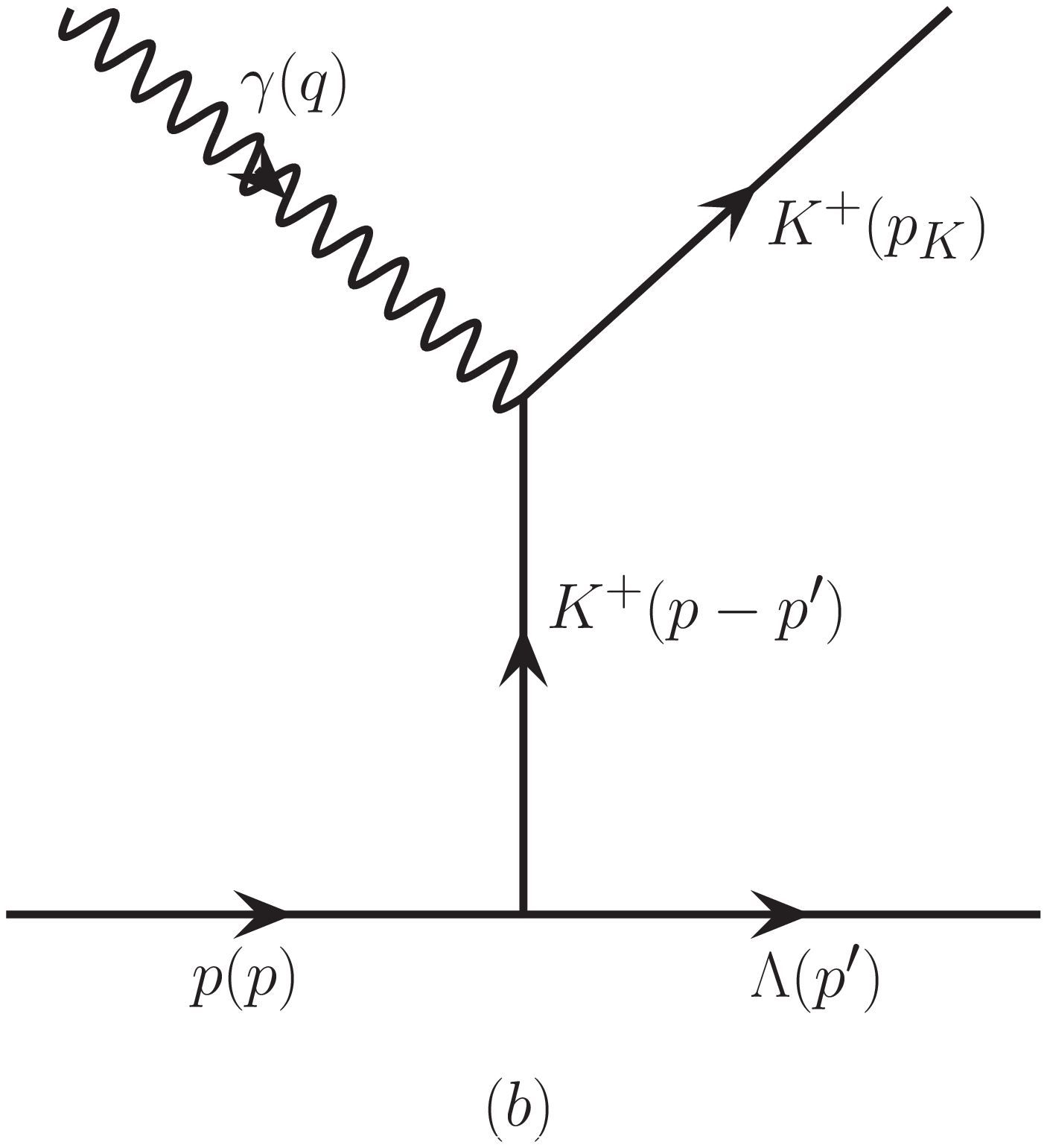}
    \hspace{5mm}
    \includegraphics[height=3cm,width=3.9cm]{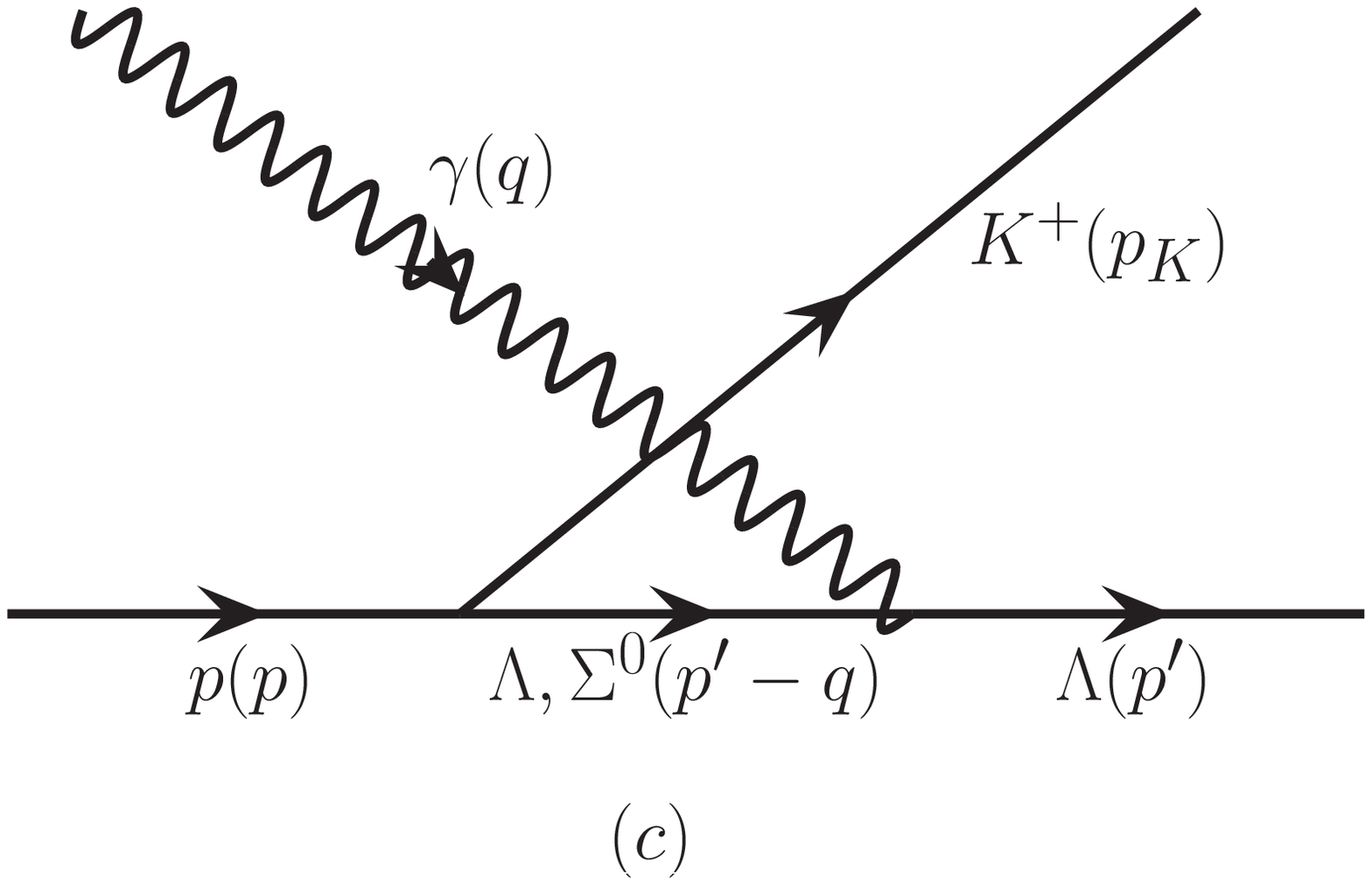}
    \vspace{5mm}
    
    \includegraphics[height=3cm,width=4cm]{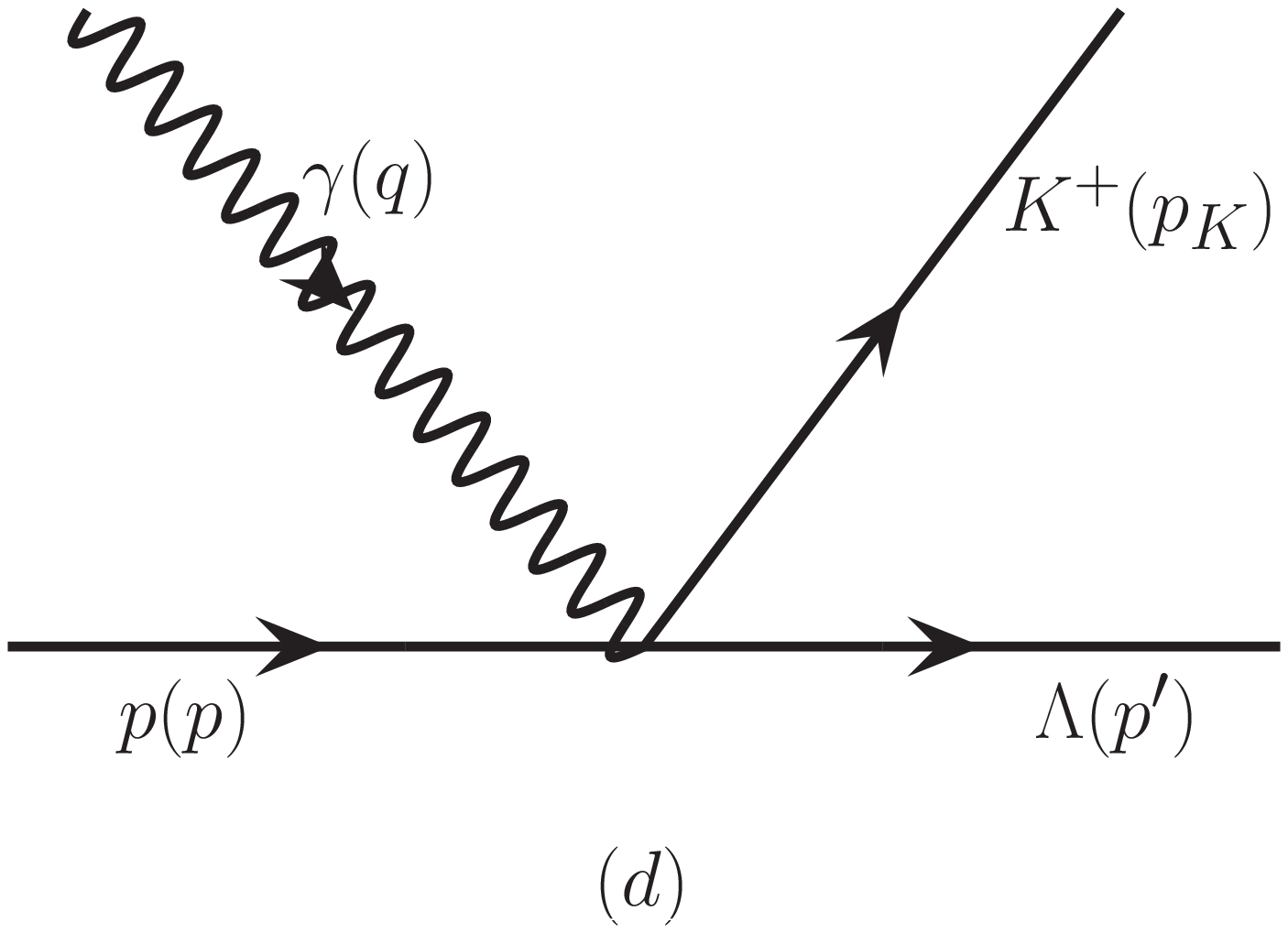}
    \hspace{5mm}
    
    \includegraphics[height=3.5cm,width=3.9cm]{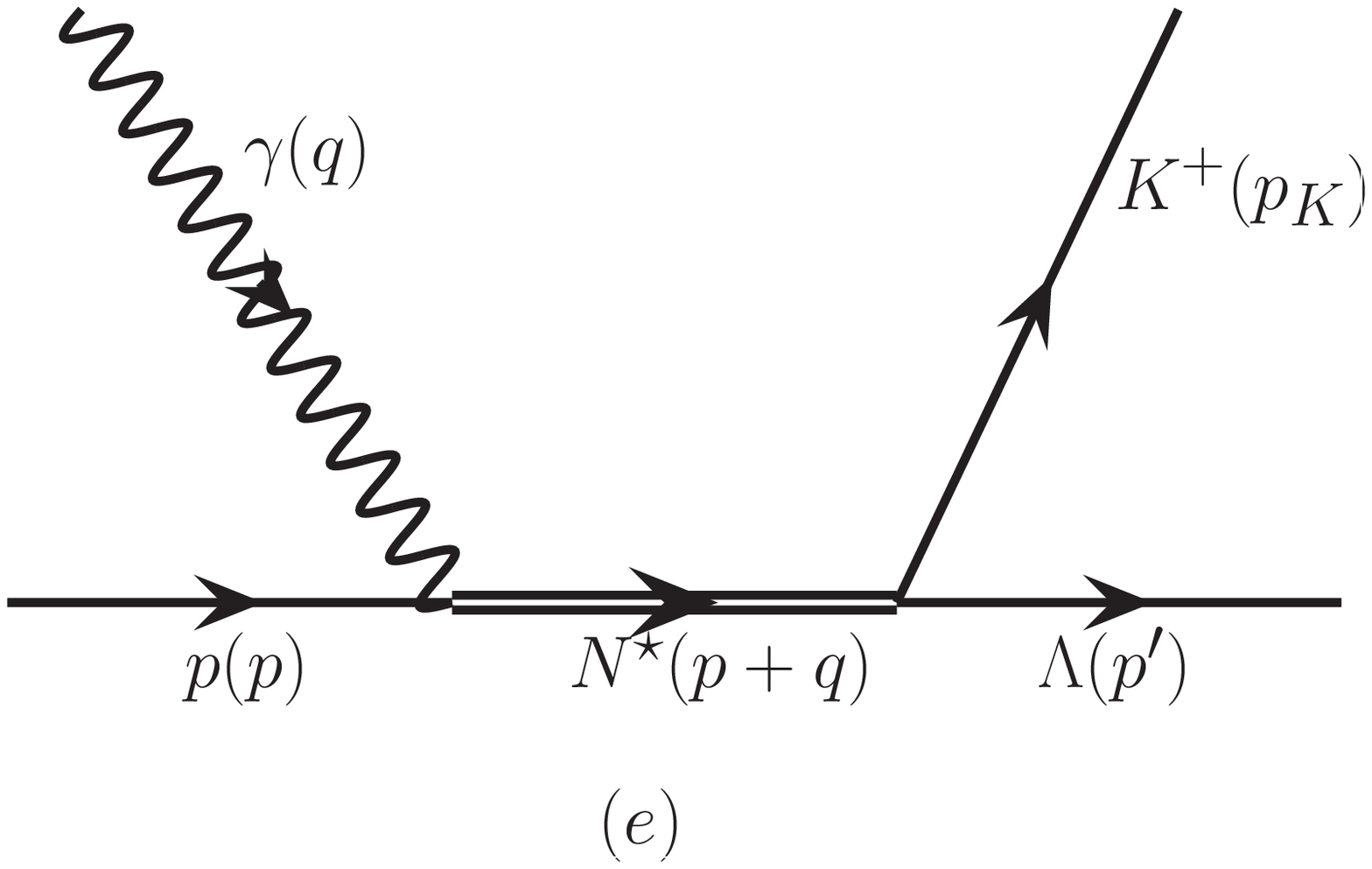}
    \hspace{5mm}
    \includegraphics[height=3cm,width=3.3cm]{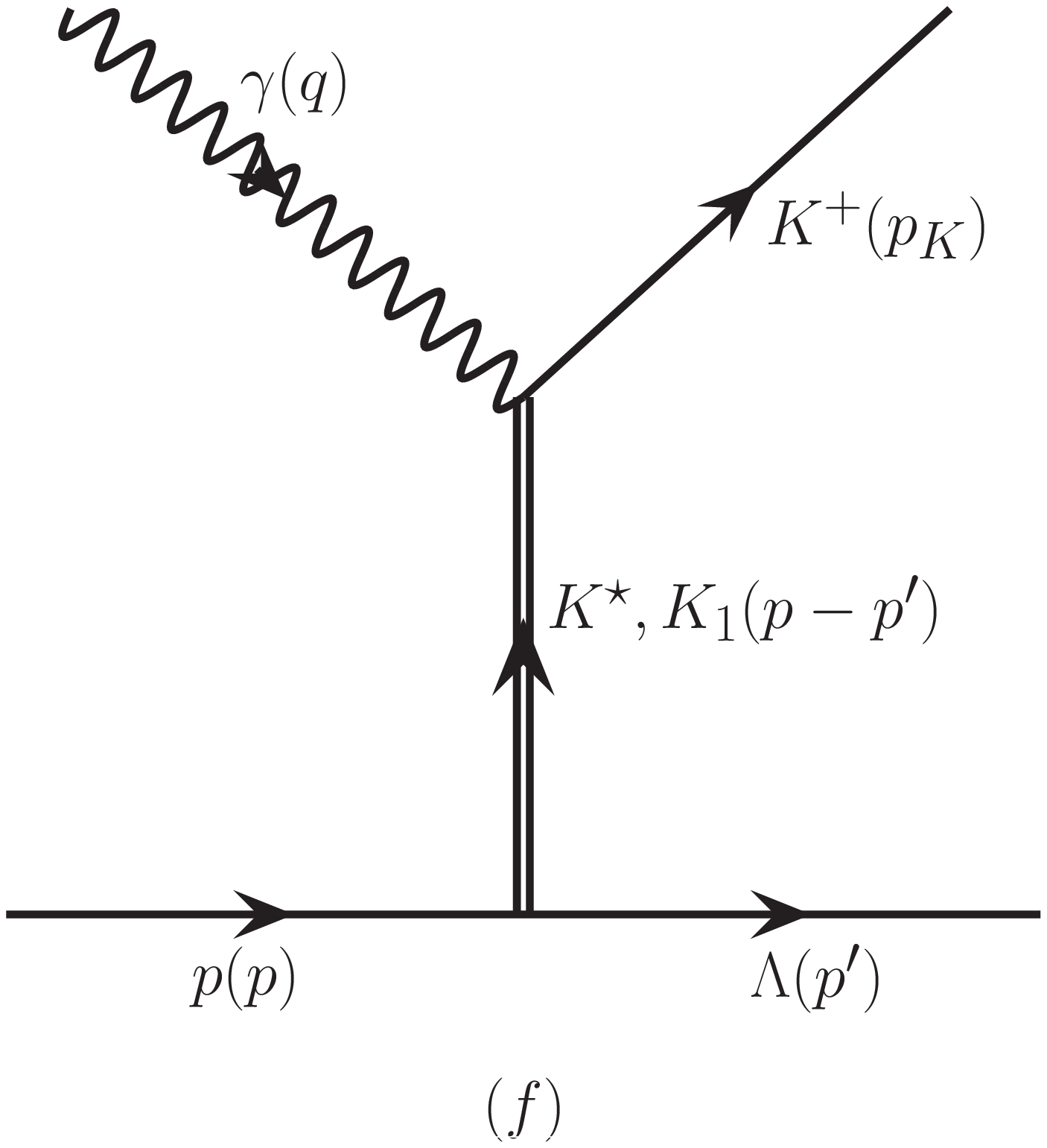}
    \hspace{5mm}
    \includegraphics[height=3cm,width=3.9cm]{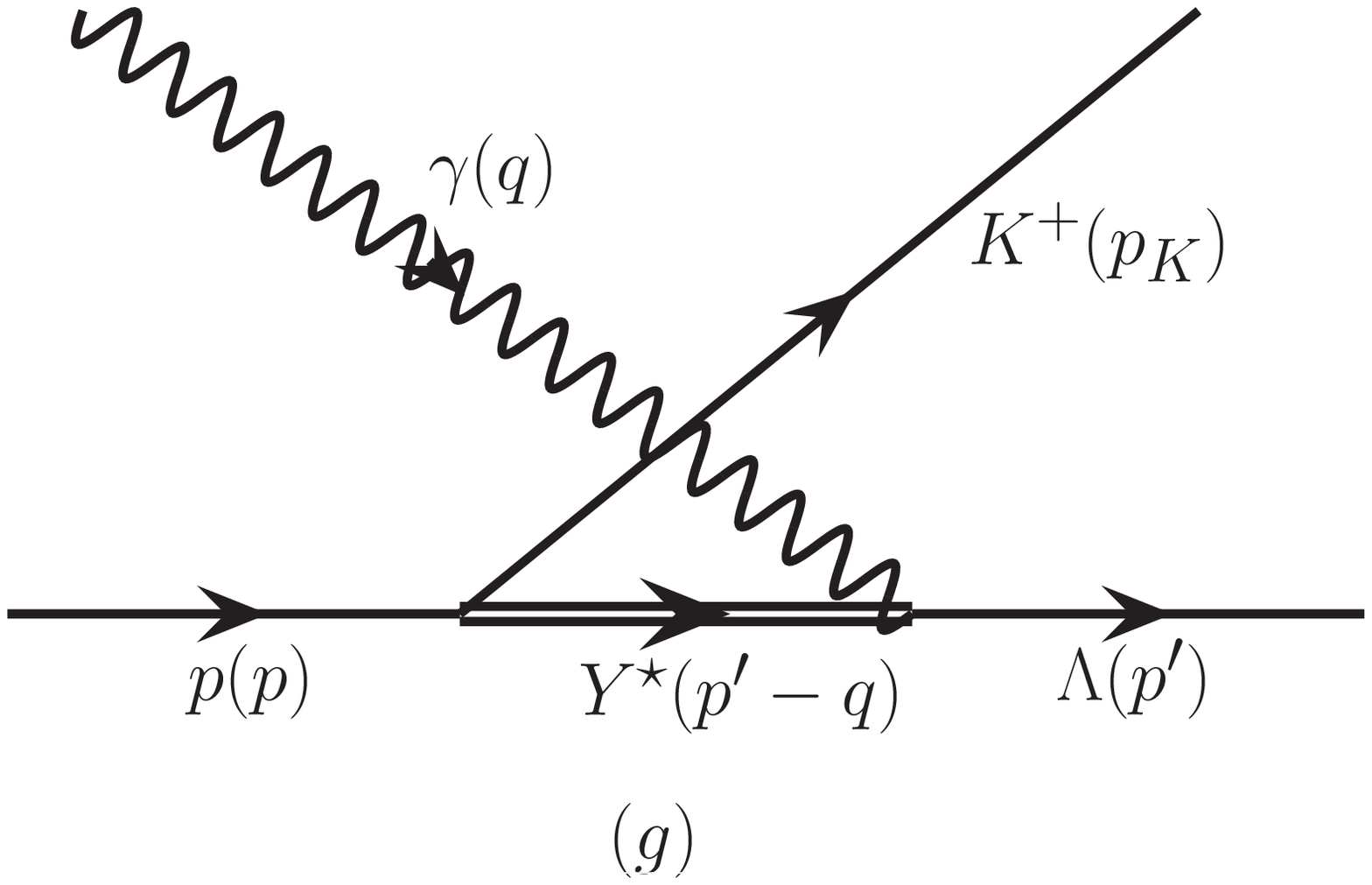}
  \caption{Feynman diagram for the various channels possible for the process $ \gamma (q) + p(p) \rightarrow K^{+}(p_{k}) + 
  \Lambda(p^{\prime})$. (a)~$s$ channel, (b)~$t$ channel, (c)~$u$ channel and (d)~contact term constitute the non-resonant 
  terms. (e)~nucleon resonances in the $s$ channel, (f)~kaon resonances in the $t$ channel and (g)~hyperon resonances in the 
  $u$ channel. The quantities in the bracket represent four momenta of the corresponding 
  particles.}\label{fyn_dia}
   \end{center}
 \end{figure}

The hadronic currents for the various non-resonant terms shown in Fig~\ref{fyn_dia}(a)--(d) are obtained using the non-linear 
sigma model described in the above sections. The expressions of the hadronic currents for the different channels are obtained 
using the Lagrangians given in Eqs.~(\ref{ppgamma})--(\ref{gammaKK}) and are expressed as~\cite{Hernandez:2007qq, 
Alam:2015gaa}:
\begin{eqnarray}\label{j:s}
J^\mu \arrowvert_{s} &=&ie A_{s}~F_{s}(s) \bar u(p^\prime) \slashed{p}_k \gamma_5 \frac{ \slashed{p} + \slashed{q} + M}
  {s -M^2} \left(\gamma^\mu e_{p} +i \frac{\kappa_{p}}{2 M} \sigma^{\mu \nu} q_\nu \right) u(p), \\
  \label{j:t}
  J^\mu \arrowvert_{t}&=& ie A_{t}~ F_{t}(t) \bar u(p^\prime)\left[(\slashed{p} - \slashed{p}^{\prime}) \cdot \gamma_{5} \right] 
  u(p) \frac{(2 p_{k}^{\mu} - q^{\mu})}{t - M_{k}^{2}} , \\
  \label{j:ulam}
J^\mu \arrowvert_{u \Lambda} &=&ie A_{u}^{\Lambda} ~F_{u}^{\Lambda} (u) \bar u(p^\prime) \left(\gamma^\mu e_{\Lambda} + i 
\frac{\kappa_{\Lambda}}{2 M_{\Lambda}} \sigma^{\mu \nu} q_\nu \right) \frac{ \slashed{p}^{\prime} -\slashed{q} + M_{\Lambda}}
{u - M_{\Lambda}^2} \slashed{p}_k \gamma_5 u(p), \\
\label{j:usig}
J^\mu \arrowvert_{u \Sigma^{0}} &=&ie A_{u}^{\Sigma^{0}} ~F_{u}^{\Sigma^{0}} (u) \bar u(p^\prime) \left(\gamma^\mu 
e_{\Sigma^{0}} + i \frac{\kappa_{\Sigma^{0}}}{2 M_{\Sigma^{0}}} \sigma^{\mu \nu} q_\nu \right) 
\frac{\slashed{p}^{\prime} -\slashed{q} + M_{\Sigma^{0}}} {u -M_{\Sigma^{0}}^2}\nonumber \\
&\times& \slashed{p}_k \gamma_5 u(p), \\
\label{j:CT}
J^\mu \arrowvert_{CT} &=&-i e A_{CT} ~F_{CT} \bar u(p^\prime) \; \gamma^\mu \gamma_5 \; u(p),
\end{eqnarray}
where $CT$ stands for the contact term and $s,~t,~u$ are the Mandelstam variables defined as
\begin{eqnarray}
 t= (p - p^{\prime})^{2}, \qquad \qquad \qquad u = (p^{\prime} - q)^{2},
\end{eqnarray}
and $s$ is defined in Eq.~(\ref{s}). $A_{i}$'s; $i=s,t,u,CT$ are the coupling strengths of $s$, $t$, $u$ channels and the 
contact term, respectively, and are obtained as
\begin{eqnarray}\label{eq:coupling}
 A_{s} = A_{t} = A_{u}^{\Lambda} = A_{CT} &=& - \left(\frac{D + 3F}{2 \sqrt{3} f_{\pi}}\right)  = -6.85~\text{GeV}^{-1},
 \end{eqnarray}
 \begin{eqnarray}
 A_{u}^{\Sigma^{0}} &=& 
 \left(\frac{D-F}{2 f_{\pi}} \right) = 1.85~\text{GeV}^{-1}.
\end{eqnarray}
All these couplings of non-resonant terms are generated by the chiral symmetry and are fixed by the low energy electroweak 
phenomenology consistent with experimental data.
 
The values of $e$ and $\kappa$ for proton, lambda and sigma are 
\begin{eqnarray}
 e_{p} = 1, ~~~~~\qquad \qquad &e_{\Lambda} = 0,& \qquad \qquad e_{\Sigma^{0}} = 0, \nonumber \\
 \label{eq:em_coup}
 \kappa_{p} = 1.793, \qquad \qquad &\kappa_{\Lambda} = -0.613,& \qquad \qquad \kappa_{\Sigma^{0}} = 1.61.
\end{eqnarray}

In order to take into account the hadronic structure, the form factors $F_{s} (s)$, $F_{t} (t)$, $F_{u} (u)$ and $F_{CT}$, are 
introduced at the strong vertices. Various parameterizations of these form factors are available in the 
literature~\cite{Skoupil:2016ast}, however, we use the most general dipole form, parameterized as~\cite{Mart:2017xtf}:
\begin{equation}\label{FF_Born}
F_{x} (x) = \frac{\Lambda_{B}^{4}}{\Lambda_{B}^{4} + (x - M_{x}^{2})^{2}}, \qquad \qquad \quad x=s,t,u
\end{equation}
where $\Lambda_{B}=0.505$~GeV is the cut-off parameter taken to be same for all the background terms, whose value is fitted to the 
experimental data, $x$ represents the Mandelstam variables $s,~t,~u$ and $M_{x} = M,~M_{k},~M_{Y},$ corresponds to the mass of 
the baryons or mesons exchanged in the $s,~t,~u$ channels. 

One of the most important property of the electromagnetic current is the gauge invariance which corresponds to the current 
conservation. The total hadronic current for the non-resonant terms is given by
\begin{eqnarray}\label{Jmu_Born}
 J^{\mu} = J^\mu \arrowvert_{s} +J^\mu \arrowvert_{t} +J^\mu \arrowvert_{u \Lambda} +J^\mu \arrowvert_{u \Sigma^{0}} + 
 J^\mu \arrowvert_{CT}.
\end{eqnarray}
The condition to fulfill gauge invariance is
\begin{eqnarray}\label{gauge1}
 q_{\mu} J^{\mu} = 0.
\end{eqnarray}

In the absence of the hadronic form factors ($F_{s} = F_{t} = F_{u} = F_{CT} = 1$), if we consider only the $s,~t,~u$ channel 
Born terms in the expression of the hadronic current  
\begin{equation}\label{j:PV}
 J^{\mu} = J^\mu \arrowvert_{s} +J^\mu \arrowvert_{t} +J^\mu \arrowvert_{u \Lambda} +J^\mu \arrowvert_{u 
\Sigma^{0}} ,
\end{equation}
then the condition given in Eq.~(\ref{gauge1}) is applied to $J^{\mu}$ as defined in Eq.~(\ref{j:PV}) in which the individual 
currents are defined in Eqs.~(\ref{j:s})--(\ref{j:usig}). Using the coupling strengths obtained in our model from 
Eqs.~(\ref{eq:coupling}) and (\ref{eq:em_coup}), we obtain
\begin{equation}
 q_{\mu} J^{\mu} = - \frac{D+F}{2\sqrt{3} f_{\pi}} \bar{u} (p^{\prime}) \left[(\slashed{p}_{k} + \slashed{p}^{\prime} - 
 \slashed{p}) \gamma_{5} \right] u(p).
\end{equation}
The above expression shows that in the presence of only $s$, $t$, $u$ channel contributions, the hadronic current is not gauge 
invariant. However, when the contribution from the contact term {\it i.e.} $J^{\mu} \arrowvert_{CT}$ is added, we obtain 
$q_{\mu} J^{\mu} =0$ and $J^{\mu}$ satisfies the gauge invariance. The present model, thus, predicts the strength of the 
coupling of the contact term in such a way that the gauge invariance is satisfied in a natural way. On the other hand, in most 
of the effective Lagrangians used in the other isobar models with pseudoscalar and/or pseudovector interactions, the coupling 
strengths are modulated to obtain the gauge invariance.

As the hadronic form factors are taken into account in the hadronic current, the condition for gauge invariance gives
\begin{equation}\label{GI}
 q_{\mu} J^{\mu} = - \frac{D+F}{2\sqrt{3} f_{\pi}} \bar{u} (p^{\prime}) \left[(\slashed{p}_{k} F_{s} + (\slashed{p}^{\prime} - 
 \slashed{p})F_{t} - \slashed{q} F_{CT}) \gamma_{5} \right] u(p).
\end{equation}
From the above equation, it is evident that due to the presence of hadronic form factor, the hadronic current is not gauge 
invariant. Therefore, in order to restore gauge invariance, the following term is added to Eq.~(\ref{GI})
\begin{equation}\label{GI1}
 q_{\mu} J^{\mu}_{add} = - \frac{D+F}{2\sqrt{3} f_{\pi}} \bar{u} (p^{\prime}) \left[ \slashed{p}_{k} \left(F_{CT} - F_{s} \right) + (\slashed{p}^{\prime} - \slashed{p})(F_{CT} - F_{t})\right] \gamma_{5} u(p).
\end{equation}
Thus, the presence of the additional terms given in Eq.~(\ref{GI1}) implies that the gauge invariance can be achieved if the hadronic current $J^\mu$ defined through Eq.~(\ref{Jmu_Born}) is supplemented by adding an additional term $J^\mu_{add}$ given by 
\begin{equation}\label{GI2}
 J^{\mu}_{add} = - \frac{D+F}{2\sqrt{3} f_{\pi}} \bar{u} (p^{\prime}) \left[\frac{2 \slashed{p}_{k} p^{\mu}}{s - M^{2}} (F_{CT} - F_{s}) + \frac{2p_{k}^{\mu}}{t - M_{k}^{2}}(\slashed{p} - \slashed{p}^{\prime}) (F_{CT} - F_{t}) \right] u(p).
\end{equation}


In order to take into account the effect of the form factor for the contact term, there are different prescriptions available in the literature, for example that of 
 Ohta~\cite{Ohta:1989ji}, Haberzettl {\it et 
al.}~\cite{Haberzettl:1998eq}, Davidson and Workman\cite{DW}, etc. In the present work, we have followed the prescription of Davidson and Workman~\cite{DW},
where 
 $F_{CT}$ is given by:  
\begin{equation}\label{FF_CT}
 F_{CT} = F_{s}(s) + F_{t}(t) - F_{s}(s) \times F_{t}(t).
\end{equation}
\subsection{Resonance contribution}\label{RES}
In this section, we discuss the contributions of the different nucleon, kaon and hyperon resonances. 

\subsubsection{Spin $\frac12$ nucleon resonances}\label{spin12N}
The hadronic current for the spin $\frac12$ resonance state is given by  
\begin{eqnarray}\label{had_curr_1/2}
j^{\mu}_{\frac{1}{2}}=\bar{u}(p') \Gamma^\mu_{\frac12} u(p), 
\end{eqnarray}
where $u(p)$ and $\bar u(p^\prime)$ are, respectively, the Dirac spinor and adjoint Dirac spinor for spin $\frac{1}{2}$ 
particles and $\Gamma^\mu_\frac12$ is the vertex function. For a positive parity state, $\Gamma^{\mu}_{\frac{1}{2}^+}$ is given 
by 
\begin{align}\label{eq:vec_half_pos}
  \Gamma^{\mu}_{\frac{1}{2}^+} &= {V}^{\mu}_\frac{1}{2},
  \end{align}
and for a negative parity resonance, $\Gamma^{\mu}_{\frac{1}{2}^-}$ is given by  
\begin{align}\label{eq:vec_half_neg}
  \Gamma^{\mu}_{\frac{1}{2}^-} &= {V}^{\mu}_\frac{1}{2} \gamma_5 ,
  \end{align}
where $V^{\mu}_{\frac{1}{2}}$ represents the vector current parameterized in terms of $F_{2}^{R^{+}} $, as
 \begin{align}\label{eq:vectorspinhalf1}
  V^{\mu}_{\frac{1}{2}} & =\left[\frac{F_2^{R^{+}}}{2 M} 
  i \sigma^{\mu\alpha} q_\alpha \right].
\end{align}
The coupling $F^{R^{+}}_{2}$ is derived from the helicity amplitudes extracted from the real photon scattering 
experiments. The explicit relation between the coupling $F_2^{R^{+}}$ and the helicity amplitude $A_{\frac{1}{2}}^{p}$ is given 
by~\cite{Leitner:2008ue}
\begin{eqnarray}\label{eq:hel_spin_12}
A_\frac{1}{2}^{p}&=& \sqrt{\frac{2 \pi \alpha}{M} \frac{(M_R \mp M)^2}{M_R^2 - M^2}} \left[ \frac{M_R \pm M}{2 M} F_2^{R^{+}} 
\right] ,
\end{eqnarray}
where the upper~(lower) sign stands for the positive~(negative) parity resonance. $M_R$ is the mass of corresponding resonance. 
The value of the helicity amplitude $A_{\frac{1}{2}}^{p}$ for $S_{11}(1650)$ resonance is taken from MAID~\cite{Tiator:2011pw} 
while for the other spin $\frac{1}{2}$ nucleon resonances, these values are taken from PDG~\cite{PDG} and are quoted in 
Table~\ref{tab:param-p2}.

The most general form of the hadronic currents for the $s$ channel processes where a resonance state $R^{\frac12}$ with spin 
$\frac{1}{2}$ is produced and decays to a kaon and a lambda in the final state, are written as~\cite{Hernandez:2013jka, 
Gonzalez-Jimenez:2016qqq} 
\begin{eqnarray}\label{eq:res1/2_had_current}
j^\mu\big|_{R}^{\frac12 \pm}&=& 
ie~ \bar u({p}\,') \frac{g_{R^{\frac12}K \Lambda}}{M_{K}} 
 \slashchar{p}_{k} \Gamma_{s} \frac{\slashchar{p}+\slashchar{q}+M_{R}}{s-M_{R}^2+ iM_{R} \Gamma_{R}} \Gamma^\mu_{\frac12 
 \pm} u({p}\,),
\end{eqnarray}
where $\Gamma_{R}$ is the decay width of the resonance, $\Gamma_{s} = 1(\gamma_{5})$ stands for the positive~(negative) parity 
resonances. $\Gamma_{\frac{1}{2}^{+}}$ and $\Gamma_{\frac{1}{2}^{-}}$ are, respectively, the vertex function for the positive and negative parity resonances, as defined in Eqs.~(\ref{eq:vec_half_pos}) and (\ref{eq:vec_half_neg}).
$ g_{R^\frac12 K \Lambda}$ is the coupling strength for the process $ R^\frac12 \to K \Lambda$, given in 
Table~\ref{tab:included_resonances}. 

Due to the lack of experimental data, there is a large uncertainty associated with $RK\Lambda$ coupling at the ${R^{\frac{1}{2}}} 
\to K \Lambda$ vertex. We determine the $RK\Lambda$ coupling using the value of branching ratio and decay width of these 
resonances from PDG~\cite{PDG} and use the expression for the decay rate which is obtained by writing the most general form of 
$RK\Lambda$ Lagrangian~\cite{Leitner:2008ue},
\begin{align}\label{eq:spin12_lag}
 \mathcal{L}_{R_{\frac{1}{2}} K \Lambda} &= \frac{g_{ R\frac12 K \Lambda} }{M_{K}}\bar{\Psi}_{R_{\frac{1}{2}}} \; 
 \Gamma^{\mu}_{s} \;
  \partial_\mu K^i \tau_i \,\Psi ,
\end{align}
where $g_{R\frac12 K \Lambda}$ is the $RK \Lambda$ coupling strength. $\Psi$ is the nucleon field and ${\Psi}_{R_{\frac{1}{2}}}$ 
is the spin $\frac12$ resonance field. $K^i$ is the kaon field and $\tau$ is the isospin factor for the isospin $\frac12$ 
states. The interaction vertex $\Gamma^{\mu}_{s} = \gamma^\mu \gamma^5$~($\gamma^\mu$) stands for positive~(negative) parity 
resonance states. 

Using the above Lagrangian, one may obtain the expression for the decay width in the resonance rest frame as
\begin{align}\label{eq:12_width}
 \Gamma_{R_{\frac{1}{2}} \rightarrow K \Lambda} &= \frac{\mathcal{C}}{4\pi} \left(\frac{g_{R\frac12 K \Lambda }}{M_{K}}
 \right)^2 \left(M_R \pm M_{\Lambda}\right)^2 \frac{E_{\Lambda} \mp M_{\Lambda}}{M_R} |\vec{p}^{\,\mathrm{cm}}_{k}|,
\end{align}
where the upper~(lower) sign represents the positive~(negative) parity resonance. The parameter $\mathcal{C}$ depends upon the 
charged state of $R$, $K\Lambda$ and is obtained from the isospin analysis and found out to be $1$. $|\vec p^{\,cm}_{k}|$ is the 
outgoing kaon momentum measured from resonance rest frame and is given by, 
\begin{equation}\label{eq:pi_mom}
|\vec{p}^{\,\mathrm{cm}}_{k}| = \frac{\sqrt{(W^2-M_{K}^2-M_{\Lambda}^2)^2 - 4 M_{K}^2 M_{\Lambda}^2}}{2 M_R}  
\end{equation}
and $E_\Lambda$, the lambda energy is
\begin{equation}\label{eq:elam}
  E_\Lambda=\frac{W^2+M_{\Lambda}^2-M_{K}^2}{2 M_R},
\end{equation}
where $W$ is the total center of mass energy carried by the resonance. 

Using Eq.~(\ref{eq:12_width}), the coupling for $ {R \frac12} \to K \Lambda$ is obtained and given in 
Table-\ref{tab:included_resonances} for various spin $\frac{1}{2}$ resonances.

\subsubsection{Spin $\frac32$ nucleon resonances}\label{spin32N}
 Next, we discuss spin $\frac{3}{2}$ resonances exchanged in the $s$ channel process. The general structure of the electromagnetic hadronic current for spin $\frac{3}{2}$ resonances describing the $\gamma NR_{\frac{3}{2}}$ excitations as well as the effective Lagrangian for describing the $R_{\frac{3}{2}} K \Lambda$ vertex is written in terms of the spin $\frac{3}{2}$ field
$\Psi_{\mu} (p)$ using the Rarita-Schwinger formalism~\cite{Rarita:1941mf}. It is well known that the Rarita-Schwinger formalism is not unique for describing the spin $\frac{3}{2}$
field~(as well as for the higher spin fields) and has a problem associated with the lower spin degrees of freedom. This leads to some 
 ambiguities in describing the propagation of the off-shell spin $\frac{3}{2}$ fields using a propagator specially in the presence of interactions 
 like the electromagnetic and strong interactions. The problem has been discussed extensively in literature for many years ever since the
 field theory of higher spins was developed using either the vector-spinor formalism~\cite{Rarita:1941mf} or the multi-spinor
 formalism~\cite{Fierz}. Consequently there are various prescriptions for treating the propagator and the effective Lagrangians for the interacting fields of higher spin
 in a consistent way for describing the interaction of spin $\frac{3}{2}$ fields. One of the most popular prescriptions given 
 Pascalutsa and Timmermans~\cite{Pascalutsa:1999zz} has been investigated further in the latest works of Mart~\cite{Mart:2019jtb} 
 and Vrancx {\it et al.}~\cite{Vrancx:2011qv} and  many other references cited there. However, in the present work, 
 we follow the prescription used by us~\cite{RafiAlam:2010kf, Alam:2011xq, Alam:2012ry, Alam:2015gaa, Alam:2015zla, 
 SajjadAthar:2007gb, Singh:1998ha, SajjadAthar:2009rc, Alam:2013vwa, Alam:2013xoa, Athar:2007wd, Ahmad:2006cy} and 
 many others~\cite{Hernandez:2007qq, Hernandez:2013jka, Leitner:2008ue, Gonzalez-Jimenez:2016qqq, SLA, AlvarezRuso:1999dy, Valverde:2008jj}
 in the past to study the photo, electro and weak interaction induced pion, eta  and kaon productions.

The general structure for the hadronic 
current for spin three-half resonance excitation is determined by the following expression
\begin{eqnarray}\label{eq:had_current_3/2}
J_{\mu}^{\frac{3}{2}}=\bar{\psi}^{\nu}(p') \Gamma_{\nu \mu}^{\frac32} u(p), 
\end{eqnarray}
where $u(p)$ is the Dirac spinor for the nucleon, ${\psi}^{\mu}(p)$ is the Rarita-Schwinger spinor for spin three-half particle 
and $\Gamma_{\nu \mu}^{\frac32}$ has the following general structure for the positive and negative parity resonance states~\cite{Hernandez:2007qq, Leitner:2008ue}: 
\begin{eqnarray}\label{eq:vec_3half_pos}
  \Gamma_{\nu \mu }^{\frac{3}{2}^+} &=& {V}_{\nu \mu }^\frac{3}{2} \gamma_5 \nonumber\\
  \Gamma_{\nu \mu }^{\frac{3}{2}^-} &=& {V}_{\nu \mu }^\frac{3}{2} ,
\end{eqnarray}
where $V_{\mu \nu}^{\frac32}$ is the vector current for spin three-half resonances and is given by~\cite{LlewellynSmith:1971uhs, Fogli:1979cz}
\begin{eqnarray}\label{eq:vec_axial}
  V_{\nu \mu }^{\frac{3}{2}} &=& \left[ \frac{{C}_3^{p}}{M} (g_{\mu \nu} \slashchar{q} \, - q_{\nu} \gamma_{\mu})+
  \frac{{ C}_4^{p}}{M^2} (g_{\mu \nu} q\cdot p' - q_{\nu} p'_{\mu}) 
  + \frac{{ C}_5^{p}}{M^2} (g_{\mu \nu} q\cdot p - q_{\nu} p_{\mu})\right]    ,
\end{eqnarray}
with ${C}^{p}_i$ being the $\gamma NR$ couplings. The couplings $C^{p}_i;~i=3,4,5$ are related with the helicity amplitudes 
$A_{\frac{1}{2}}$, $A_{\frac{3}{2}}$ and $S_{\frac{1}{2}}$ by the following relations~\cite{Leitner:2008ue}:
\begin{eqnarray}\label{a32}
A_{\frac{3}{2}}^{p} &=& \sqrt{\frac{\pi \alpha}{M} \frac{(M_R \mp M)^2}{M_R^2-M^2}}  
\left[ \frac{C^{p}_3}{M} (M \pm M_R) \pm \frac{C^{p}_4}{M^2} \frac{M_R^2-M^2}{2} \right. \nonumber \\ 
&\pm& \left.
\frac{C^{p}_5}{M^2} \frac{M_R^2-M^2}{2} \right] ,
\end{eqnarray}
\begin{eqnarray}
\label{a12}
 A_{\frac{1}{2}}^{p} &=&\sqrt{\frac{\pi \alpha}{3 M} \frac{(M_R \mp M)^2}{M_R^2-M^2}}  
 \left[\frac{C^{p}_3}{M} \frac{M^2+M M_R }{M_R} - \frac{C^{p}_4}{M^2} \frac{M_R^2-M^2 }{2} \right. \nonumber \\
 &-&\left. \frac{C^{p}_5}{M^2} 
 \frac{M_R^2-M^2}{2} \right] ,\\ 
 \label{s12}
S_{\frac{1}{2}}^{p} &=& \pm \sqrt{\frac{\pi \alpha}{6 M} \frac{(M_R \mp M)^2}{M_R^2-M^2}} 
\frac{\sqrt{ (M_R^2-M^2)^2 }}{M_R^2} \left[ \frac{C^{p}_3}{M} M_R + \frac{C^{p}_4}{M^2} M_R^2 \right. \nonumber \\
&+& \left. \frac{C^{p}_5}{M^2}
\frac{M_R^2+M^2}{2} \right],
\end{eqnarray}
where $A_{\frac32, \frac12}$ and $S_{\frac12}$ are the amplitudes corresponding to the transverse and longitudinal 
polarizations of the photon, respectively. Since in the present work, we have considered $K\Lambda$ production induced by the 
real photon, therefore, the amplitude corresponding to the longitudinal polarization vanishes. Thus, in the numerical 
calculations, we have taken $S_{\frac{1}{2}} =0$. The fitted values of $A_{\frac{1}{2}}$ and $A_{\frac{3}{2}}$ have been taken 
from MAID~\cite{Tiator:2011pw} and PDG~\cite{PDG} for $P_{13}(1720)$ and $P_{13}(1900)$ resonance, respectively, and are quoted 
in Table~\ref{tab:param-p2}. The upper~(lower) sign in Eqs.~(\ref{a32})--(\ref{s12}) represents the positive~(negative) parity 
resonance states.

The most general expression of the hadronic current for the $s$ channel where a resonance state with spin $\frac{3}{2}$, 
$R^{\frac32}$ (with positive or negative parity) is produced and decays to a kaon and a lambda in the final state may be 
written as~\cite{Hernandez:2013jka, Gonzalez-Jimenez:2016qqq}:
\begin{eqnarray}\label{eq:res_had_current_pos}
j^\mu\big|_{R}^{\frac32 \pm} &=& ie~ \frac{g_{RK \Lambda}}{M_{K}} 
   \frac{p_{k}^{\alpha}\Gamma_{s}}{s - M_R^2+ i M_R \Gamma_R}
   \bar u({p}\,') P_{\alpha\beta}^{3/2}(p_R) \Gamma^{\beta\mu}_{\frac32 \pm}
   u({p}\,),\quad p_R=p+q,
\end{eqnarray}
where $\Gamma_{s} = 1 (\gamma_{5})$ for positive~(negative) parity resonances, $ g_{RK \Lambda}$ is the coupling strength for 
$R \to K \Lambda$~($R$ can be any spin $\frac{3}{2}$ resonance given in Table~\ref{tab:included_resonances}), determined from 
partial decay widths. $M_R$ is the mass of the resonance and $\Gamma_R$ is its decay width. For the sake of unitarity restoration, we have  considered the energy dependent decay width of the nucleon resonances,
 which will be discussed in Section~\ref{sec:width}.

In Eq.~(\ref{eq:res_had_current_pos}), $P_{\alpha\beta}^{3/2}$ is spin three-half projection operator and is given by
\begin{equation}
P_{\alpha\beta}^{3/2} (p^{\prime})=- \left(\slashchar{p^\prime} \, + M_R \right) \left( g_{\alpha \beta}
- \frac{2}{3} \frac{p'_{\alpha } p'_{\beta}}{M_R^2} 
+ \frac{1}{3} \frac{p'_{\alpha } \gamma_{\beta} - p'_{\beta } \gamma_{\alpha}}{M_R} 
- \frac{1}{3} \gamma_{\alpha} \gamma_{\beta} \right).
\end{equation}

The coupling strength $ g_{RK \Lambda}$ is determined using the data of branching ratio and decay width of these resonances 
from PDG~\cite{PDG}. 

The expression for the decay rate is obtained by writing the most general form of $RK \Lambda$ Lagrangian~\cite{Leitner:2008ue},
\begin{align}
 \label{eq:spin32_lag} 
 \mathcal{L}_{R_{\frac{3}{2}} K \Lambda} &= \frac{g_{RK \Lambda}}{M_K} \bar{\Psi}_{R_{\frac{3}{2}}}^{\mu} \; 
 \Gamma_{s} \;\partial_\mu K^i \tau_i \,\, \Psi  
\end{align}
where $ g_{RK \Lambda}$ is the $RK \Lambda$ coupling strength. $\Psi$ is the nucleon field and ${\Psi}_{R_{\frac{3}{2}}}^{\mu}$ are 
the fields associated with the spin $\frac32$ resonances. $K^i$ is the kaon field and $\tau$ is isopin factor. The interaction 
vertex $\Gamma_{s}$ is $1$ for positive parity state and $\gamma_5$ for negative parity state. Using the above 
Lagrangian, one may obtain the expression for the decay width in the resonance rest frame as
\begin{align}\label{width32}
\Gamma_{R_{\frac{3}{2}} \rightarrow K \Lambda } &= \frac{\mathcal{C}}{12\pi} \left(\frac{g_{RK\Lambda}}{M_K}\right)^2 
\frac{E_{\Lambda} \pm M_{\Lambda}}{M_R} |\vec p^{\,cm}_{k}|^3,
\end{align}
where the upper~(lower) sign represents the positive~(negative) parity resonance state. The parameter $\mathcal{C}$ is obtained 
from the isospin analysis and found out to be $1$ for isospin $\frac12$ state. The expressions for $|\vec{p}^{\,cm}_{k}|$ and 
$E_{\Lambda}$ are given in Eqs.~(\ref{eq:pi_mom}) and (\ref{eq:elam}), respectively.

Using the above expressions for decay width, the couplings for $ {R \frac32} \to K \Lambda$ are obtained and given in 
Table-\ref{tab:included_resonances} for the spin $\frac{3}{2}$ resonances considered in this work. 

\begin{table*}
  \caption{Properties of the resonances included in the present model, with Breit-Wigner mass $M_R$, spin $J$, isospin $I$, 
  parity $P$, the total decay width $\Gamma$, the branching ratio into $K \Lambda$ and $g_{RK \Lambda}$ stands for the coupling 
  strength at the $RK \Lambda$ vertex.} \label{tab:included_resonances}
  \begin{center}
    \begin{tabular*}{129mm}{@{\extracolsep{\fill}}c c c c c c c c}
      \noalign{\vspace{-8pt}}
      \hline \hline
      Resonances               & $M_R$ [GeV] & I \quad    & J\quad   &    P   & $\Gamma$  &  $K \Lambda$ branching  
      &   $g_{RK\Lambda}$  \\
      $R_{2I2J}$ &&&&&(GeV)&ratio ($\%$)  \\ \hline
      $S_{11}$(1650)  &       $1.655 \pm 0.015$ & $\frac{1}{2}$ &$ \frac{1}{2}$ &$ -$ &   $ 0.135 \pm 0.035$ &   $ 10 \pm 5$  & 
      $0.45$ \\ \hline
      
      $P_{11}$(1710)  &       $1.700 \pm 0.020$ & $\frac{1}{2}$ &$ \frac{1}{2}$ &$ +$ &   $ 0.120 \pm 0.040$ &  $ 15\pm 10$ & 
      $-0.61 $ \\ \hline
      
      $P_{13}$(1720)  &       $1.675 \pm 0.015$ & $\frac{1}{2}$ &$ \frac{3}{2}$ &$ +$ &   $ 0.250 \pm^{0.150}_{0.100}$ &  $ 4.5 
      \pm 0.5$ &     $1.73$     \\ \hline
      
      $P_{11}$(1880)  &       $1.860 \pm 0.040$ & $\frac{1}{2}$ &$ \frac{1}{2}$ &$ +$ &   $ 0.230 \pm 0.050$ &  $ 20 \pm 8$ &  
      $0.52$    \\ \hline
      
      $S_{11}$(1895)  &       $1.910 \pm 0.020$ & $\frac{1}{2}$ &$ \frac{1}{2}$ &$ -$ &   $ 0.110 \pm 0.030$ &   $ 18 \pm 5$ & 
      $ 0.28$       \\ \hline
      
      $P_{13}$(1900)  &       $1.920 \pm 0.020$ & $\frac{1}{2}$ &$ \frac{3}{2}$ &$ +$ &   $ 0.150 \pm 0.050$ &   $ 11 \pm 9$ &  
      $-0.62$       \\ \hline      \hline
    \end{tabular*}
  \end{center}
\end{table*}

\subsubsection{Energy dependent decay widths of the nucleon resonances}\label{sec:width}
As already discussed in the introduction that the unitarity can be restored, even at the tree level, 
 if widths for the various nucleon resonances are taken to be energy dependent. In the present work, we have considered the energy dependent decay widths to be of the following form~\cite{Skoupil:2018vdh}
\begin{equation}
\Gamma_R(W)=\Gamma_{R}\frac{W}{M_{R}}\sum_i\left[x_i \left(\frac{|\vec{q}_i|}{|\vec{q}_i^{R}|}\right)^{2l+1}\!\frac{D_l(|\vec{q}_i|)}{D_l(|\vec{q}_i^{\,R}|)}\right],
\label{eq:Gamma-s}
\end{equation}
where the sum $i$ runs over all possible meson-baryon decay modes, with the relative orbital momentum $l$. $\Gamma_{R}$ and $x_i$ denote the total decay width, and the branching ratio of a resonance into different meson-baryon channels~\cite{PDG}, respectively. The momenta $|\vec{q}_i|$ and $|\vec{q}_i^{R}|$ have the following form:
\begin{eqnarray}
|\vec{q}_i^{R}|&=& \sqrt{\frac{(M_{R}^2-M_{B}^2+M_{m}^2)^2}{4M_{R}^{2}}-M_{m}^2},
\end{eqnarray}
\begin{eqnarray}
|\vec{q}_i|&=& \sqrt{\frac{(W^{2}-M_B^2+M_m^2)^2}{4W^{2}}-M_m^2},\;\;\text{and}\\
D_l(x) &=& \texttt{exp}\left(-\frac{x^2}{3\alpha^2}\right),
\end{eqnarray}
which is consistent with the value of $\alpha=400$~MeV taken in Ref.~\refcite{Skoupil:2018vdh}, and $x= |\vec{q}_i|$ or $|\vec{q}_i^{R}|$. 
 
In the case of energy dependent widths, $\Lambda_{B}$ and $\Lambda_{R}$ are taken to be $\Lambda_{B} = 0.505$ GeV and $\Lambda_{R} = 1.32$ GeV, respectively, while in the case of fixed widths, these values are $\Lambda_{B} = 0.525$ GeV and $\Lambda_{R} = 1.1$ GeV.

\subsubsection{Spin $\frac{1}{2}$ hyperon resonances}\label{spin12Y}
Along with the nucleon resonance exchange contributions in the $s$ channel, we have also taken into account the hyperon 
resonances exchanged in the $u$ channel. In the present work, we have taken two lambda resonances, $\Lambda^{*} (1405)$ and 
$\Lambda^{*} (1800)$ with $J^{P} = \frac{1}{2}^{-}$ in the $u$ channel. The Lagrangians for the strong and the electromagnetic 
vertices in the case of $\Lambda^{*}$ exchange are given as~\cite{SLA, Feuster:1998cj, Benmerrouche:1994uc, Feuster:1996ww}:
\begin{eqnarray}
 {\cal L}_{\gamma \Lambda \Lambda^{*}} &=& e \frac{\kappa_{\Lambda \Lambda^{*}}}{2(M_{\Lambda^{*}} + M_{\Lambda})} 
 \bar{\psi}_{{\Lambda}^{*}} \sigma_{\mu \nu} \Gamma_{s} \psi_{\Lambda} F^{\mu \nu} + h.c., \\
 {\cal L}_{p K \Lambda^{*}} &=& \frac{g_{p K \Lambda^{*}}}{f_{\pi}} (\partial^{\mu} K^{\dagger}) \bar{\psi}_{p} \Gamma_{\mu} 
 \psi_{\Lambda^{*}} + h.c.,
\end{eqnarray}
with $\kappa_{\Lambda \Lambda^{*}}$, the transition magnetic moment between $\Lambda$ and $\Lambda^{*}$. $g_{p K \Lambda^{*}}$ 
is the coupling strength at $p K \Lambda^{*}$ vertex. $\Gamma_{s} = 1 (\gamma_{5})$ and $\Gamma_{\mu} = \gamma_{\mu} (\gamma_{\mu} 
\gamma_{5})$ for the positive~(negative) parity resonances.

\begin{table}
\caption{Values of the helicity amplitude $A_{\frac{1}{2}}$ and $A_{\frac{3}{2}}$ for the different nucleon resonances. The 
values for $S_{11}(1650)$ and $P_{13}(1720)$ are taken from MAID~\cite{Tiator:2011pw}. For the rest of the resonances, the 
parameterization of $A_{\frac{1}{2}}$ and $A_{\frac{3}{2}}$ are not available in MAID and are taken from PDG~\cite{PDG}.}
\label{tab:param-p2} 
\begin{center}
\begin{tabular*}{119mm}{@{\extracolsep{\fill}}ccc}
\hline \hline
 Resonance & \multicolumn{2}{c}{Helicity amplitude} \\
   & $A_{\frac{1}{2}}$ ($10^{-3}$ GeV$^{-1/2}$)  & $A_{\frac{3}{2}}$  ($10^{-3}$ GeV$^{-1/2}$)  \\
\hline
$S_{11}(1650)$ &  $ 33.3$ & -   \\
$P_{11}(1710)$ &  $ 50$ & - \\
$P_{13}(1720)$ &  $ 73$ & $-11.5$  \\
$P_{11}(1880)$ &  $ 21$ & -  \\
$S_{11}(1895)$ &  $ -16$ & -  \\
$P_{13}(1900)$ &  $ 24$ & $-67$  \\
\hline \hline
\end{tabular*}
\end{center}
\end{table}

Using the above Lagrangians, the hadronic current for the $\Lambda^{*}$ resonance exchange may be written as
\begin{eqnarray}
 J_{\mu} \big|_{\Lambda^{*}\pm} &=& ie \bar{u} (p^{\prime}) \frac{g}{M_{\Lambda} + M_{\Lambda^{*}}} \sigma_{\mu \nu} 
 q^{\nu} \Gamma_{s}\left( \frac{\slashed{p}^{\prime} - \slashed{q} + M_{\Lambda^{*}}}{u - {M_{\Lambda^{*}}^2} + i 
 M_{\Lambda^{*}} \Gamma_{\Lambda^{*}}} \right)  \nonumber \\
&\times& \slashed{p}_{k} \gamma_{5} \Gamma u(p),
\end{eqnarray}
with $g= \kappa_{\Lambda \Lambda^{*}} g_{p K \Lambda^{*}}/{f_{\pi}}$, $M_{\Lambda^{*}}$ and $\Gamma_{\Lambda^{*}}$ being the 
mass and the decay width of $\Lambda^{*}$. Unlike the nucleon resonances where the strong and electromagnetic couplings are 
determined phenomenologically by the partial decay width and the helicity amplitudes, respectively, the experimental data is not 
adequate in the case of hyperon resonances, to determine these couplings. Therefore, the parameter $g$ is treated as a free 
parameter to be fitted to the experimental data. The values of the different parameters of the $\Lambda^{*}$ taken in the 
present model are summarized in Table~\ref{hyperon_resonances}.  

\subsubsection{Spin 1 kaon resonances}\label{spin1K}
In the present work, we have considered two kaon resonances in the $t$ channel: a vector meson $K^{*} (892)$ and an axial 
vector meson $K_{1} (1270)$. The Lagrangians for the electromagnetic and strong vertices, when a vector kaon is exchanged in 
the $t$ channel, are given by~\cite{SLA, Feuster:1998cj, Benmerrouche:1994uc, Feuster:1996ww}:
\begin{eqnarray}\label{Kstar:em}
{\cal L}_{\gamma K K^{*}} &=& i \frac{e \kappa_{K K^{*}}}{4 \mu} \epsilon_{\mu \nu \lambda \sigma} F^{\mu \nu} V^{\lambda 
\sigma} K, \\
\label{Kstar:had}
{\cal L}_{K^{*} \Lambda p} &=& - \left(g_{K^{*} \Lambda p}^{v} \bar{\psi}_{\Lambda} \gamma_{\mu} \psi_{p} V^{\mu} - 
\frac{g_{K^{*} \Lambda p}^{t}}{2(M+M_{\Lambda})} \bar{\psi}_{\Lambda} \sigma_{\mu \nu}V^{\mu \nu} \psi_{p} \right) + h.c., 
\end{eqnarray}
where $\kappa_{K K^{*}}$ is the coupling strength of the $\gamma K K^{*}$ vertex, $\mu$ is an arbitrary mass factor which is 
introduced to make the Lagrangian dimensionless. $\mu$ is chosen to be 1 GeV. The vector meson tensor $V^{\mu \nu}$ is defined 
as $V^{\mu \nu} = \partial^{\nu} V^{\mu} - \partial^{\mu} V^{\nu}$, with $V^{\mu}$, the vector kaon field. $g_{K^{*} \Lambda 
p}^{v}$ and $g_{K^{*} \Lambda p}^{t}$ are the vector and the tensor couplings, respectively, at the strong $K^{*} \Lambda p$ 
vertex.
\begin{table*}
  \caption{Properties of the hyperon and the kaon resonances included in the present model, with mass $M_R$, spin $J$, isospin 
  $I$, parity $P$, the total decay width $\Gamma$, the coupling parameter $g$ for the hyperon resonances, and the vector 
  $G_{K}^{v}$ and tensor $G_{K}^{t}$ couplings for the kaon resonances. It is to be noted that these couplings $g$, $G_{K}^{v}$ 
  and $G_{K}^{t}$ contains both the electromagnetic as well as the strong coupling strengths.}\label{hyperon_resonances}
\small{
  \begin{center}
    \begin{tabular*}{132mm}{@{\extracolsep{\fill}}c c c c c c c c c}
      \noalign{\vspace{-8pt}}
      \hline \hline
      Resonances               & $M_R$ [GeV] & J\quad   & I \quad    &   P   & $\Gamma$  &  $g$  & $G_{K}^{v}$
      &   $G_{K}^{t}$  \\
      &&&&&(GeV)& &  \\ \hline
      $\Lambda^{*}$ (1405)  &  $1.405 \pm^{0.0013}_{0.001}$ & $\frac{1}{2}$ &$ 0$ &$ -$ & $ 0.0505 \pm 0.002$ &  $-10.18$  &  
      -&-
      \\ \hline
      
      $\Lambda^{*}$ (1800)  &  $1.800 \pm_{0.050}^{0.080}$ & $\frac{1}{2}$ &$ 0$ &$ -$ &  $ 0.300 \pm 0.100$ &  $ -4.0$ &  -& -
      \\ \hline
      
      $K^{*}$(892) & $0.89166 \pm 0.00026$ & $1$ &$ \frac{1}{2}$ &$ -$ &   $ 0.0508 \pm 0.0009$ & - &  $-0.18$ & $0.02$ \\
      \hline
      
      $K_{1}$(1270)  & $1.272 \pm 0.007$ & $1 $ &$ \frac{1}{2}$ &$ +$ &   $ 0.090 \pm 0.020$ & - & $0.28$ & $-0.28$   \\  \hline
      \hline
    \end{tabular*}
  \end{center}
}
\end{table*}

Using the Lagrangians given in Eqs.~(\ref{Kstar:em}) and (\ref{Kstar:had}), the hadronic current for the $K^{*}$ exchange is 
obtained as
\begin{eqnarray}
 J_{\mu} \big|_{K^{*}} &=& ie \bar{u} (p^{\prime}) \epsilon_{\mu \nu \rho \sigma} q^{\rho} (p^{\prime} - p)^{\sigma} \left(
 \frac{ -g^{\nu \alpha} + (p - p^{\prime})^{\nu} (p - p^{\prime})^{\alpha}/{M_{K^{*}}^{2}}}{t - {M_{K^{*}}^{2}} + i 
 M_{K^{*}} \Gamma_{K^{*}}} \right) \nonumber \\
 &\times&
 \left[G_{K^{*}}^{v} \gamma_{\alpha} + \frac{G_{K^{*}}^{t}}{M + M_{\Lambda}} 
 (\slashed{p}^{\prime} - \slashed{p}) \gamma_{\alpha} \right] u(p),
 \end{eqnarray}
with $G_{K^{*}}^{v} = \kappa_{K K^{*}} g_{K^{*} \Lambda p}^{v}/\mu$ and $G_{K^{*}}^{t} = \kappa_{K K^{*}} g_{K^{*} \Lambda 
p}^{t}/\mu$. $M_{K^{*}}$ and $\Gamma_{K^{*}}$ are the mass and width of the $K^{*}$ resonance, respectively. Due to the lack of 
the experimental data on the $K^{*}$ and $K_{1}$ resonances, the values of $G_{K^{*}}^{v}$ and $G_{K^{*}}^{t}$ can not be 
determined phenomenologically and are treated as free parameters to be fitted to the experimental data of the $K \Lambda$ 
production and are quoted in Table~\ref{hyperon_resonances}. 

The Lagrangian for the electromagnetic and strong vertices, when an axial vector kaon resonance is exchanged in the $t$ channel, 
is given by~\cite{SLA, Feuster:1998cj, Benmerrouche:1994uc, Feuster:1996ww}:
\begin{eqnarray}\label{K1:em}
{\cal L}_{\gamma K K_{1}} &=& i \frac{e \kappa_{K K_{1}}}{\mu} \partial_{\mu} A_{\nu} {\cal V}_{p}^{\mu\nu} K ,\\
\label{K1:had}
{\cal L}_{K_{1} \Lambda p} &=& - \left(g_{K_{1} \Lambda p}^{v} \bar{\psi}_{\Lambda} \gamma_{\mu} \gamma_{5} \psi_{p} 
{\cal V}_{p}^{\mu} - \frac{g_{K_{1} \Lambda p}^{t}}{2(M+M_{\Lambda})} \bar{\psi}_{\Lambda} \sigma_{\mu \nu} \gamma_{5} 
{\cal V}_{p}^{\mu \nu} \psi_{p} \right) + h.c. ,
\end{eqnarray}
where $\kappa_{K K_{1}}$ is the coupling strength of the electromagnetic $\gamma K K_{1}$ vertex. The axial vector meson tensor 
${\cal V}_{p}^{\mu \nu}$ is defined as ${\cal V}_{p}^{\mu \nu} = \partial^{\nu} {\cal V}_{p}^{\mu} - \partial^{\mu} {\cal 
V}_{p}^{\nu}$, with ${\cal V}_{p}^{\mu}$, the axial vector kaon field. $g_{K_{1} \Lambda p}^{v}$ and $g_{K_{1} \Lambda p}^{t}$ 
are the vector and the tensor couplings, respectively, at the strong $K_{1} \Lambda p$ vertex.

The hadronic current for the axial vector kaon $K_{1}$ exchange in the $t$ channel is obtained, using Eqs.~(\ref{K1:em}) and 
(\ref{K1:had}), as
\begin{eqnarray}
 J_{\mu} \big|_{K_{1}} &=& ie \bar{u} (p^{\prime}) [g_{\alpha \mu} q \cdot (p - p^{\prime}) - q_{\alpha} (p - 
 p^{\prime})_{\mu}] \left(\frac{-g^{\alpha \rho} + (p - p^{\prime})^{\alpha} (p - p^{\prime})^{\rho}/{M_{K_{1}}^{2}}}
 {t - {M_{K_{1}}^{2}} + i M_{K_{1}} \Gamma_{K_{1}}} \right) \nonumber \\
 &\times& \left[G_{K_{1}}^{v} \gamma_{\rho} \gamma_{5} + \frac{G_{K_{1}}^{t}}{M + M_{\Lambda}} (\slashed{p}^{\prime} - 
 \slashed{p}) \gamma_{\rho} \gamma_{5} \right] u(p),
 \end{eqnarray}
with $G_{K_{1}}^{v} = \kappa_{K K^{*}} g_{K_{1} \Lambda p}^{v}/\mu$ and $G_{K_{1}}^{t} = \kappa_{K K^{*}} g_{K_{1} \Lambda 
p}^{t}/\mu$. $M_{K_{1}}$ and $\Gamma_{K_{1}}$ are the mass and width of the $K_{1}$ resonance, respectively. The values of 
$G_{K_{1}}^{v}$ and $G_{K_{1}}^{t}$ are treated as free parameters to be fitted to the experimental data of the $K \Lambda$ 
production and are quoted in Table~\ref{hyperon_resonances}.

In analogy with the non-resonant terms, in the case of resonances we have considered the following form factors at the strong 
vertex, in order to take into account the hadronic structure:
\begin{equation}
F^{*}_{x} (x) = \frac{\Lambda_{R}^{4}}{\Lambda_{R}^{4} + (x - M_{x}^{2})^{2}},
\end{equation}
where $\Lambda_{R}$ is the cut-off parameter whose value is fitted to the experimental data, $x$ represents the Mandelstam 
variables $s,~t,~u$ and $M_{x} = M_{R},~M_{K^{*}},~M_{K_{1}},~M_{Y^{*}},$ corresponding to the mass of the nucleon, kaon or 
hyperon resonances exchanged in the $s,~t$, and $u$ channels. In the case of nucleon resonances, the value of the cut-off 
parameter $\Lambda_{R}$ is fitted to be $\Lambda_{R}=1.32$ GeV, while in the case of kaon and hyperon resonances, the value of 
$\Lambda_{R}$ is taken to be $\Lambda_{R} = \Lambda_{B} =$ 0.505 GeV.

\section{Results and discussions}\label{results}

We have used Eq.~(\ref{dsig}) to numerically evaluate the differential cross section $\left.\frac{d\sigma}{d\cos 
\theta_{k}}\right|_{CM}$ and the total cross section $\sigma$ is obtained by integrating Eq.~(\ref{dsig}) over the 
polar angle {\it i.e.} 
\begin{equation}\label{sigma}
 \sigma = \int_{\cos \theta_{k}^{min}}^{\cos \theta_{k}^{max}} \frac{1}{32 \pi s} \frac{|\vec{p}\;^{\prime}|}{|\vec{p}|} 
\overline{\sum_{r}} \sum |\mathcal{M}^{r}|^2 d\cos \theta_{k},
\end{equation}
where $\cos \theta_{k}^{min}~(\cos \theta_{k}^{max})$ is taken to be $-1~(+1)$ in order to cover the full range of the scattering angle.

In the expression for the transition amplitude $\mathcal{M}^{r}$ in the aforementioned equation, we have taken the contributions 
from the background and the resonance terms and added them coherently. Therefore, the hadronic current of the full model is 
expressed as:
\begin{eqnarray}\label{j:full}
 J^{\mu}|_{Full} &=& J^{\mu}|_{s} + J^{\mu}|_{t} + J^{\mu}|_{u\Lambda} + J^{\mu}|_{u\Sigma} + J^{\mu}|_{CT} +J^{\mu}|_{add} + J^{\mu}|_{R} +
 J^{\mu}|_{\Lambda^{*}}  \nonumber \\
 &+& J^{\mu}|_{K^{*}} + J^{\mu}|_{K_{1}},
\end{eqnarray}
where $J^{\mu}|_{add}$~(given in Eq.~(\ref{GI2})) ensures the gauge invariance of the total hadronic current, and $J^{\mu}|_{R}$ and $J^{\mu}|_{\Lambda^{*}}$ are expressed as
\begin{eqnarray}
 J^{\mu}|_{R} &=& J^{\mu}|_{S_{11}(1650)} + J^{\mu}|_{P_{11} (1710)} + J^{\mu}|_{P_{13} (1720)} + J^{\mu}|_{P_{11} (1880)} \nonumber \\
 &+& 
 J^{\mu}|_{S_{11} (1895)} + J^{\mu}|_{P_{13} (1900)},\\
 J^{\mu}|_{\Lambda^{*}} &=& J^{\mu}|_{\Lambda^{*} (1405)} + J^{\mu}|_{\Lambda^{*} (1800)}.
\end{eqnarray}
The expressions of $J^{\mu}$ appearing in the above equations are given explicitly in Section~\ref{Formalism}.

The background terms consist of the non-resonant {\it i.e.} $s$, $t$, $u$ and contact terms as well as the kaon and hyperon 
resonances exchanged in the $t$ and $u$ channels. The nucleon resonances exchanged in the $s$ channel constitute the resonance 
contribution. 
 This nomenclature for the resonance and the background terms is used because all the terms used in calculating the background contribution
 do not resonate in the physical region while the $s$ channel resonances do so.
The strong and electromagnetic couplings of the non-resonant terms are predicted by the non-linear sigma model 
with the chiral SU(3) symmetry. For the $s$, $t$ and $u$ channels, a dipole parameterization of the hadronic form 
factors~(Eq.~(\ref{FF_Born})) is used while for the contact term, the prescription given by Davidson and 
Workman~\cite{DW}~(Eq.~(\ref{FF_CT})) is used. In the case of nucleon resonances, the energy dependent decay widths~(unless stated otherwise) of the different resonances as discussed in Section~\ref{sec:width}, are taken into account.
The 
electromagnetic~($\gamma NR$) and strong~($RK\Lambda$) couplings of the $s$ channel resonances~($R$) are deduced, respectively, 
from the helicity amplitudes of the $\gamma N \rightarrow R$ transitions and the partial decay width of the resonances~($R$) to $K\Lambda$ channel 
and those of the $u$ channel~($Y^{*}$) and $t$ channel~($K^{*},~K_{1}$) resonances are fitted to reproduce the 
current experimental data available in this energy region. For the numerical calculations, we have taken the cut-off parameter for the background and resonance terms, respectively, to
be $\Lambda_{B} =0.505$~GeV and $\Lambda_{R} =1.32$~GeV for the energy dependent decay widths of the nucleon resonances, while, for comparison, we have also taken the fixed widths, for which the best fit 
for the total scattering cross section~(Fig.~\ref{fig_data}) is obtained with $\Lambda_{B} =0.525$~GeV and $\Lambda_{R} =1.1$~GeV, whereas in other calculations, these parameters are taken as $\Lambda_{B} = 1.235$~GeV 
and $\Lambda_{R} = 1.864$~GeV in Ref.~\refcite{Bydzovsky:2019hgn}, and $\Lambda_{B} = 0.70$~GeV and $\Lambda_{R} = 1.31$~GeV in Ref.~\refcite{Suciawo:2017wtv}. The numerical results are presented for the total and differential cross sections 
and are compared with the available experimental data from CLAS and SAPHIR as well as with some of 
the recent theoretical models.

In the following, we present the results of the total cross section $\sigma$ as a function of CM energy $W$ in 
section~\ref{results_total} and the results of the differential  cross section $\left.\frac{d\sigma} {d\cos 
\theta_{k}} \right|_{CM}$ as a function of $\cos \theta_{k}^{CM}$ for fixed $W$ as well as a function of 
 $W$ for fixed $\cos\theta_{k}^{CM}$ in section~\ref{results_diff}. 

\subsection{Total cross section}\label{results_total} 
\subsubsection{Discussion of theoretical results}

\begin{figure}
\begin{center}
 \includegraphics[height=7cm,width=13cm]{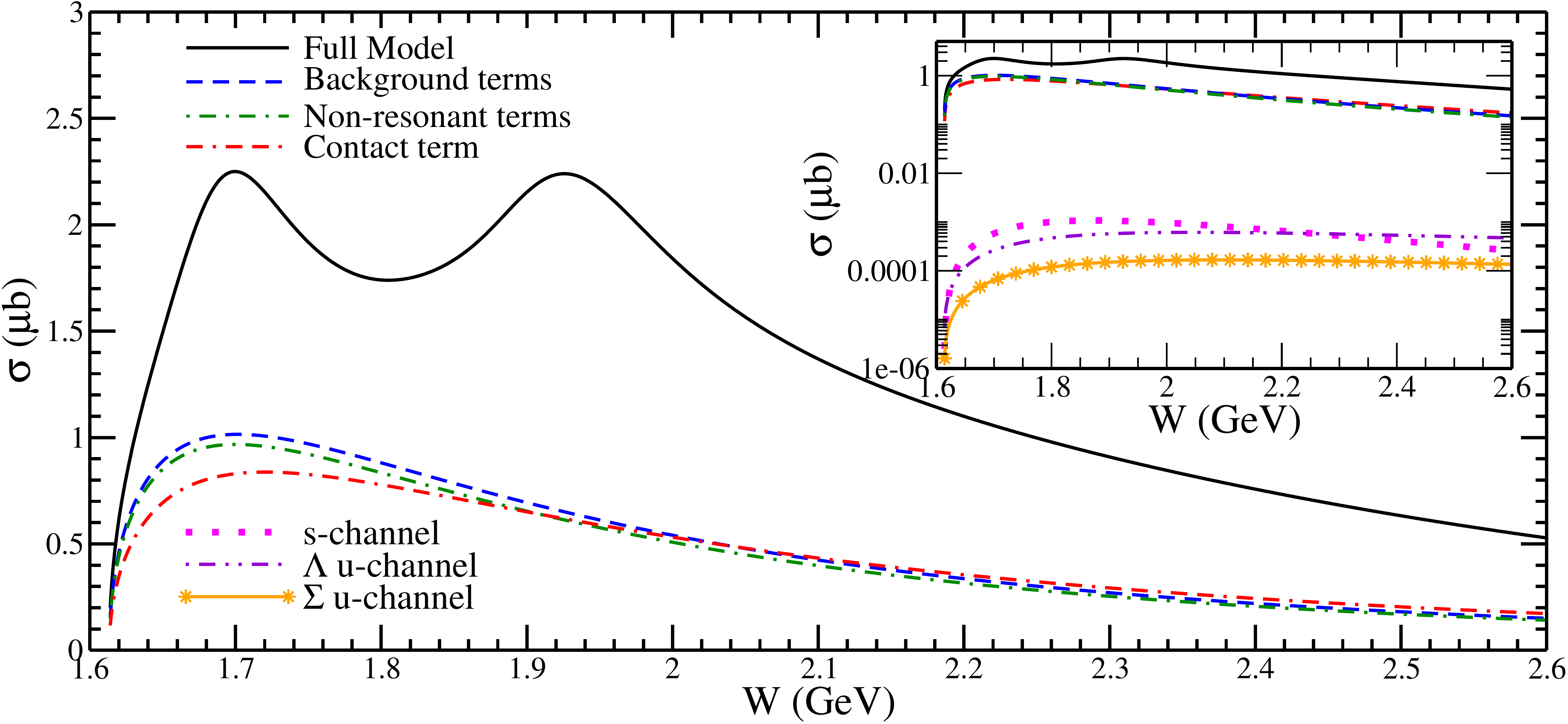}
\caption{Total cross section $\sigma$ as a function of CM energy $W$ for the process $ \gamma + p \rightarrow K^{+} + 
\Lambda$. Solid line represents the results of the full model of the present work by taking into account the $W$ dependent decay widths of the nucleon resonances as discussed in Section~\ref{sec:width}. Dashed line, dashed-dotted line, 
double-dashed-dotted line represents the results of the background terms, non-resonant terms and the contact term, respectively. 
In addition to the aforementioned four cases, in the inset~(note the log scale), the individual contribution from the nucleon exchanged in the $s$ 
channel, $\Lambda$ exchanged in the $u$ channel and $\Sigma$ exchanged in the $u$ channel have been shown, respectively, by the 
dotted line, double-dotted-dashed line and solid line with star symbol.}
\label{fig_born}
\end{center}
\end{figure}
In Fig.~\ref{fig_born}, we have presented the results of $\sigma$ {\it vs.} $W$ for the $K\Lambda$ production induced by the 
photon beam off the proton target. The results have been presented separately showing the contributions
 from the background terms as well as the total contributions by
including the well established nucleon resonances in the $s$ channel with spin- 1/2 and 3/2 lying below $W=2$~GeV, which have been
 considered in this paper~(Table~\ref{tab:included_resonances}). The 
individual contribution from the non-resonant terms i.e. individual contribution of $s$, $t$ and $u$ channel Born terms as well 
as the contact term, are also shown separately. It may be observed that the 
contact term is the dominant one among the non-resonant terms. At low and intermediate $W$ {\it i.e.} from threshold up to 2 
GeV, the contact term has smaller contribution than the total non-resonant contribution, however, at high $W$, beyond 2 GeV, 
the contact term has a larger contribution than the contribution from the total non-resonant terms. For example, in the peak 
region, $W=1.7$ GeV, individually the contact term is $\sim 15\%$ smaller than the non-resonant terms while at $W=2.6$ GeV, the 
contact term contributes $\sim 20 \%$ more than the non-resonant terms. The contributions of the hyperon and the kaon resonances 
are small but increase the value of the total cross section. In the inset of this figure, we have also shown the incoherent 
contribution from the $s$ and $u$ channel Born terms. It may be observed from the figure that the incoherent contributions of 
the $s$ and $u$ channels are very small as compared to the results of the full model, however, their interference with the 
contact term and with the different resonances considered in the $s$ channel when all the amplitudes are added coherently, has 
a significant contribution to the cross section~(not explicitly shown here). It is worth mentioning that in the present model, the contribution of the non-resonant terms including the contact term is relatively small as compared to the other isobar model using SU(3) symmetry, even though the value of the coupling constants are similar. 
 This is mainly because of the smaller value of the cut-off parameter $\Lambda_{B}$. Moreover, the smaller
  value of $\Lambda_{B}$ used in the strong form factors of $t$ and $u$ channel diagrams mediated by the kaon and hyperon resonances also suppresses their contribution. Both these effects prevent the cross section from rising and obtain better agreement with the experiments in the region of low energy.

\begin{figure}
\begin{center}
 \includegraphics[height=7cm,width=13cm]{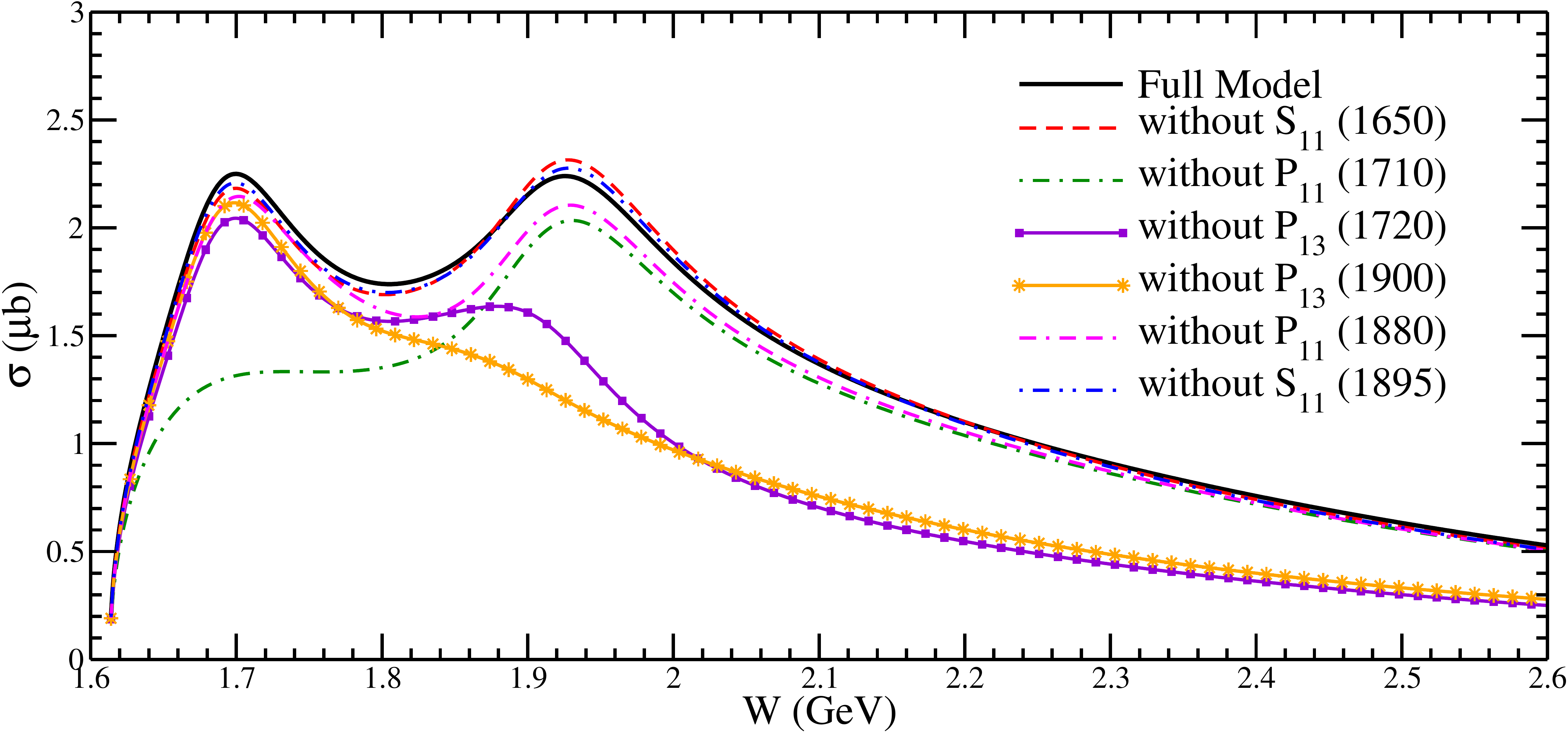}
\caption{$\sigma$ {\it vs.} $W$ for the process $ \gamma + p \rightarrow K^{+} + \Lambda$. Solid line represents the results of 
the full model of the present work, dashed line, dashed-dotted line, solid line with square, solid line with star, 
double-dashed-dotted line and double-dotted-dashed line represents the results of the full model when $S_{11}(1650)$, $P_{11} 
(1710)$, $P_{13}(1720)$, $P_{13}(1900)$, $P_{11}(1880)$ and $S_{11} (1895)$ resonance, respectively, is not taken into 
account.}\label{fig2}
\end{center}
\end{figure}

 The impact of each resonance considered in this 
work on the total contribution has been explicitly discussed in Fig.~\ref{fig2}, where we have depicted the effect of individual contributions from the various $s$ channel resonances considered 
in the present work. The incoherent contribution of the individual resonances is comparatively low as compared to the total cross section 
but the interference of the $s$ channel resonance with the background terms when all the amplitudes are added coherently, contributes significantly to the total cross section. In this figure, we have presented the results of the full model when a particular 
resonance is switched off. The comparison of the results of the full model with the result when a particular resonance is 
switched off shows the significance~(without the experimental data) of that particular resonance. 

One may observe from the figure that $P_{11} (1710)$ resonance has a significant 
effect on the total cross section in the region $W=1.61 - 2.3$ GeV which becomes small for $W>2.3$ GeV. The absence of $P_{11} (1710)$ 
reduces the first peak by $\sim 42\%$ while the second peak is reduced by $\sim 9\%$. In the dip region, the total cross section in the absence of $P_{11}(1710)$ is reduced by $\sim 22\%$. The contribution of 
$P_{13}(1720)$ and $P_{13}(1900)$ resonances are 
important for the $K\Lambda$ production at all values of $W$. It must be noted that when the 
contribution from $P_{13} (1720)$ or $P_{13}(1900)$ resonance is excluded, the results beyond $W=2$~GeV are suppressed significantly. 
Although the 
first peak and the dip region are not much affected~(reduced by about $10\%$) by the absence of the $P_{13} (1720)$ resonance, the second peak is suppressed by $\sim 20\%$ as well 
as it is shifted from $W=1.92$ GeV to $W=1.88$ GeV. In the absence of $P_{13} (1900)$ resonance, the cross section in the first peak is reduced by about $6\%$, while it is reduced by $13\%$ in the dip region and the second peak in the energy region of $W$ around 1.9 GeV does not appear. It must be pointed out that beyond $W>1.7$ GeV, 
$P_{13}(1900)$ has the most dominant contribution followed by $P_{13} (1720)$ resonance. At $W=2.6$~GeV, by switching $P_{13}(1720)$ or $P_{13}(1900)$ resonance off, the total cross section is reduced by $\sim 53\%$ and $47\%$, respectively.
The effect of $P_{11}(1880)$ is small in the entire range of $W$ in which the cross section reduces to $\sim 5-10\%$, if this resonance is switched off. 
The two $S_{11}$ resonances, $viz.$ 
$S_{11}(1650)$ and $S_{11} (1895)$, have very small effect on the total cross 
section in the entire range of $W$. 

\subsubsection{Comparison with the experimental data}
\begin{figure}
\begin{center}
 \includegraphics[height=7cm,width=13cm]{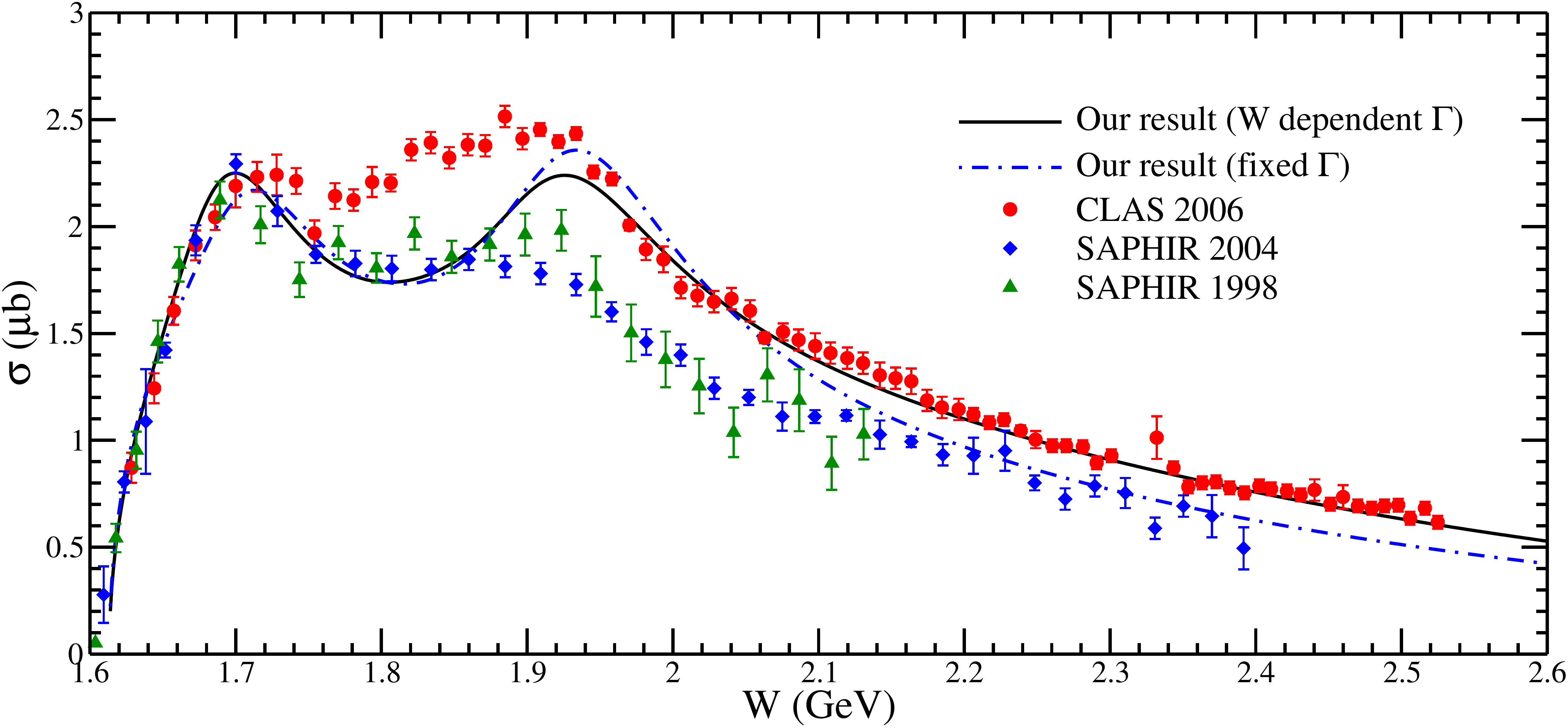}
\caption{$\sigma$ {\it vs.} $W$ for the process $ \gamma + p \rightarrow K^{+} + \Lambda$. Solid and dashed-dotted lines, respectively, show the results of the present model taking the $W$ dependent decay widths of the nucleon resonances as discussed in Sect.~\ref{sec:width} and the fixed values of the decay widths as listed in Table~\ref{tab:included_resonances}.
The experimental data has been taken 
from the CLAS 2006~\cite{CLAS05}~(solid circle), SAPHIR 2004~\cite{SAPHIR}~(solid diamond) and SAPHIR 1998~\cite{Tran:1998qw}~(solid 
triangle).}\label{fig_data}
\end{center}
\end{figure}

\begin{figure}
\begin{center}
 \includegraphics[height=7cm,width=13cm]{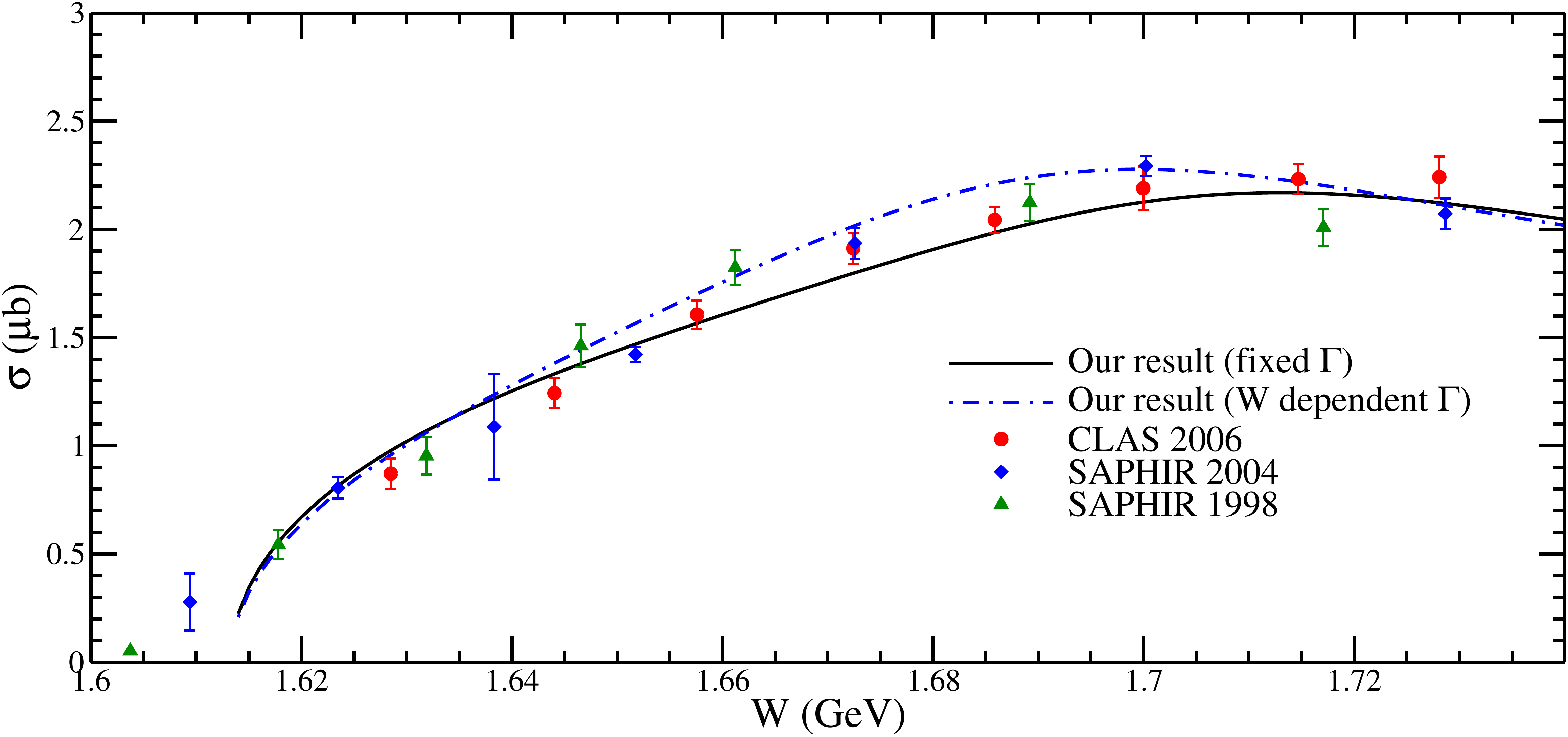}
\caption{$\sigma$ {\it vs.} $W$ for the process $ \gamma + p \rightarrow K^{+} + \Lambda$ in the threshold region. Lines and 
points have the same meaning as in Fig.~\ref{fig_data}.}\label{fig_threshold}
\end{center}
\end{figure}

 In Figs.~\ref{fig_data} and \ref{fig_threshold}, we have shown the results for the total scattering cross section $\sigma$ {\it vs.} $W$ obtained 
 for the full model~(Eq.~(\ref{j:full})) by taking both energy dependent as well as fixed decay widths of the nucleon resonances into account and compared them with the experimental results available from CLAS 2006~\cite{CLAS05}, SAPHIR 2004~\cite{SAPHIR} and 
SAPHIR 1998~\cite{Tran:1998qw}. To show the significance of the present model in the threshold region, we have presented in 
Fig.~\ref{fig_threshold} our results in the range 1.6 GeV $< W <$ 1.72 GeV. Before we compare our results with the experimental 
results from CLAS and SAPHIR, the main features of the data on the total cross section can be classified with some distinct features 
in three kinematic regions summarized as:
\begin{itemize}
 \item [(i)] $W<$ 1.72 GeV\\
 In the region of $W <$ 1.72 GeV, all the experimental data show a continuous rise with $W$ in the kinematic region from 
 threshold up to 1.72 GeV.\\
 
 \item [(ii)] 1.72 GeV $<W<$ 1.92 GeV\\
 In this region of $W$, the data from SAPHIR 1998 and SAPHIR 2004 are fairly consistent with each other, both being lower than 
 the data from CLAS 2006. All the three data show double peaks at around $W \simeq 1.7$ and 1.9 GeV with a minimum around $W=1.75$~GeV in the CLAS 2006 data and at $W = 
 1.8$ GeV in the SAPHIR data. In both of the SAPHIR data, the later peak at $W=1.9$ GeV is lower than the earlier peak at $W=1.7$ GeV while the CLAS 
 2006 data show a moderate two peak structure at $W=1.7$ and 1.9 GeV, in which the minimum is considerably mild and shifted to 
 lower $W$, {i.e.}, at $W=1.74$ GeV. Moreover, unlike the SAPHIR data, the second peak at $W=1.9$~GeV is higher than the first peak at $W=1.7$~GeV.
   
 \item [(iii)] $W \ge 1.92$ GeV\\
 In the region of $W \ge 1.92$ GeV, the CLAS 2006 data are fairly in agreement with both the SAPHIR data in shape but are significantly higher than both the SAPHIR data
 in the entire range of 1.92 GeV $\le W \le $ 2.4 GeV while both the SAPHIR data are 
 reasonably consistent with each other.
\end{itemize}

We see from Fig.~\ref{fig_data} that our results with energy dependent decay widths are in good agreement with the SAPHIR 1998 and SAPHIR 2004 data~(which are consistent with each other) in the range 
$W=1.61-1.9$ GeV as well as with CLAS data in the range $W = 1.61-1.73$ GeV and $W=1.94-2.52$ GeV. Although, the results for energy dependent as well as fixed widths of the resonances are almost consistent with each other in the region of $W$ from threshold up to 1.9 GeV, in the region of $W$ from 1.9~GeV to 2.1~GeV, the present results obtined with fixed widths are consistent with CLAS 2006 data while in the region of $W=2.1$~GeV to 2.4~GeV, these results are consistent with SAPHIR 1998 and SAPHIR 2004 data. For the region $W=1.75 - 1.9$ GeV, our results show a prominent dip at $W \approx 1.8$ GeV, which is consistent with SAPHIR but 
not with CLAS. However, in the energy region of $W>1.9$~GeV, our results with energy dependent decay widths are in agreement with the CLAS data and are higher than both the SAPHIR data.
It must be pointed out that we have considered only those nucleon resonances which are well established and are 
present in PDG and have known branching ratios for decay in $K\Lambda$ mode. Therefore, the region of $W \approx 1.8$ GeV may 
indicate the existence of some other resonances which are yet to be observed experimentally. \\

\begin{figure}
\begin{center}
 \includegraphics[height=7cm,width=13cm]{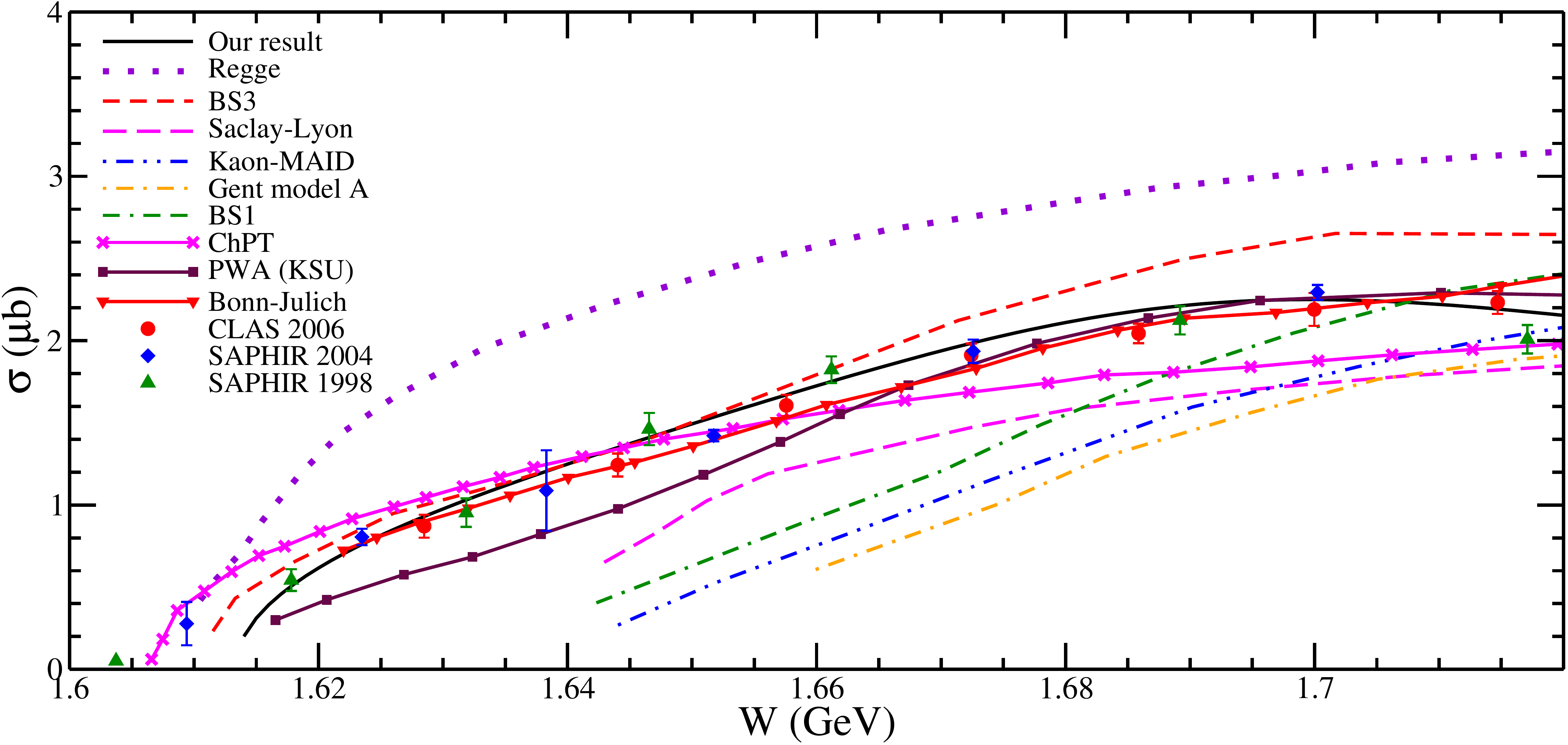}
\caption{$\sigma$ {\it vs.} $W$ for the process $ \gamma + p \rightarrow K^{+} + \Lambda$ in the threshold region. Solid line represents the results of 
the present work taking the $W$ dependent decay widths of the nucleon resonances, which are compared with other theoretical models like Regge model~\cite{Guidal:1997hy,  Guidal:1999qi}~(dotted line), 
BS3 model~\cite{Skoupil:2018vdh}~(short dashed line), Saclay-Lyon model~\cite{SL, SLA}~(long dashed line), Kaon-MAID 
model~\cite{KM2}~(double-dotted-dashed line), Ghent model A~\cite{Jan01A, Jan01B, Jan02}~(dashed-dotted line), BS1 
model~\cite{Skoupil:2016ast}~(double-dashed-dotted line), partial wave analysis~(PWA) available from Kent State University~(KSU)~\cite{Hunt:2018mrt}~(squares with solid line), chiral perturbation theory~(ChPT)~\cite{Chpt}~(cross with solid line) and Bonn-Julich model~\cite{Ronchen:2018ury}~(down triangle with solid line).}\label{fig1}
\end{center}
\end{figure}

When our results are compared with the experimental data, we observe that:
\begin{itemize}
 \item [i)] In the threshold region~(Fig.~\ref{fig_threshold}) before the first peak, our numerical results are in a very good 
 agreement with the experimental results available from the CLAS 2006~\cite{CLAS05}, SAPHIR 2004~\cite{SAPHIR} and SAPHIR 
 1998~\cite{Tran:1998qw}, where the data from all the three experiments are also in agreement among themselves. To fix the unknown parameters~($\Lambda_{B}$, $\Lambda_{R}$ and the couplings of the $t$ and $u$ channel resonances) in our model, we have performed a Chi-square fit with the results of the present model using the energy dependent decay widths and obtained the best $\chi^{2}/N_{d.o.f}$ to be 1.3.

 \item [ii)] Beyond the second peak region, there is a disagreement between the CLAS and the SAPHIR data and our numerical results both with energy dependent as well as with fixed decay widths are 
 consistent with CLAS data.
 
 \item [iii)] At high $W$ region, CLAS and SAPHIR data are in a reasonably agreement with each other in shape but not in the absolute values and the present 
 results in our model with energy dependent decay widths explain the CLAS data very well in shape as well as in absolute magnitude while the present results with fixed widths explain the SAPHIR data very well in shape as well as in absolute magnitude.  
 
 \item [iv)] As the CM energy increases from $W = 1.75$ to $W=1.9$ GeV, where CLAS and SAPHIR data are not consistent with 
 each other, the results of the present work are in good agreement with the SAPHIR data.
 
 \item [v)] We emphasize that the present model {with energy dependent decay widths} reproduces all the experimental results in the threshold region $W<1.75 
 $~(Figs.~\ref{fig_threshold} and \ref{fig1}) and the data from CLAS 2006 in the region 1.9 GeV $<W<$ 2.54 GeV.

\end{itemize}

\subsubsection{Comparison with other theoretical results}

\begin{figure}
\begin{center}
 \includegraphics[height=7cm,width=13cm]{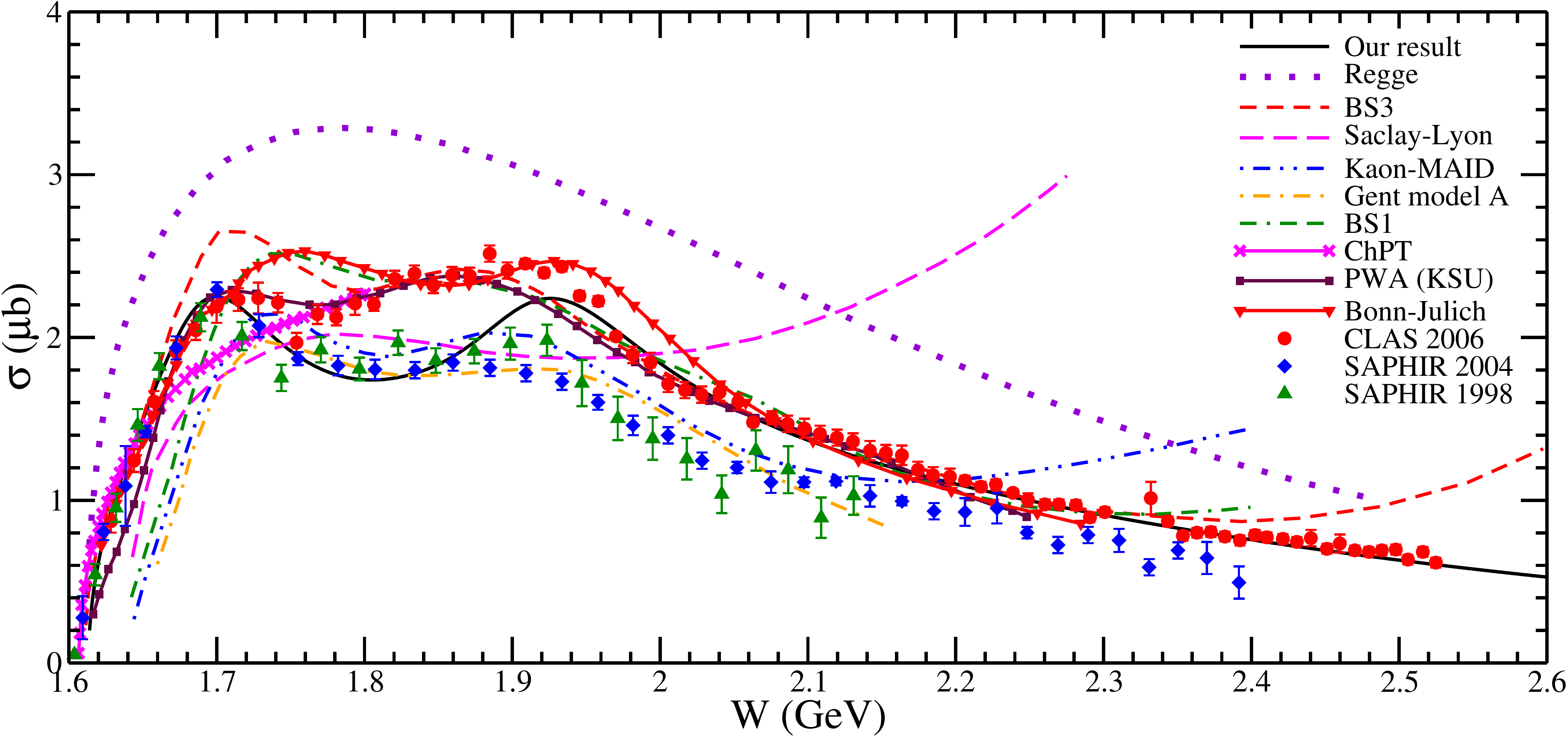}
\caption{$\sigma$ {\it vs.} $W$ for the process $ \gamma + p \rightarrow K^{+} + \Lambda$. Lines and points have the same meaning as in Fig.~\ref{fig1}}\label{fig_theory}
\end{center}
\end{figure}

To compare our results with some of the theoretical results available in the literature in Figs.~\ref{fig1} and \ref{fig_theory}, we have 
presented the results for $\sigma$ {\it vs.} $W$ for the process $ \gamma + p \rightarrow K^{+} + \Lambda$, where we have shown 
the results from the different models like Regge model~\cite{Guidal:1997hy,  Guidal:1999qi}, model based on chiral perturbation theory~\cite{Chpt}, BS3 model~\cite{Skoupil:2018vdh}, Saclay-Lyon model~\cite{SL, SLA}, Kaon-MAID model~\cite{KM2}, Ghent model 
A~\cite{Jan01A, Jan01B, Jan02}, BS1 model~\cite{Skoupil:2016ast}, Bonn-Julich model~\cite{Ronchen:2018ury} and model based on the partial wave analysis~\cite{Hunt:2018mrt}. For completeness, we have also shown the experimental data 
from CLAS 2006~\cite{CLAS05}, SAPHIR 2004~\cite{SAPHIR} and SAPHIR 1998~\cite{Tran:1998qw}. 

In Fig.~\ref{fig1}, we have compared the results with energy dependent decay widths of the present model with the other theoretical and experimental results available in the 
literature, in the threshold region. A very good agreement between the  numerical results obtained using the present model with the experimental results from CLAS 
2006~\cite{CLAS05}, SAPHIR 2004~\cite{SAPHIR} and SAPHIR 1998~\cite{Tran:1998qw} may be observed in the region of $W<1.72$ GeV. Other than the present model, the threshold region is explained only by the model based on the chiral perturbation theory~(ChPT) up to $W=1.66$~GeV and by the model based on the partial wave analysis, {\it i.e.}, the KSU model beyond $W=1.66$ GeV.
Moreover, the Bonn-Julich model explains the experimental data quite well in the threshold region.

In Fig.~\ref{fig_theory}, we have compared our results with the other theoretical and experimental results in the entire range of $W$.
It may be observed from Fig.~\ref{fig_theory} that the Regge model over predicts the 
experimental data at all values the of $W$ from 1.61 to 2.6 GeV. There is only one broad peak at $W\sim 1.8$ GeV in the Regge model and 
the model does not explain the experimental data from CLAS as well as from SAPHIR which show two peaks at $W=1.7$ and 1.9 GeV. 
The model based on the chiral perturbation theory explains the experimental data very well in the threshold region but the model is not applicable at 
high $W$. However, the model based on the partial wave analysis explains the experimental data available from the CLAS very well in the intermediate and high $W$ region, the threshold region~(Fig.~\ref{fig1}) is not well explained by this model. 
Except the Ghent model A, all the other models like Saclay-Lyon, Kaon-MAID, BS1 and BS3 show that, at high $W$, the 
cross section increases with $W$, which is not supported by the experimental data. However, the Ghent model A as well as the 
present work~(even at $W = 2.6$ GeV) do not show any such increase. As pointed out in Ref.~\refcite{Skoupil:2016ast}, this increase 
comes mainly from the background part of the amplitude. Also, this increase of the total cross section is model 
dependent. For example, the Saclay-Lyon~\cite{SL, SLA} model starts increasing from $W\sim2$ GeV while the Kaon-MAID model 
starts increasing beyond $W\sim 2.1$ GeV. In the BS1 and BS3 models, the cross section increases for $W>2.3$ GeV.

\begin{figure}
\begin{center}
 \includegraphics[height=0.9\textheight,width=0.95\textwidth]{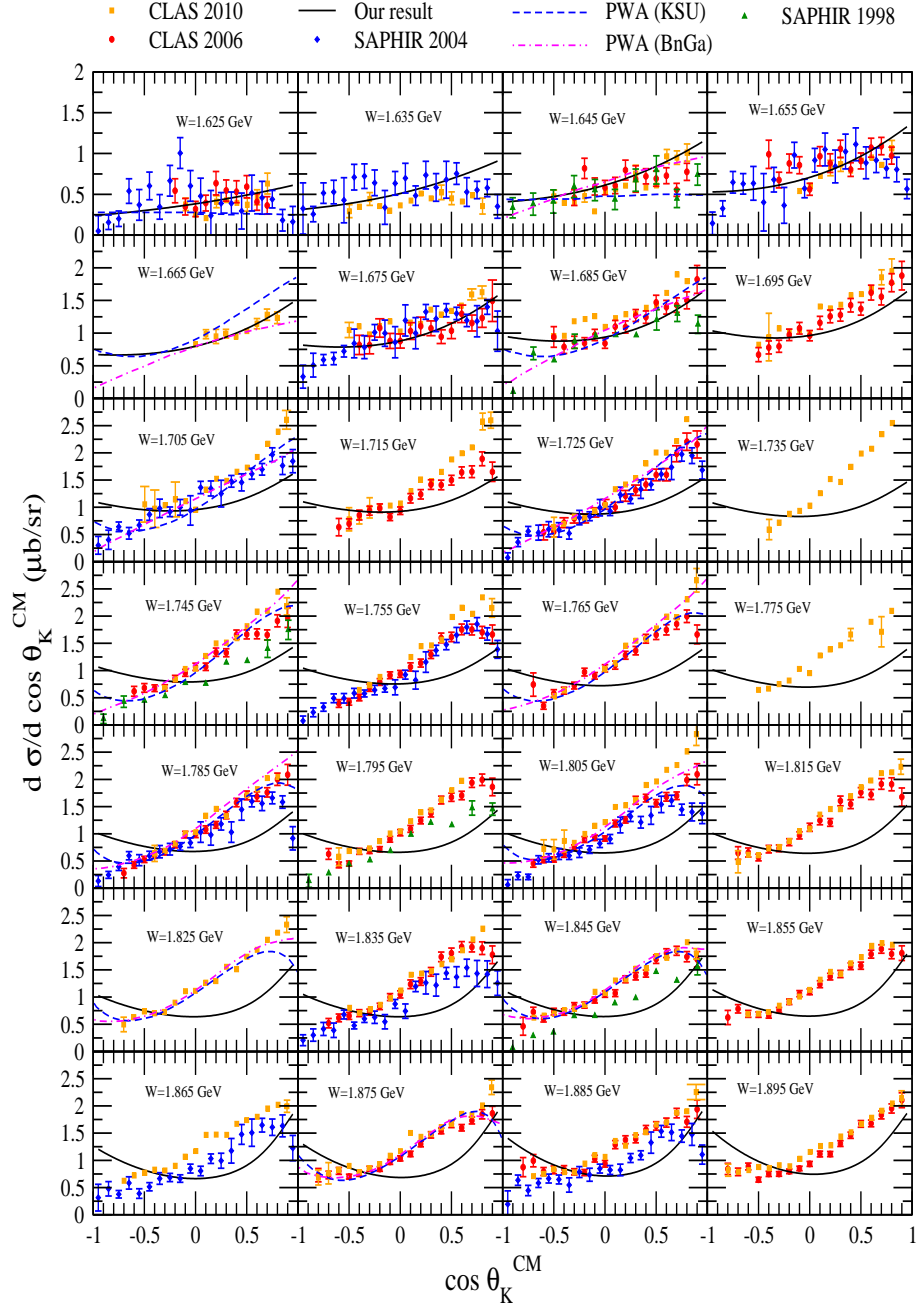}
\caption{$d\sigma/d \cos \theta_{k}^{CM}$ {\it vs.} $\cos \theta_{k}^{CM}$ at fixed $W$ ranging from $1.625-1.895$ GeV, for the 
process $\gamma + p \rightarrow K^{+} + \Lambda$. The experimental data has been taken from CLAS 2010~\cite{CLAS10}~(solid 
square), CLAS 2006~\cite{CLAS05}~(solid circle), SAPHIR 2004~\cite{SAPHIR}~(solid diamond) and SAPHIR 
1998~\cite{Tran:1998qw}~(solid triangle). Solid line represents the results of the full model taking $W$ dependent decay widths of the nucleon resonance. Dashed~(Dashed-dotted) line shows the results obtained using the partial wave analysis done by the Kent State University~(KSU)~(Bonn-Gatchina~(BnGa)) model~\cite{Hunt:2018mrt}. }\label{fig3}
\end{center}
\end{figure}

In comparison with other theoretical models, the success of the present model in describing the data except in the energy region 
of 1.75 GeV $<W<$ 1.9 GeV is due to the physical input parameters used in calculating the contribution from the resonance terms which 
dominate in this energy region. Moreover, the use of a small value of the cut-off parameter in the form factors in the calculation of the non-resonant Born and the contact terms as well as the $t$ and $u$ channel resonances suppresses the contribution of the background terms preventing the rise of the cross section in the low energy region as compared to other models. 
We would like 
to emphasize that the present model is very economical version of isobar models used in the literature as it uses a minimal number of resonances and highlights the importance of a few resonances like $S_{11} (1650)$, $P_{11} (1710)$, $P_{13} (1720)$, $P_{11} (1880)$, $S_{11} (1895)$ and $P_{13} (1900)$ in 
 explaining the total cross section data. The results of the present model are in agreement with many elaborate calculations available in literature~(Bonn-Gatchina~\cite{Anisovich:2014yza, Anisovich:2017bsk, Anisovich:2017ygb}, Bonn-Julich~\cite{Ronchen:2018ury}, KSU~\cite{Hunt:2018mrt}, Skoupil-Bydzovsky~\cite{Skoupil:2016ast, Skoupil:2018vdh,  Bydzovsky:2019hgn}, Mart~\cite{Mart:2019jtb,Mart:2019mtq, Mart:2017mwj, Mart:2017xtf}, Ghent model~\cite{GhentRPR07, GhentRPR071, GhentRPR072}).

\subsection{Differential cross section}\label{results_diff}
\begin{figure}
 \includegraphics[height=6cm,width=0.95\textwidth]{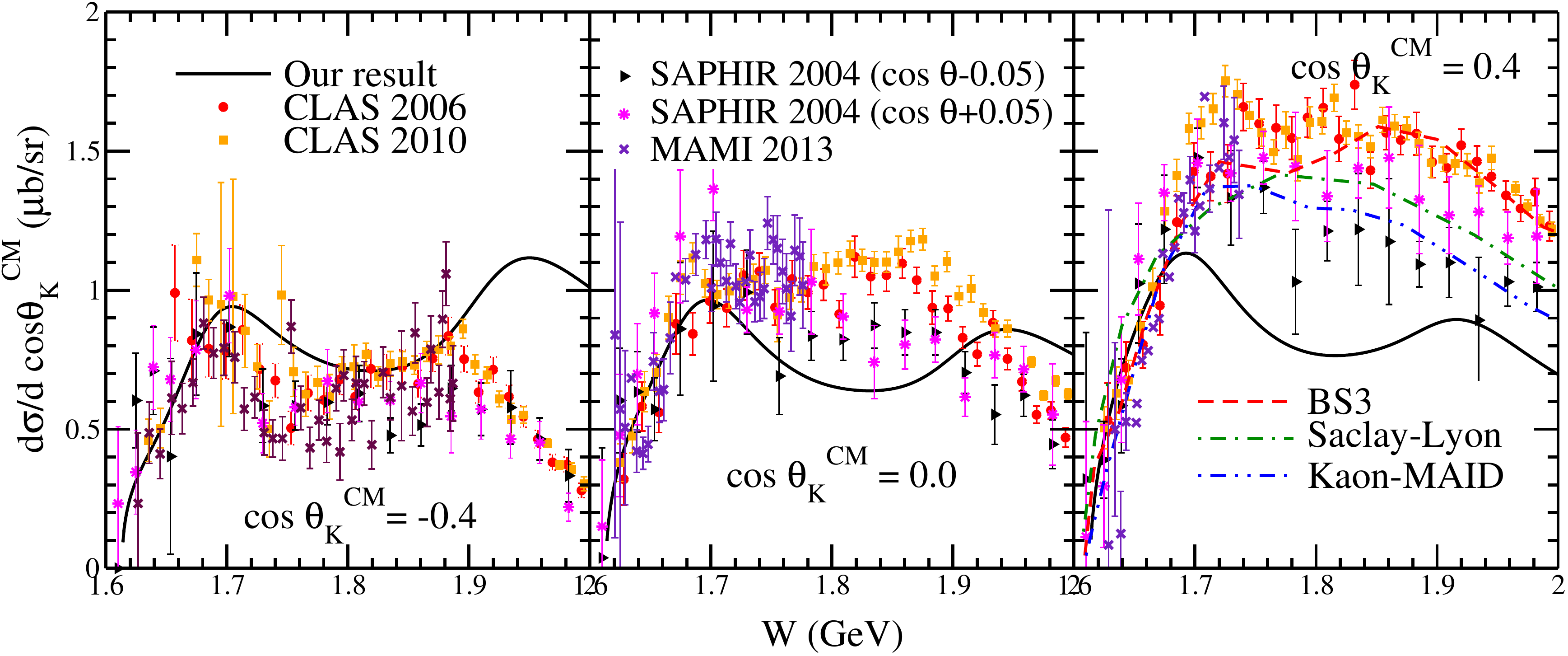}
 \caption{$d\sigma/d \cos \theta_{k}^{CM}$ {\it vs.} $W$ at fixed $\cos \theta_{k}^{CM}=-0.4,~0$ and 0.4, for the 
process $\gamma + p \rightarrow K^{+} + \Lambda$. The experimental data has been taken from CLAS 2010~\cite{CLAS10}~(solid 
square), CLAS 2004~\cite{CLAS05}~(solid circle), MAMI 2013~\cite{Jude:2013jzs}~(cross symbol), SAPHIR 2004 ($\cos \theta - 0.05$)~\cite{SAPHIR}~(right triangle) and SAPHIR 2004 ($\cos \theta + 0.05$)~(diamond). Solid line represents the results of the full model of the present work taking the energy dependent decay widths of the nucleon into account. Dashed line, dashed-dotted line, dashed-double-dotted line show the results of BS3~\cite{Skoupil:2018vdh}, Saclay-Lyon~\cite{SL, SLA} and Kaon-MAID~\cite{KM2} models, respectively. }\label{fig:energy_dep}
\end{figure}

We have presented our results for the differential cross sections and compared them with the experimental data 
available from CLAS, SAPHIR and MAMI as well as with the models based on the partial wave analysis~\cite{Hunt:2018mrt}, Saclay-Lyon model~\cite{SL, SLA}, Kaon-MAID model~\cite{KM2} and BS3 model~\cite{Skoupil:2018vdh}. 

In Fig.~\ref{fig3}, we have presented the results for $d\sigma/d \cos \theta_{k}^{CM}$ as a 
function of $\cos \theta_{k}^{CM}$ at fixed $W$ ranging from $1.625-1.895$ GeV in the interval of 10 MeV obtained using the energy dependent decay width of the various resonances considered in the present work, for the $K\Lambda$ 
photoproduction process. 
The present results are also compared with the experimental results available from 
CLAS~2010~\cite{CLAS10}, CLAS~2006~\cite{CLAS05}, SAPHIR~2004~\cite{SAPHIR} and SAPHIR~1998~\cite{Tran:1998qw} as well as with the theoretical results obtained by the Kent State University~(KSU)~\cite{Hunt:2018mrt} and the Bonn-Gatchina~(BnGa)~\cite{Hunt:2018mrt} groups, using the partial wave analysis.
In the low $W$ 
region, {\it i.e.}, from the threshold up to $W = 1.695 $~GeV, our results are in a good agreement with the available 
experimental data as well as with the models based on the partial wave analysis. In the intermediate and high region of $W$, our results in the backward region are fairly in agreement with KSU and BnGa models, however, this is not the case at the forward angle region.
In the region of $W$ from $1.705-1.895$~GeV, our results in the forward region are in agreement with SAPHIR 
data but not with CLAS data.  CLAS 2006 and CLAS 2010 data are not in agreement with each other and the results obtained using 
the model discussed in this work favor SAPHIR experimental observations.

To show the energy dependence of the differential cross section, in Fig.~\ref{fig:energy_dep}, 
we have presented our results for $d\sigma/d\Omega_{k}^{CM}$ {\it vs.} $W$ at $\cos \theta_{k}^{CM} =-0.4$, 0 and $+0.4$, obtained 
 using the fixed decay width of the resonances. We have compared our results with the experimental data available from SAPHIR 2004~\cite{SAPHIR}, CLAS 2006~\cite{CLAS05}, CLAS 2010~\cite{CLAS10} and MAMI 2013~\cite{Jude:2013jzs} as well as with the theoretical models like BS3 model~\cite{Skoupil:2018vdh}, Saclay-Lyon model~\cite{SL, SLA} and Kaon-MAID model~\cite{KM2}.
 It may be observed from the figure that in the threshold region, at all values of $\cos \theta_{k}^{CM}$, 
 our results are in a very good agreement with the available experimental data.
In the backward region, our results are in a reasonable agreement with the experimental data up to $W =1.9$~GeV. However, our results in the forward region emphasize the need for a missing resonance in order to explain the experimental data.
  
\section{Summary and conclusions}\label{summary}
In this work, we have presented a version of the isobar model based on the chiral SU(3) symmetry, to study the photoproduction 
of $K\Lambda$ from the proton. The results are presented for the total cross section as a function of CM energy $W$ 
and the differential cross sections for various values of $W$ in the region of few GeV of photon energy $E_{\gamma} 
\le 3$ GeV. The results are compared with the experimental data available from CLAS~\cite{CLAS05, CLAS10} and 
SAPHIR~\cite{SAPHIR, Tran:1998qw} and are found to be in a good agreement with the experimental data, except for the region 1.75 
GeV$<W< 1.9$ GeV. The results of the present model for the total cross sections are also compared with the results reported using various 
theoretical models available in the literature like Regge model~\cite{Guidal:1997hy, Guidal:1999qi}, chiral unitary model~\cite{Chpt}, 
Saclay-Lyon model~\cite{SL, SLA}, Kaon-MAID model~\cite{KM2}, Bonn-Julich model~\cite{Ronchen:2018ury}, Bonn-Gatchina model~\cite{Anisovich:2014yza, Anisovich:2017bsk, Anisovich:2017ygb}, KSU model~\cite{Hunt:2018mrt},  BS1~\cite{Skoupil:2016ast} and BS3~\cite{Skoupil:2018vdh} models.
 
In this model, the non-resonant terms are obtained using the non-linear $\sigma$ model in which the contact term appears quite 
naturally with the strength of its couplings predicted by the model. The different diagrams contributing to the non-resonant 
terms are the $s$, $t$ and $u$ channels and the contact term in which the various meson-nucleon-hyperon couplings {\it viz.} 
$g_{KN\Lambda}$ and $g_{KN\Sigma}$, are uniquely predicted in the model. The hadronic form factors at the strong vertices are 
introduced to account for the hadronic structure and a dipole parameterization is used, using a cut-off parameter $\Lambda_{B}$ 
taken to be the same for all the diagrams. For the contact term, the prescription of the form factor given by Davidson and 
Workman~\cite{DW} is used.

We have also considered nucleon, kaon and hyperon resonance exchanges in the $s$, $t$ and $u$ channels. The nucleon resonances 
with spin $\le \frac{3}{2}$ and mass in the range $1.6-1.9$ GeV, which are well established, represented by $****$ and $***$ 
states in the PDG having branching ratio in $K \Lambda$ are included. The electromagnetic couplings of the various nucleon 
resonances are obtained in terms of the experimental helicity amplitudes given in MAID~\cite{Tiator:2011pw} and PDG~\cite{PDG}. 
The strong couplings at $RK \Lambda$ vertices are obtained from the observed partial decay width of the resonance decaying to 
$K\Lambda$. The unitary corrections are partially implemented by using the energy dependent widths for the various nucleon resonances~\cite{Bydzovsky:2019hgn}.
The kaon resonances {\it viz.} $K^{*}$ and $K_{1}$ having spin 1, are considered in the $t$ channel and spin 
$\frac{1}{2}$ lambda resonances {\it viz.} $\Lambda^{*}(1405)$ and $\Lambda^{*}(1800)$ are considered in the $u$ channel. Since 
the experimental information about the kaon and hyperon resonances is not adequate to determine its electromagnetic and strong 
couplings phenomenologically, therefore, the couplings for these resonances are fitted to reproduce the experimental data 
available from CLAS and SAPHIR. 
\vspace{2mm}
 
We summarize the results of the present study in the following:
\begin{itemize}
 \item [(i)] Our results explain very well the threshold region and the role of the non-resonant terms are quite significant in 
 the threshold region up to $W \sim 1.75$ GeV. Moreover, the contact term, which occurs quite naturally in our model, has the 
 most dominant contribution and plays important role in explaining the data in this region.
 
 \item [(ii)] The results of the full model for the total cross section are in good agreement with the experimental 
 data of SAPHIR as well as CLAS experiments, in a wide region of $W$ considered in this work, {\it i.e.,} from the threshold up 
 to 1.75 GeV and from 1.9 to 2.6 GeV, except for a narrow range of $W$ {\it i.e.} 1.75--1.9 GeV.
 
 \item [(iii)] Among the resonance contribution, at low $W$, the contribution of $P_{11}(1710)$ is found to be significant 
 although it is not considered in most of the isobar models. We find that both  $P_{13} (1720)$ and $P_{13} (1900)$ resonances make  significant contributions to the cross section for $W \gtrsim 
 1.8$ GeV.   
The $S_{11}$ 
 resonances {\it viz.} $S_{11} (1650)$ and $S_{11} (1895)$ have small contribution to the cross section in the entire range of $W$.
 
 \item[(iv)] We would like to emphasise the important role of $P_{13}(1720)$ and $P_{13}(1900)$ resonances at higher energies specially in the region of $W>2$~GeV. When these resonances are taken into 
 account, the results are closer to CLAS data. The contribution of  both resonances are almost equally important in the entire region of $W$ considered in this paper. However, in the region of 1.8 GeV $<W<2.0$~GeV, the $P_{13}(1900)$ resonance gives a significant enhancement in the total cross section leading to the appearance of second peak around $W=1.9$~GeV seen in the data of CLAS 2006 and SAPHIR 1998.

 \item [(v)] Our results for the angular distribution are in fair agreement with the experimental data especially in the 
 threshold region.
 
 \item [(vi)] When we take an energy dependent width in order to restore the unitarity partially, it has been found that
 the effect (energy independent {\it vs.} energy dependent width) on the results of the total cross section
 is generally small but could be up to $5-20\%$ in the region of $W>2$ GeV leading to a better agreement with the CLAS 2006 data.
 
\end{itemize}

The present study of the $K\Lambda$ production induced by photons may be quite useful in the planned experiments at TJNAF, 
SPring-8 and MAMI in this energy region. In future, we plan to extend this model to study the electromagnetic and weak productions of 
associated particles~($KY$) induced by electrons and (anti)neutrinos relevant for future experiments at ESRF, MAMI, ELSA and 
TJNAF in the case of electrons and at MINERvA, NOvA, T2KK and DUNE in the case of (anti)neutrinos.

\section*{Acknowledgements}
We acknowledge useful discussions with I. Ruiz Simo. Z. Ahmad Dar, M. Sajjad Athar and S. K. Singh are thankful to Department 
of Science and Technology (DST), Government of India for providing financial assistance under Grant No. EMR/2016/002285.


%
%
%
%
%
%


\begin{thebibliography}{0}
\bibitem{Kawaguchi}
M.~Kawaguchi and M.~J.~Moravcsik, 
Phys. Rev. {\bf 107}, 563 (1957).





\bibitem{McDaniel:1959zz} 
  B.~D.~McDaniel, A.~Silverman, R.~R.~Wilson and G.~Cortellessa,
    Phys.\ Rev.\  {\bf 115}, 1039 (1959).
    




\bibitem{McDaniel:1959zz1} 
    A. Silverman, R. R. Wilson and W. M. Woodward, 
    Phys.\ Rev.\  {\bf 108}, 501 (1957).
    




\bibitem{Donoho:1958zz} 
  P.~L.~Donoho and R.~L.~Walker,
    Phys.\ Rev.\  {\bf 112}, 981 (1958).
    




\bibitem{Capps} 
R. H. Capps, 
Phys. Rev. {\bf 114}, 920 (1959).





\bibitem{Groom} 
D.~E.~Groom and J.~H.~Marshall, 
Phys. Rev. {\bf 159}, 1213 (1967).





\bibitem{ABBHHM}
 R. Erbe {\it et al.} (ABBHHM Collaboration), Phys. Rev. {\bf 188}, 2060 (1969); data tabulated in Numerical Data and Functional 
 Relationships in Science and Technology, edited by H. Schopper, Landolt-B$\ddot{\text{o}}$rnstein, New Series I/12b 
 (Springer-Verlag, New York, 1988), p. 355.





\bibitem{Tran:1998qw} 
  M.~Q.~Tran {\it et al.} [SAPHIR Collaboration],
    Phys.\ Lett.\ B {\bf 445}, 20 (1998).





\bibitem{SAPHIR} K.-H.~Glander \emph{et al.}, 
               Eur. Phys. J. A {\bf 19}, 251 (2004).                             





\bibitem{CLAS05} R.~Bradford \emph{et al.}, 
               Phys. Rev. C \textbf{73} 035202 (2006).

               




\bibitem{CLAS051}
               J.~W.~C.~McNabb {\it et al.}, Phys. Rev. C {\bf 69}, 042201(R) (2004).





\bibitem{CLAS10} M.~E.~McCracken \emph{et al.}, 
               Phys. Rev. C \textbf{81} 025201 (2010).





\bibitem{Sumihama:2005er} 
  M.~Sumihama {\it et al.} [LEPS Collaboration],
    Phys.\ Rev.\ C {\bf 73}, 035214 (2006).





\bibitem{Lleres:2007tx} A.~Lleres \emph{et al.}, Eur. Phys. J. A {\bf 31}, 79 (2007).





\bibitem{Lleres:2007tx1}
               A.~Lleres \emph{et al.}, Eur. Phys. J. A {\bf 39}, 149 (2009). 





\bibitem{Jude:2013jzs} 
  T.~C.~Jude {\it et al.} [Crystal Ball at MAMI Collaboration],
    Phys.\ Lett.\ B {\bf 735}, 112 (2014).





\bibitem{Tiator:2011pw} 
  L.~Tiator, D.~Drechsel, S.~S.~Kamalov and M.~Vanderhaeghen,
    Eur.\ Phys.\ J.\ ST {\bf 198}, 141 (2011).





\bibitem{ZPA}
G. Knochlein, D. Drechsel and L. Tiator, 
Z. für Physik A, {\bf 352}, 327 (1995).




\bibitem{Tiator:2018heh}
L.~Tiator, M.~Gorchtein, V.~Kashevarov, K.~Nikonov, M.~Ostrick, M.~Hadžimehmedović, R.~Omerović, H.~Osmanović, J.~Stahov and A.~Švarc,
Eur. Phys. J. A \textbf{54}, no.12, 210 (2018).




\bibitem{Chiang:2002vq}
W.~T.~Chiang, S.~N.~Yang, L.~Tiator, M.~Vanderhaeghen and D.~Drechsel,
Phys. Rev. C \textbf{68}, 045202 (2003).




\bibitem{Mart:2017xtf} 
  T.~Mart and A.~Rusli,
    PTEP {\bf 2017}, 123D04 (2017).
    




\bibitem{Skoupil:2016ast} 
  D.~Skoupil and P.~Bydzovsky,
    Phys.\ Rev.\ C {\bf 93}, 025204 (2016).
    




\bibitem{Skoupil:2018vdh}   D.~Skoupil and P.~Bydzovsky,
    Phys.\ Rev.\ C {\bf 97},  025202 (2018).





\bibitem{Jan01A} S.~Janssen, J.~Ryckebusch, D.~Debruyne, and 
             T.~Van Cauteren, Phys. Rev. C {\bf 65}, 015201 (2001).





\bibitem{MartSu} T.~Mart and A.~Sulaksono, 
           Phys. Rev. C {\bf 74}, 055203 (2006). 





\bibitem{Mart:2017mwj}
T.~Mart and S.~Sakinah,
Phys. Rev. C \textbf{95},  045205 (2017).





\bibitem{Mart:2019mtq}
T.~Mart,
Phys. Rev. D \textbf{100},  056008 (2019).





\bibitem{CapRob00} S.~Capstick and W.~Roberts, 
                  Prog. Part. Nucl. Phys. {\bf 45}, S241 (2000).





\bibitem{Bonn} U.~Loring, B.~C.~Metsch, and H.~R.~Petry, 
              Eur. Phys. J. A {\bf 10}, 395 (2001).
              




\bibitem{Bonn1}  
U.~Loring, B.~C.~Metsch, and H.~R.~Petry, 
              Eur. Phys. J. A {\bf 10}, 447 (2001).





\bibitem{Ambrozewicz:2006zj} 
  P.~Ambrozewicz {\it et al.} [CLAS Collaboration],
    Phys.\ Rev.\ C {\bf 75}, 045203 (2007).
    




\bibitem{Gabrielyan:2014zun} 
  M.~Gabrielyan {\it et al.} [CLAS Collaboration],
    Phys.\ Rev.\ C {\bf 90}, 035202 (2014).





\bibitem{Carman:2012qj} 
  D.~S.~Carman {\it et al.} [CLAS Collaboration],
    Phys.\ Rev.\ C {\bf 87}, 025204 (2013).
    




\bibitem{Achenbach:2011rf} 
  P.~Achenbach {\it et al.} [A1 Collaboration],
   Eur.\ Phys.\ J.\ A {\bf 48}, 14 (2012).    





\bibitem{Gogami:2013dua} 
  T.~Gogami {\it et al.},
    Few Body Syst.\  {\bf 54}, 1227 (2013).





\bibitem{SL} J.~C.~David, C.~Fayard, G.-H.~Lamot, and B.~Saghai,
               Phys. Rev. C {\bf 53}, 2613 (1996).





\bibitem{Williams:1992tp} 
  R.~A.~Williams, C.~R.~Ji and S.~R.~Cotanch,
    Phys.\ Rev.\ C {\bf 46}, 1617 (1992).
    




\bibitem{SLA} T.~Mizutani, C.~Fayard, G.-H.~Lamot, and B.~Saghai, 
               Phys. Rev. C {\bf 58}, 75 (1998).





\bibitem{Mart:2010ch} 
  T.~Mart,
    Phys.\ Rev.\ C {\bf 82}, 025209 (2010).





\bibitem{Janssen:2003kk} 
  S.~Janssen, J.~Ryckebusch and T.~Van Cauteren,
    Phys.\ Rev.\ C {\bf 67}, 052201 (2003).





\bibitem{Solomey:2005rs} 
  N.~Solomey [MINERvA Collaboration],
    Nucl.\ Phys.\ Proc.\ Suppl.\  {\bf 142}, 74 (2005).





\bibitem{MINERvA:2006aa} 
  [MINERvA Collaboration], FERMILAB-DESIGN-2006-01, MINERVA-DOCUMENT-700.





\bibitem{Shrock:1975an} 
  R.~E.~Shrock,
    Phys.\ Rev.\ D {\bf 12}, 2049 (1975).





\bibitem{Mechlenburg:1976} W. Mechlenburg. Acta Phys. Aust. {\bf 48}, (1976).





\bibitem{Amer:1977fy} 
  A.~A.~Amer,
    Phys.\ Rev.\ D {\bf 18}, 2290 (1978).





\bibitem{Dewan:1981ab} 
  H.~K.~Dewan,
    Phys.\ Rev.\ D {\bf 24}, 2369 (1981).





\bibitem{Adera:2010zz} 
  G.~B.~Adera, B.~I.~S.~Van Der Ventel, D.~D.~van Niekerk and T.~Mart,
    Phys.\ Rev.\ C {\bf 82}, 025501 (2010).





\bibitem{ZpL95}Zhenping Li, Phys. Rev. C {\bf 52}, 1648 (1995).





\bibitem{ZpL951}
              Zhenping Li, Hongxing Ye, Minghui Lu, 
              Phys. Rev. C {\bf 56}, 1099 (1997).





\bibitem{LLP95}D.~Lu, R.~H.~Landau, and S.~C.~Phatak, 
              Phys. Rev. C {\bf 52}, 1662 (1995).              





\bibitem{FHZ91}G.~R.~Farrar, K.~Huleihel, and H.~Zhang, 
              Nucl. Phys. B {\bf 349}, 655 (1991).





\bibitem{JuliaDiaz:2005qj} 
  B.~Julia-Diaz {\it et al.},
    Nucl.\ Phys.\ A {\bf 755}, 463 (2005).





\bibitem{Chpt} S.~Steininger and U.-G.~Meissner, Phys. Lett. B
              {\bf 391}, 446 (1997).  





\bibitem{Borasoy} B.~Borasoy, P.~C.~Bruns, U.-G.~Meissner, and R.~Nissler, 
                Eur. Phys. J. A {\bf 34}, 161 (2007).
                




\bibitem{Giessen} V.~Shklyar, H.~Lenske, and U.~Mosel, 
                Phys. Rev. C {\bf 72}, 015210 (2005).
                




\bibitem{Giessen1}                
                R.~Shyam, O.~Scholten, and H.~Lenske, 
                Phys. Rev. C {\bf 81}, 015204 (2010).





\bibitem{JDiaz} B.~Julia-Diaz, B.~Saghai, T.-S.~H.~Lee, and F.~Tabakin, 
               Phys. Rev. C {\bf 73}, 055204 (2006).





\bibitem{Anisovich:2007bq} A.~V.~Anisovich {\it et al.}, 
              Eur. Phys. J. A {\bf 34}, 243 (2007).





\bibitem{JuliaDiaz:2005ju} 
  B.~Julia-Diaz, B.~Saghai, T.-S.~H.~Lee and F.~Tabakin,
    nucl-th/0512010.





\bibitem{Thom} H.~Thom, Phys. Rev. {\bf 151}, 1322 (1966).


               




\bibitem{Mart:1999ed} 
 T.~Mart and C.~Bennhold,
   Phys.\ Rev.\ C {\bf 61}, 012201 (2000).





\bibitem{Mart:2000jv} 
 T.~Mart,
   Phys.\ Rev.\ C {\bf 62}, 038201 (2000).





\bibitem{KM2}               
               T.~Mart, C.~Bennhold, H.~Haberzettl, and L.~Tiator,
               http://www.kph.uni-mainz.de/MAID/kaon/kaonmaid.html. 





\bibitem{Jan01B} S.~Janssen {\it et al.}, Eur. Phys. J. A {\bf 11}, 105 (2001).





\bibitem{Jan02} S.~Janssen, D.~G.~Ireland, J.~Ryckebusch, 
               Phys. Lett. B {\bf 562}, 51 (2003).         





\bibitem{SF} M.~Sotona and S.~Frullani, Prog.\ Theor.\ Phys.\ Suppl.\
        {\bf 117}, 151 (1994).





\bibitem{PSPV} S.~S.~Hsiao, D.~H.~Lu, and S.~N.~Yang, 
       Phys. Rev. C {\bf 61}, 068201 (2000).
       




\bibitem{PSPV1}       
       B.~Han, M.~Cheoun, K.~Kim, and I.-T.~Cheon, 
       Nucl. Phys. A {\bf 691}, 713 (2001). 





\bibitem{Maxwell} O.~V.~Maxwell, Phys. Rev. C {\bf 76}, 014621 (2007).





\bibitem{Maxwell1}
          A.~de la Puente, O.~V.~Maxwell, and B.~A.~Raue, 
          Phys. Rev. C {\bf 80}, 065205 (2009).     





\bibitem{Adelseck:1986fb} 
  R.~A.~Adelseck, C.~Bennhold and L.~E.~Wright,
    Phys.\ Rev.\ C {\bf 32}, 1681 (1985).





\bibitem{Adelseck:1989zt} 
  R.~A.~Adelseck and L.~E.~Wright,
    Phys.\ Rev.\ C {\bf 39}, 580 (1989).





\bibitem{Adelseck:1990ch} 
  R.~A.~Adelseck and B.~Saghai,
    Phys.\ Rev.\ C {\bf 42}, 108 (1990).





\bibitem{Saghai:1994mm} 
  B.~Saghai and F.~Tabakin,
    AIP Conf.\ Proc.\  {\bf 339}, 521 (1995).





\bibitem{Williams:1991tw} 
  R.~A.~Williams, C.~R.~Ji and S.~R.~Cotanch,
    Phys.\ Rev.\ C {\bf 43}, 452 (1991).





\bibitem{Williams:1990hh} 
  R. A.~Williams, C.~R.~Ji and S.~R.~Cotanch,
    Phys.\ Rev.\ D {\bf 41}, 1449 (1990).





\bibitem{GhentRPR07} T.~Corthals, J.~Ryckebusch, and T.~Van Cauteren, 
        Phys. Rev. C {\bf 73}, 045207 (2006).
        




\bibitem{GhentRPR071}        
         T.~Corthals, T.~Van Cauteren, J.~Ryckebusch, and  
        D.~G.~Ireland, Phys. Rev. C {\bf 75}, 045204 (2007).
        




\bibitem{GhentRPR072}        
        T.~Corthals {\it et al.}, 
        Phys. Lett. {\bf B656}, 186 (2007). 






\bibitem{Bydzovsky:2019hgn} 
  P.~Bydzovsky and D.~Skoupil,
 Phys. Rev. C {\bf 100}, 035202 (2019).





\bibitem{Guidal:1997hy} 
  M.~Guidal, J.~M.~Laget and M.~Vanderhaeghen,
    Nucl.\ Phys.\ A {\bf 627}, 645 (1997).





\bibitem{Adler:1968}
    S. L. Adler and R. F. Dashen, Current Algebras and applications to particle physics,
    W. A. Benjamin (1968).





\bibitem{Sakinah:2019zbd} 
  S.~Sakinah, S.~Clymton and T.~Mart,
    Acta Phys.\ Polon.\ B {\bf 50}, 1389 (2019).
  




\bibitem{Guidal:1999qi} 
  M.~Guidal, J.~M.~Laget and M.~Vanderhaeghen,
    Phys.\ Rev.\ C {\bf 61}, 025204 (2000).





\bibitem{RPR11} L.~De Cruz, T.~Vrancx, P.~Vancraeyveld, and J.~Ryckebusch, 
         Phys. Rev. Lett. {\bf 108} 182002 (2012).
         




\bibitem{RPR111}         
         L.~De Cruz, J.~Ryckebusch, T.~Vrancx, P.~Vancraeyveld, 
         Phys. Rev. C {\bf 86}, 015212 (2012).





\bibitem{DW} R.~M.~Davidson and R.~Workman, 
                Phys. Rev. C \textbf{63}, 025210 (2001).
                




\bibitem{Hernandez:2007qq} 
  E.~Hernandez, J.~Nieves and M.~Valverde,
    Phys.\ Rev.\ D {\bf 76}, 033005 (2007).





\bibitem{Hernandez:2013jka} 
E.~Hernandez, J.~Nieves and M.~J.~Vicente Vacas,
Phys.\ Rev.\ D {\bf 87}, 113009 (2013).





\bibitem{Alam:2015gaa}
  M.~Rafi Alam, M.~Sajjad Athar, S.~Chauhan and S.~K.~Singh,
    Int.\ J.\ Mod.\ Phys.\ E {\bf 25}, 1650010 (2016).          





\bibitem{RafiAlam:2010kf}
 M.~Rafi Alam, I.~Ruiz Simo, M.~Sajjad Athar and M.~J.~Vicente Vacas,
   Phys.\ Rev.\ D {\bf 82}, 033001 (2010).





\bibitem{Alam:2012ry}
 M.~Rafi Alam, I.~Ruiz Simo, M.~Sajjad Athar and M.~J.~Vicente Vacas,
   Phys.\ Rev.\ D {\bf 87}, 053008 (2013).





\bibitem{Alam:2011xq}
 M.~Rafi~Alam, I.~Ruiz~Simo, M.~Sajjad~Athar and M.~J.~Vicente Vacas,
   Phys.\ Rev.\ D {\bf 85}, 013014 (2012).





\bibitem{Alam:2015zla} 
  M.~Rafi Alam {\it et al.}, 
    AIP Conf.\ Proc.\  {\bf 1680}, 020001 (2015).





\bibitem{Bennhold:1998ib} 
  C.~Bennhold {\it et al.}, nucl-th/9901066.






\bibitem{Bennhold:1996} C. Bennhold, T. Mart, and D. Kusno, Proceedings of the CEBAF/INT Workshop on N∗Physics, Seattle, USA, 1996 (World Scientific, Singapore, 1997, T.-
S. H. Lee and W. Roberts, editors), p.166.





\bibitem{Ohta:1989ji} 
  K.~Ohta,
    Phys.\ Rev.\ C {\bf 40}, 1335 (1989).





\bibitem{Haberzettl:1997jg} 
  H.~Haberzettl,
  Phys.\ Rev.\ C {\bf 56}, 2041 (1997).





\bibitem{Benmerrouche:1989uc} 
  M.~Benmerrouche, R.~M.~Davidson and N.~C.~Mukhopadhyay,
  Phys.\ Rev.\ C {\bf 39}, 2339 (1989).





\bibitem{Kamano:2016bgm}
  H.~Kamano, S.~X.~Nakamura, T.~S.~H.~Lee and T.~Sato,
    Phys.\ Rev.\ C {\bf 94},  015201 (2016).





\bibitem{Kamano:2009im}
  H.~Kamano, B.~Julia-Diaz, T.-S.~H.~Lee, A.~Matsuyama and T.~Sato,
    Phys.\ Rev.\ C {\bf 80}, 065203 (2009).





\bibitem{JuliaDiaz:2009ww}
  B.~Julia-Diaz, H.~Kamano, T.-S.~H.~Lee, A.~Matsuyama, T.~Sato and N.~Suzuki,
    Phys.\ Rev.\ C {\bf 80}, 025207 (2009).





\bibitem{Chiang:2004ye}
  W.~T.~Chiang, B.~Saghai, F.~Tabakin and T.~S.~H.~Lee,
    Phys.\ Rev.\ C {\bf 69}, 065208 (2004).





\bibitem{Kamano:2012id}
  H.~Kamano, S.~X.~Nakamura, T.-S.~H.~Lee and T.~Sato,
    Phys.\ Rev.\ D {\bf 86}, 097503 (2012).





\bibitem{Nakamura:2015rta}
  S.~X.~Nakamura, H.~Kamano and T.~Sato,
    Phys.\ Rev.\ D {\bf 92},  074024 (2015).
  




\bibitem{Watson:1952ji}
  K.~M.~Watson,
    Phys.\ Rev.\  {\bf 88}, 1163 (1952).





\bibitem{Sato:2003rq}
  T.~Sato, D.~Uno and T.~S.~H.~Lee,
    Phys.\ Rev.\ C {\bf 67}, 065201 (2003)





\bibitem{Alvarez-Ruso:2015eva}
  L.~Alvarez-Ruso, E.~Hernández, J.~Nieves and M.~J.~Vicente Vacas,
    Phys.\ Rev.\ D {\bf 93},  014016 (2016).





\bibitem{Anisovich:2012ct} 
  A.~V.~Anisovich {\it et al.},
    Eur.\ Phys.\ J.\ A {\bf 48}, 88 (2012),   https://pwa.hiskp.uni-bonn.de/





\bibitem{Anisovich:2014yza}
A.~Anisovich {\it et al.},
Eur. Phys. J. A \textbf{50}, 129 (2014).





\bibitem{Anisovich:2017bsk}
A.~Anisovich {\it et al.},
Phys. Rev. Lett. \textbf{119}, 062004 (2017).





\bibitem{Anisovich:2017ygb}
A.~Anisovich {\it et al.},
Eur. Phys. J. A \textbf{53}, 242 (2017).





\bibitem{Nikonov:2007br}
V.~Nikonov {\it et al.},
Phys. Lett. B \textbf{662}, 245 (2008).





\bibitem{Ronchen:2018ury}
D.~Ronchen, M.~Doring and U.~Meissner,
Eur. Phys. J. A \textbf{54}, 110 (2018).





\bibitem{Wang:2017cfp} 
  K.~L.~Wang, L.~Y.~Xiao and X.~H.~Zhong,
    Phys.\ Rev.\ C {\bf 95},  055204 (2017).





\bibitem{Hunt:2018mrt}
B.~Hunt and D.~Manley,
Phys. Rev. C \textbf{99}, 055204 (2019).





\bibitem{PDG} 
 M.~Tanabashi {\it et al.} [Particle Data Group],
   Phys.\ Rev.\ D {\bf 98}, 030001 (2018).





\bibitem{Scherer:2002tk}
S.~Scherer, 
{Adv. Nucl. Phys.} {\bf 27\/}, 277 (2003).





\bibitem{Scherer:2012xha}
S.~Scherer and M.~R. Schindler, 
  {Lect. Notes Phys.} {\bf 830\/}, 1 (2012).





\bibitem{Koch:1995vp} 
  V.~Koch,
    nucl-th/9512029.





\bibitem{Koch:1997ei} 
  V.~Koch,
    Int.\ J.\ Mod.\ Phys.\ E {\bf 6}, 203 (1997).





\bibitem{Kubis:2007iy}
B.~Kubis, 
{arXiv:hep-ph/0703274}.





\bibitem{Cabibbo:2003cu}
N.~Cabibbo, E.~C. Swallow, and R.~Winston, 
  {Ann. Rev. Nucl. Part. Sci.} {\bf 53\/}, 39 (2003).





\bibitem{Haberzettl:1998eq} 
  H.~Haberzettl, C.~Bennhold, T.~Mart and T.~Feuster,
    Phys.\ Rev.\ C {\bf 58}, R40 (1998).





\bibitem{Leitner:2008ue} 
  T.~Leitner, O.~Buss, L.~Alvarez-Ruso and U.~Mosel,
    Phys.\ Rev.\ C {\bf 79}, 034601 (2009).





\bibitem{Gonzalez-Jimenez:2016qqq} 
R.~González-Jiménez {\it et al.},
Phys.\ Rev.\ D {\bf 95}, 113007 (2017).





\bibitem{Rarita:1941mf}
W.~Rarita and J.~Schwinger,
Phys. Rev. \textbf{60}, 61 (1941).





\bibitem{Fierz}
M. Fierz and W. E. Pauli,  Proceedings of the Royal Society of London Series A Mathematical and Physical Sciences {\bf 173}, 211 (1939).





\bibitem{Pascalutsa:1999zz}
V.~Pascalutsa and R.~Timmermans,
Phys. Rev. C \textbf{60}, 042201 (1999).





\bibitem{Mart:2019jtb}
T.~Mart, J.~Kristiano and S.~Clymton,
Phys. Rev. C \textbf{100},  035207 (2019).





\bibitem{Vrancx:2011qv}
T.~Vrancx, L.~De Cruz, J.~Ryckebusch and P.~Vancraeyveld,
Phys. Rev. C \textbf{84}, 045201 (2011).





\bibitem{SajjadAthar:2007gb}
M.~Sajjad Athar, S.~Ahmad and S. K.~Singh,
Nucl. Phys. A \textbf{782}, 179 (2007).





\bibitem{Singh:1998ha}
S. K.~Singh, M.~Vicente-Vacas and E.~Oset,
Phys. Lett. B \textbf{416}, 23 (1998).





\bibitem{SajjadAthar:2009rc}
M.~Sajjad Athar, S.~Chauhan and S. K.~Singh,
J. Phys. G \textbf{37}, 015005 (2010).





\bibitem{Alam:2013vwa}
M.~Rafi Alam, M.~Sajjad Athar, L.~Alvarez-Ruso, I.~Ruiz Simo, M.~J.~Vicente Vacas and S. K.~Singh,
[arXiv:1311.2293 [hep-ph]].




\bibitem{Alam:2013xoa}
M.~Rafi Alam, L.~Alvarez-Ruso, M.~Sajjad Athar and M.~Vicente Vacas,
AIP Conf. Proc. \textbf{1663}, 120014 (2015).





\bibitem{Athar:2007wd}
M. Sajjad~Athar, S.~Ahmad and S. K.~Singh,
Phys. Rev. D \textbf{75}, 093003 (2007).





\bibitem{Ahmad:2006cy}
S.~Ahmad, M.~Sajjad Athar and S. K.~Singh,
Phys. Rev. D \textbf{74}, 073008 (2006).





\bibitem{AlvarezRuso:1999dy}
L.~Alvarez-Ruso, E.~Oset, S.~K. Singh and M.~Vicente-Vacas,
Nucl. Phys. A \textbf{663}, 837 (2000).





\bibitem{Valverde:2008jj}
M.~Valverde, J.~Nieves, E.~Hernandez, S.~K. Singh and M.~Vicente Vacas,
Mod. Phys. Lett. A \textbf{23}, 2309 (2008).





\bibitem{LlewellynSmith:1971uhs}
  C.~H.~Llewellyn Smith,
    Phys.\ Rept.\  {\bf 3}, 261 (1972).
  
  


\bibitem{Fogli:1979cz}
G.~L.~Fogli and G.~Nardulli,
Nucl. Phys. B \textbf{160}, 116-150 (1979).



\bibitem{Feuster:1998cj} 
  T.~Feuster and U.~Mosel,
    Phys.\ Rev.\ C {\bf 59}, 460 (1999).





\bibitem{Benmerrouche:1994uc} 
  M.~Benmerrouche, N.~C.~Mukhopadhyay and J.~F.~Zhang,
    Phys.\ Rev.\ D {\bf 51}, 3237 (1995).





\bibitem{Feuster:1996ww} 
  T.~Feuster and U.~Mosel,
    Nucl.\ Phys.\ A {\bf 612}, 375 (1997).
    

\bibitem{Suciawo:2017wtv}
W.~Suciawo, S.~Clymton and T.~Mart,
J. Phys. Conf. Ser. \textbf{856}, 012011 (2017).




\end{thebibliography}
\end{document}